\documentclass[journal]{IEEEtran}
\usepackage{amsmath,amssymb}
\usepackage[dvips]{graphicx}
\usepackage{amsfonts}
\usepackage[mathscr]{eucal}

\usepackage{latexsym}
\usepackage{amsthm}
\usepackage{exscale}
\usepackage[mathscr]{eucal}
\usepackage{bm}
\usepackage[dvipsnames]{color}
\usepackage{cases}
\usepackage{color}
\usepackage{epsfig}
\usepackage{algorithm}
\usepackage{algorithmic}
\usepackage[verbose,nospace,sort]{cite}
\usepackage{tabularx,ragged2e}
\usepackage{multirow}
\usepackage{multicol}
\usepackage{balance}
\usepackage{caption}
\usepackage{subcaption}
\usepackage{setspace}
\usepackage{multirow}
\usepackage{multicol,afterpage,wrapfig}
\usepackage{pifont}
\usepackage{array}

\graphicspath{{./figs/}}

\newcommand{\me}{\mathrm{me}}
\newcommand{\mr}{\mathrm{mr}}

\newcommand{\vbd}{\mathbf v}
\newcommand{\q}{\mathbf q}
\newcommand{\LoS}{\mathrm{LoS}}
\newcommand{\NLoS}{\mathrm{NLoS}}
\newcommand{\twoD}{\mathrm{2D}}
\newcommand{\PL}{\mathrm{PL}}
\newcommand{\Prb}{\mathrm{P}}
\newcommand{\cov}{\mathrm{cov}}
\newcommand{\EE}{\mathrm{EE}}
\newcommand{\com}{\mathrm{com}}
\newcommand{\out}{\mathrm{out}}
\newcommand{\lb}{\mathrm{lb}}

\begin{document}
\title{Accessing From The Sky: A Tutorial on UAV Communications for 5G and Beyond}
\author{\IEEEauthorblockN{Yong~Zeng,~\IEEEmembership{Member,~IEEE,} Qingqing Wu,~\IEEEmembership{Member,~IEEE,} and~Rui~Zhang,~\IEEEmembership{Fellow,~IEEE}}
\thanks{Y. Zeng is with the School of Electrical and Information Engineering, The University of Sydney, Australia 2006 (e-mail: yong.zeng@sydney.edu.au). Q. Wu and R. Zhang are with the Department of Electrical and Computer Engineering, National University of Singapore, Singapore 117583 (e-mail:\{elewuqq, elezhang\}@nus.edu.sg).
(Corresponding author: Q. Wu.) }
}

\IEEEspecialpapernotice{(Invited Paper)}

\markboth{\emph{ S\MakeLowercase{ubmitted to} P\MakeLowercase{roceedings of the} IEEE}}%
{Shell \MakeLowercase{\textit{et al.}}: Bare Demo of IEEEtran.cls for IEEE Journals}


\maketitle

\begin{abstract}
Unmanned aerial vehicles (UAVs) have found numerous applications and are expected to bring fertile  business opportunities in the next decade. Among  various enabling technologies for UAVs, wireless communication is essential and has drawn significantly growing attention in recent years. Compared to  the conventional terrestrial communications, UAVs' communications face new challenges due to their high altitude above the ground and great flexibility of movement in the three-dimensional (3D) space. Several critical issues  arise, including the line-of-sight (LoS) dominant UAV-ground channels and resultant strong aerial-terrestrial network interference, the distinct communication  quality of service (QoS) requirements for UAV  control messages versus  payload data, the stringent constraints imposed by the size, weight and power (SWAP) limitations of UAVs, as well as the exploitation of the new  design degree of freedom (DoF) brought  by the highly controllable 3D  UAV mobility. In this paper, we give a tutorial overview of the recent advances in UAV communications to address the above issues, with an  emphasis on how to integrate UAVs into the forthcoming fifth-generation (5G) and future cellular networks. In particular, we partition our discussions into two promising research and application frameworks of UAV communications, namely {\emph{ UAV-assisted wireless communications}} and {\emph {cellular-connected UAVs}}, where UAVs serve as aerial communication platforms and users, respectively. Furthermore, we point out promising directions for future research and investigation.
\end{abstract}

\begin{center}\bf Keywords\end{center}

{\small{ Unmanned aerial vehicle (UAV), wireless communication,  cellular network, channel model, antenna model, energy efficiency, air-ground interference, 3D placement and trajectory,  optimization.} }

\section{Introduction}

Unmanned aerial vehicles (UAVs), also commonly known as drones,   are aircrafts piloted by remote control or embedded computer programs without human onboard. Historically, UAVs were mainly used in military applications deployed in hostile territory for remote surveillance and armed attack, to reduce the pilot losses. In recent years, the  enthusiasm for using UAVs in civilian and commercial applications has skyrocketed, thanks to the advancement of UAVs' manufacturing technologies and their reducing  cost, making them more easily accessible to the public. Nowadays, UAVs have found numerous applications in a proliferation of fields, such as aerial inspection, photography, precision agriculture, traffic control, search and rescue, package delivery,  telecommunications, etc. In June 2016, the U.S. Federal Aviation Administration (FAA) released the operational rules for routine civilian use of small unmanned aircraft systems (UAS) with aircraft weight less than 55 pounds (25 Kg) \cite{936}.  In November 2017, FAA further launched a national program  in Washington, namely ``Drone Integration Pilot Program'',   to explore the expanded use of drones,  including beyond-visual-line-of-sight (BVLoS) flights, night-time operations, flights above people, etc. \cite{USprogramm_UAV}.  It is anticipated that these new guidelines and programs will spur the further growth of global UAV industry in the coming years. The scale of the industry of UAVs is potentially enormous with realistic predictions in the realm of 80 billion dollars for the U.S. economy alone, expected to create  tens of thousands of new jobs within the next decade \cite{1011}. Therefore, UAVs have emerged  as a promising technology to offer fertile business opportunities in the next decade.

 In practice, there are many types of UAVs due to their numerous and diversified applications. While there is no single standard for UAV classification, UAVs can be practically  assorted into different categories according to different criteria such as   functionality,  weight/payload, size, endurance, wing configuration, control methods, cruising range, flying altitude, maximum speed, energy supplying methods, etc.
For example, in terms of wing configuration, fixed-wing and rotary-wing UAVs  are the two main  types of UAVs that have been widely used in practice.
Typically, fixed-wing UAVs have higher maximum flying  speed and can carry greater payloads for traveling  longer distances as compared to rotary-wing UAVs,  while their disadvantages lie in that a runway or launcher is needed for takeoff/landing as well as that  hovering at a fixed position   is impossible. In contrast, rotary-wing UAVs are able to takeoff/land vertically and remain static at a hovering  location. The above different characteristics of these two types of UAVs thus have   a great impact on their respectively suitable use cases.
A detailed classification for different types of UAVs has been provided in \cite{616}. In general, selecting a suitable type of UAVs is crucial for  accomplishing  their    mission efficiently, which needs to take into account their specifications as well as the  requirements of practical applications.

\subsection{Wireless Communication with UAVs: Basic Requirements}

\begin{table*}[t]
\centering
\caption{UAV communication requirements specified by 3GPP \cite{1012}.}
\label{table:trafficRequirements3GPP}
\begin{tabular}{|l|l|l|l|l|}
\hline
& {\bf Data Type}           & {\bf Data Rate}     & {\bf Reliability}                                  & {\bf Latency}                                                               \\ \hline
\begin{tabular}[c]{@{}l@{}}{\it DL} (Ground \\ station to UAV)\end{tabular}                  & Command and control & 60-100 Kbps   & $10^{-3}$ packet error rate & 50 ms                                                                  \\ \hline
\multirow{2}{*}{\begin{tabular}[c]{@{}l@{}} {\it UL} (UAV to\\ ground station)\end{tabular}} & Command and control & 60-100 Kbps   & $10^{-3}$ packet error rate & --                                                                   \\ \cline{2-5}
                                                                                       & Application data    & Up to 50 Mbps & --                                          & \begin{tabular}[c]{@{}l@{}}Similar to\\ terrestrial user\end{tabular} \\ \hline
\end{tabular}
\end{table*}

\begin{table*}[tp] \small
\caption{Communication requirements for typical  UAV applications  \cite{China_mobile_UAV}.}\label{table1}
\centering
\begin{tabular}{|l|l|l|l|}
\hline
{\bf UAV Application}                                                     & \begin{tabular}[c]{@{}l@{}}{\bf Height coverage}\\ {\bf in meter (m)}\end{tabular} & \begin{tabular}[c]{@{}l@{}}{\bf Payload traffic latency}\\ {\bf in millisecond (ms)}\end{tabular} & \begin{tabular}[c]{@{}l@{}}{\bf Payload data rate}\\ {\bf (DL/UL)}\end{tabular} \\ \hline
\it Drone delivery                                                      & 100 m                                                                  & 500 ms                                                                                & 300 Kbps/200 Kbps                                                   \\ \hline
\it Drone filming                                                       & 100 m                                                                  & 500 ms                                                                                & 300 Kbps/30 Mbps                                                    \\ \hline
\it Access point                                                        & 500 m                                                                  & 500 ms                                                                                & 50 Mbps/50 Mbps                                                     \\ \hline
\it Surveillance                                                        & 100 m                                                                  & 3000 ms                                                                               & 300 Kbps/10 Mbps                                                    \\ \hline
\begin{tabular}[c]{@{}l@{}}\it Infrastructure\\ \it inspection\end{tabular} & 100 m                                                                  & 3000 ms                                                                               & 300 Kbps/10 Mbps                                                    \\ \hline
\it Drone fleet show                                                    & 200 m                                                                  & 100 ms                                                                                & 200 Kbps/200 Kbps                                                   \\ \hline
\begin{tabular}[c]{@{}l@{}}\it Precision\\ \it agriculture\end{tabular}     & 300 m                                                                  & 500 ms                                                                                & 300 Kbps/200 Kbps                                                   \\ \hline
\it Search and rescue                                                   & 100 m                                                                  & 500 ms                                                                                & 300 Kbps/6 Mbps                                                     \\ \hline
\end{tabular}
\end{table*}

An essential enabling technology of UAS is wireless communication. On one hand, UAVs need to exchange safety-critical information with various parties such as remote pilots, nearby aerial vehicles, and air traffic controller, to ensure the safe, reliable, and efficient  flight operation. This is commonly known as the {\it control and non-payload communication (CNPC)} \cite{942}. On the other hand, depending on their  missions, UAVs may need to timely transmit and/or receive mission-related data such as aerial image, high-speed  video, and data packets for relaying, to/from various ground entities such as UAV operators, end users, or ground gateways. This is known as {\it payload communication}.


Enabling reliable and secure CNPC links is a necessity for the large-scale deployment and wide usage of UAVs. The International Telecommunication Union (ITU) has classified the required CNPC to ensure safe UAV operations into three categories \cite{942}, including:
\begin{itemize}
\item {\it Communication for UAV command and control}: This includes the telemetry report (e.g., flight status) from the UAV to the ground pilot, the real-time telecommand signaling from ground to UAVs for non-autonomous UAVs, and regular flight command update (such as waypoint update) for (semi-) autonomous UAVs.
\item {\it Communication for air traffic control (ATC)  relay}: It is critical to ensure that UAVs do not cause any safety threat to traditional manned aircraft, especially for operations approaching areas with high density of aircraft. To this end, a link between air traffic controller and the ground control station via the UAV, called ATC relay, is required.
\item {\it Communication supporting ``sense and avoid''}: The ability to support ``sense and avoid'' ensures that the UAV maintains sufficient safety distance with nearby aerial vehicles, terrain and obstacles.
\end{itemize}

The specific communication and spectrum requirements in general differ for CNPC and payload communications. Recently, the 3rd Generation Partnership Project (3GPP) has specified the communication requirements for these two types of links \cite{1012}, which is summarized in  Table \ref{table:trafficRequirements3GPP}. CNPC is usually of low data rate, say, in the order of Kbps (Kilobits per second), but has rather stringent requirement on  high  reliability and low latency.  For example, as shown in Table  \ref{table:trafficRequirements3GPP},  the data rate requirement for UAV command and control is only in the range of 60-100 Kbps  for both downlink (DL) and uplink (UL) directions,  but  a reliability of less than $10^{-3}$ packet error rate and a latency less than 50 milliseconds (ms) are  required.   While the communication requirements of CNPC links are similar for different types of UAVs due to their common safety consideration, those  for payload data are highly application-dependent. In Table \ref{table1}, we list several typical  UAV applications and their corresponding data communication requirements based on  \cite{China_mobile_UAV}.

Since the  lost of CNPC link may cause catastrophic consequences,  the International Civil Aviation Organization (ICAO) has determined that CNPC links for UAVs must operate over protected aviation spectrum \cite{943}, \cite{991}. Furthermore, ITU studies have revealed that to support CNPC for the forecasted number of UAVs in the coming years, 34 MHz (Mega Hertz)  terrestrial spectrum and 56 MHz satellite spectrum are needed for supporting both line-of-sight (LoS)  and beyond-LoS UAV operations\cite{942}. To meet such requirement, the C-band spectrum at 5030-5091 MHz has been made available for UAV CNPC at WRC (World Radiocommunication Conference)-12. More recently, the WRC-15 has decided that geostationary Fixed Satellite Service (FSS) networks may be used for UAS CNPC links.  

Compared to CNPC, UAV payload communication usually has much higher data rate requirement. For instance, to support the transmission of full high-definition
(FHD) video from the UAV to the ground user, the transmission rate is about several Mbps, while for 4K video, it is higher than 30 Mbps. The rate requirement for  UAV serving as aerial communication platform can be even higher, e.g., up to dozens of Gbps for data forwarding/backhauling applications.

 \begin{table*}\caption{Comparison of wireless  technologies for UAV communication.} \label{table:networking:architectures}
 \centering
\begin{tabular}{llll}
 \hline
\multicolumn{1}{|l|}{\bf Technology}                                                 & \multicolumn{1}{l|}{\bf Description}                                                                                                                   & \multicolumn{1}{l|}{\bf Advantages}                                                                                                                            & \multicolumn{1}{l|}{\bf Disadvantages}                                                                                                                                       \\ \hline
\multicolumn{1}{|l|}{\it Direct link}                                                & \multicolumn{1}{l|}{\begin{tabular}[c]{@{}l@{}} Direct point-to-point\\ communication with\\ ground node\end{tabular}}                             & \multicolumn{1}{l|}{Simple, low cost}                                                                                                                      & \multicolumn{1}{l|}{\begin{tabular}[c]{@{}l@{}}Limited range, low data rate,\\ vulnerable to interference, \\ non-scalable\end{tabular}}                                 \\ \hline
\multicolumn{1}{|l|}{\it Satellite}                                                  & \multicolumn{1}{l|}{\begin{tabular}[c]{@{}l@{}}Communication and\\ Internet access via\\ satellite\end{tabular}}                                   & \multicolumn{1}{l|}{Global coverage}                                                                                                                       & \multicolumn{1}{l|}{\begin{tabular}[c]{@{}l@{}}Costly, heavy/bulky/energy-\\ consuming communication\\ equipment, high latency, large\\ signal attenuation\end{tabular}} \\ \hline
\multicolumn{1}{|l|}{\it Ad-hoc network}                                             & \multicolumn{1}{l|}{\begin{tabular}[c]{@{}l@{}}Dynamically\\ self-organizing and\\ infrastructure-free\\ network\end{tabular}}                     & \multicolumn{1}{l|}{\begin{tabular}[c]{@{}l@{}}Robust and adaptable,\\ support for high\\ mobility\end{tabular}}                                           & \multicolumn{1}{l|}{\begin{tabular}[c]{@{}l@{}}Costly, low spectrum efficiency,\\ intermittent connectivity, \\ complex routing protocol\end{tabular}}                   \\ \hline
\multicolumn{1}{|l|}{\begin{tabular}[c]{@{}l@{}}\it Cellular\\ \it network\end{tabular}} & \multicolumn{1}{l|}{\begin{tabular}[c]{@{}l@{}}Enabling UAV\\ communications by\\ using cellular\\ infrastructure and\\ technologies\end{tabular}} & \multicolumn{1}{l|}{\begin{tabular}[c]{@{}l@{}}Almost ubiquitous\\ accessibility,\\ cost-effective, superior\\ performance and\\ scalability\end{tabular}} & \multicolumn{1}{l|}{\begin{tabular}[c]{@{}l@{}}Unavailable in remote areas, \\ potential interference with \\terrestrial  communications\end{tabular}}                             \\ \hline
\end{tabular}
\end{table*}

\subsection{Wireless Technologies for UAV Communication}
To meet both the CNPC and payload  communication requirements in  multifarious UAV applications, proper wireless technologies are needed  for achieving seamless connectivity and high reliability/throughput for both air-to-air and air-to-ground  wireless communications in the three-dimensional (3D) space. Towards this end,  four candidate  communication technologies are listed and compared in Table \ref{table:networking:architectures}, including  i) direct link; ii) satellite; iii)  ad-hoc  network; and iv) cellular network. In the following, we discuss the advantages as well as limitations of each of these technologies in detail.

 \subsubsection{Direct Link}
Due to its simplicity and low cost, the direct-link communication between UAV and its associated ground node over the unlicensed band (e.g., the Industrial Scientific Medical (ISM) 2.4 GHz band) was most commonly used for commercial UAVs in the past, where the ground node can be a joystick, remote controller, or ground station.  However, it is usually limited to LoS communication, which significantly constrains its operation range and hinders its applications in complex propagation environment.  For example, in urban areas, the communication can be easily blocked by e.g.,  trees and high-rise buildings, which results in low reliability and low rate.  Furthermore, the ground node needs to connect to a gateway for enabling Internet access of the UAV, which may cause long delay in case of wireless data backhaul.  In addition, such a simple solution is usually  insecure and vulnerable to interference and jamming.  Due to the above limitations, the simple direct-link communication cannot be a scalable solution for supporting large-scale deployment of UAVs in the future.

\subsubsection{Satellite}
Enabling UAV communications by leveraging satellites is a viable option due to their global coverage. Specifically, satellites can help relay data communicated  between widely separated UAVs and ground gateways, which  is particularly useful for UAVs above ocean and in remote areas where the terrestrial network  (WiFi or cellular) coverage is unavailable. Furthermore, satellite  signals  can also be used for navigation and localization of UAVs.  In WRC 2015,  the conditional use of satellite communication  frequencies in the Ku/Ka band has been approved  to connect drones to satellites, and some satellite companies such as Inmarsat  have launched satellite communication service for UAVs \cite{Inmarsat}. However, there are also several disadvantages of satellite-enabled  UAV communications.   
Firstly, the propagation loss and delay are quite significant due to the long distances between satellite and  low-altitude UAVs/ground stations. This thus poses great challenges for meeting ultra-reliable and delay-sensitive CNPC for UAVs. Secondly, UAVs usually have stringent size, weight and power (SWAP) constraints, and thus  may not be able to carry the heavy, bulky and energy-consuming satellite communication equipment  (e.g., dish antenna).  Thirdly,  the high operational cost  of satellite communication also hinders its wide use for densely deployed UAVs in consumer-grade applications.

\subsubsection{Ad Hoc Network}

Mobile ad-hoc network (MANET) is an infrastructure-free and  dynamically self-organizing network for enabling peer-to-peer communications among  mobile devices such as laptops, cellphones, walkie-talkies, etc. Such  devices usually  communicate over bandwidth-constrained wireless links using e.g.  IEEE 802.11 a/b/g/n. Each device in a MANET can move randomly over time; as a result, its link conditions with other devices may change frequently. Furthermore, for  supporting communications  between two far-apart  nodes,  some other nodes in between  need to help forward the data via multi-hop relaying, thus incurring more energy consumption, low spectrum efficiency, and long end-to-end delay. Vehicular ad hoc network (VANET) and flying ad hoc network (FANET) are two applications of MANET, for supporting communications among high-mobility  ground vehicles and UAVs  in 2D and 3D networks, respectively \cite{bekmezci2013flying}.
 The topology or configuration of an FANET for  UAVs  may take different  forms, such as  a mesh, ring,  star,  or even a straight line,  depending on the application  scenario. For example,  a star network topology is suitable for UAV swarm applications, where UAVs in a swam all communicate through a central hub UAV, which may also be responsible for communicating with the ground stations.  Although FANET is a robust and flexible architecture for supporting UAV communications in a small network, it is generally unable to provide a scalable solution for serving massive UAVs deployed in a wide area, due to the complexities and difficulties for realizing a reliable routing protocol over the whole network with dynamic and intermittent link connectivities among the flying UAVs.

\subsubsection{Cellular Network}
It is evident that the above technologies generally cannot support large-scale UAV communications in a cost-effective manner. On the other hand, it is also economically nonviable to build new and dedicated ground networks for achieving this goal. As such, there has been significantly growing interest recently in
leveraging the existing as well as  future-generation cellular networks for enabling UAV-ground communications \cite{952}.  Thanks to the almost ubiquitous coverage of the cellular network worldwide as well as its high-speed optical backhaul and advanced communication technologies, both CNPC and payload communication requirements for UAVs can be potentially met, regardless of the density of UAVs as well as their distances with the corresponding ground nodes.  For example, the forthcoming fifth-generation (5G) cellular  network is expected to support the peak data rate of 10 Gbits/s with only  1 ms round-trip latency, which in principle is adequate for high-rate and delay-sensitive UAV communication applications such as real-time video streaming and data relaying.

\begin{figure*}[!t]
\centering
\includegraphics[width=1\textwidth]{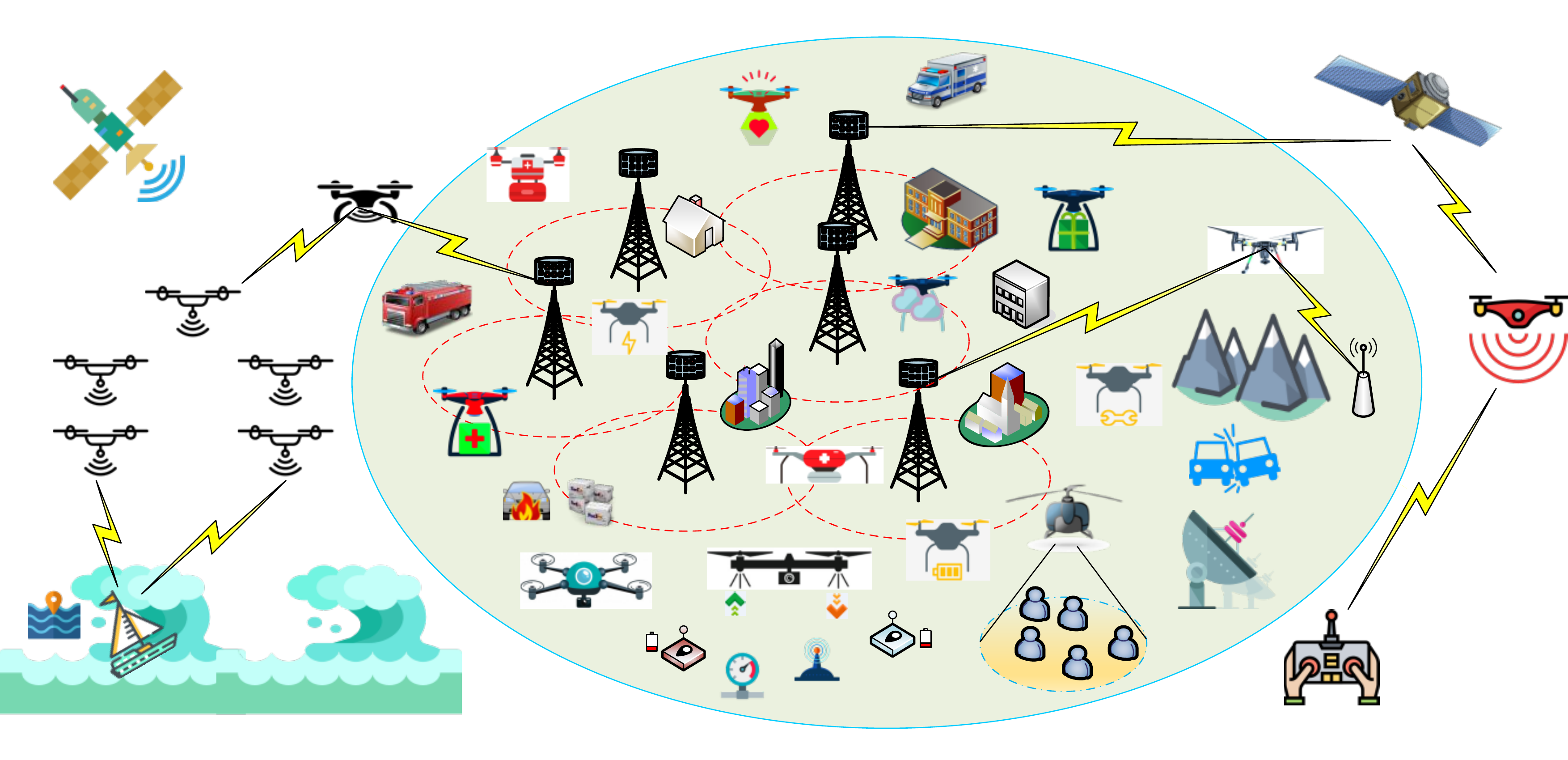}
\caption{Supporting UAV communications with an integrated network  architecture.  } \vspace{-0.4cm}\label{SecI:APP}
\end{figure*}

Despite the promising  advantages of cellular-enabled  UAV communications, there are still scenarios where the cellular services are unavailable, such as in remote areas like sea, desert, forest, etc. In such scenarios, other technologies such as   the direct link, satellite and FANET, can be used to support UAV communications beyond the terrestrial coverage of cellular network. Therefore, it is envisioned that the future wireless network for supporting large-scale UAV communications will have an integrated 3D architecture consisting of  UAV-to-UAV, UAV-to-satellite and UAV-to-ground communications, as shown in Fig. \ref{SecI:APP}, where each UAV may be enabled with one or more communication technologies to exploit the rich connectivity diversity in such a hybrid network.

\subsection{The New Paradigm: Integrating UAVs into Cellular Network}
In this subsection, we further discuss the aforementioned  new paradigm of integrating UAVs into the cellular network, to provide their full horizon of applications and benefits. In particular, we partition our discussion into two main categories.
 On one hand, UAVs are considered as new aerial users that access the cellular network from the sky for communications, which we refer to as {\it cellular-connected UAVs}. On the other hand,  UAVs are used  as new aerial communication platforms such as base stations (BSs) and relays, to assist in terrestrial wireless communications by providing data access  from the sky, thus called  {\it UAV-assisted wireless communications}.

\subsubsection{Cellular-Connected UAVs}
By incorporating UAVs as new user equipments (UEs) in the cellular network, the following benefits can be achieved \cite{952}.  Firstly, thanks to the  almost worldwide accessibility of cellular networks, cellular-connected UAV makes it possible for the ground pilot to remotely command and control the UAV with virtually unlimited operation range. Besides, it also provides an effective solution to maintain wireless connectivity between UAVs and various other stakeholders, such as the end users and the air traffic controllers, regardless of their locations. This thus opens up many new UAV applications in the future.  Secondly, with the advanced cellular technologies and authentication mechanisms, cellular-connected UAV is expected to achieve significant  performance improvement over the other technologies introduced in the previous subsection, in terms of reliability, security, and data throughput. For instance,  the current fourth-generation (4G) long term evolution (LTE) cellular network  employs scheduling-based channel access mechanism, where multiple users can be served simultaneously by assigning them orthogonal  resource blocks (RBs). In contrast, WiFi (e.g., 802.11g employed in FANET) adopts contention-based channel access with a random backoff mechanism, where users are allowed to only access channels that are sensed to be  idle. Thus, multiuser transmission with centralized scheduling/control enables the cellular network to make a more efficient  use of the spectrum than WiFi, especially when the user density is high.  In addition,  UAV-to-UAV communication can also be realized  by leveraging the available device-to-device (D2D) communications in LTE and 5G systems. 
Thirdly, cellular-based localization service can provide UAVs a new and complementary means in addition to the conventional satellite-based global positioning system (GPS) for achieving  more robust or enhanced  UAV navigation performance.  Last but not least, cellular-connected UAV is a cost-effective solution since it reuses the millions of cellular BSs worldwide without the need of building new infrastructure dedicated for UAS only. Thus, cellular-connected UAV is expected  to be a win-win technology for both UAV and cellular industries, with rich business opportunities to explore in the future.

\subsubsection{UAV-Assisted Wireless Communications}
 Thanks to the continuous cost reduction in UAV manufacturing and device miniaturization in communication equipment, it becomes more feasible to mount compact and small-size BSs or relays on UAVs to enable flying aerial platforms to assist in terrestrial wireless communications. For instance, commercial LTE BSs with light weight (e.g., less than 4 Kg) are already available in the market, which are suitable to be mounted on UAVs with moderate payload. Compared to conventional terrestrial communications with typically static BSs/relays deployed at fixed locations, UAV-assisted communications bring  the following main  advantages \cite{649}. Firstly, UAV-mounted BSs/relays can be swiftly deployed on demand. This is especially appealing for application scenarios  such as  temporary or unexpected events, emergency response, search and  rescue, etc.  Secondly, thanks to their high  altitude above the ground, UAV-BSs/relays are more likely to have LoS connection with their ground users as compared to their terrestrial counterparts, thus providing more reliable links for communication as well as multiuser scheduling and resource allocation.  Thirdly, thanks to the controllable high-mobility of UAVs, UAV-BSs/relays possess  an additional degree of freedom (DoF) for communication performance enhancement, by dynamically adjusting their locations in 3D to cater for the terrestrial communication demands.

The above benefits make UAV-assisted communication a promising new technology to support the ever-increasing and highly dynamic wireless data  traffic  in future 5G-and-beyond cellular systems. There are abundant new applications in anticipation, such as for cellular data offloading in hot-spot areas  (e.g., stadium during a sport event), information dissemination and data collection in wireless sensor and Internet of Things (IoT) networks, big data transfer between geographically separated data centers, fast service recovery after infrastructure failure, mobile data relaying in emergency situations or customized communications, etc.

\subsection{UAV Communications: What's New?}
The integration of UAVs into cellular networks, either as aerial users or communication platforms, brings new design opportunities as well as  challenges.  Both cellular-connected UAV communication and UAV-assisted wireless communication are significantly different from their terrestrial counterparts, due to the high altitude and high mobility of UAVs, the high probability  of UAV-ground LoS channels,  the distinct communication  quality of service (QoS) requirements for CNPC versus mission-related payload data, the stringent SWAP constraints  of UAVs, as well as the new design  DoF by jointly exploiting the  UAV mobility control and communication scheduling/resource allocation.  Table~\ref{Table:OppurAndChallenge} summarizes the main design opportunities and challenges of cellular communications with UAVs, which are further elaborated as follows.

\subsubsection{High Altitude} Compared with conventional terrestrial BSs/users, UAV BSs/users usually have much higher altitude. For instance, a typical height of a terrestrial BS is around 10 m for Urban Micro (UMi) deployment and 25 m for Urban Macro (UMa) deployment \cite{1012}, whereas the current regulation already allows the UAVs to fly up to 122 m \cite{936}. For cellular-connected UAVs, the high UAV altitude requires cellular BSs to offer 3D aerial coverage for UAV users, in contrast  to the conventional 2D coverage for terrestrial users. However, existing BS antennas are usually tilted downwards, either mechanically or electronically, to cater for the ground coverage as well as suppressing the  inter-cell interference. Nevertheless, preliminary field measurement results have demonstrated satisfactory aerial coverage to meet the basic communication requirements by the antenna side lobes of BSs for UAVs below 400 feet (122 m) \cite{3gpp_UAV}. However, as the altitude further increases, weak signal coverage is observed, which thus calls for new BS antenna designs and cellular communication   techniques to achieve satisfactory UAV coverage up to the maximum altitude of 300 m as currently specified by 3GPP \cite{1012}.  On the other hand, for UAV-assisted wireless communications, the high UAV altitude enables the UAV-BS/relay to achieve wider ground coverage as compared to their  terrestrial counterparts.

\subsubsection{High LoS Probability} The high UAV altitude leads to unique air-ground channel characteristics as compared to terrestrial communication channels. Specifically, compared to the terrestrial  channels that  generally suffer more severe path loss due to shadowing and multi-path fading effects, the UAV-ground channels, including both the UAV-BS and UAV-user channels, typically experience limited scattering and thus have a dominant  LoS link with high probability.
 The LoS-dominant air-ground channel  brings both opportunities and challenges to the design of  UAV communications as compared to the traditional terrestrial communications.  On one hand, it offers more reliable link performance between the UAV and its serving/served ground BSs/users, as well as a pronounced macro-diversity in terms of more flexible UAV-BS/user associations. Moreover, as LoS-dominant links have less channel variations in time and frequency, communication scheduling and resource allocation can be more efficiently implemented in a much slower pace as compared to that over  terrestrial fading channels.
  On the other hand, however, it also causes strong air-ground  interference, which is a critical issue that may severely limit the cellular network capacity with coexisting aerial and terrestrial BSs/users. For example, in the UL communication of a UAV user,  it may pose severe interference to many adjacent cells at the same frequency band due to its high-probability LoS channels with their BSs; while in the DL communication, the UAV user also suffers strong interference from these co-channel BSs. Interference mitigation is crucial  for both frameworks of cellular-connected UAVs and UAV-assisted terrestrial communications. Furthermore, the LoS-dominant air-ground links also make UAV communications more susceptible to the jamming/eavesdropping attacks by malicious ground nodes as compared to the terrestrial communications over fading channels, thus imposing a new security threat at the physical layer \cite{wu2019safeguarding}.

\subsubsection{High 3D Mobility} Different from the terrestrial networks  where the BSs/relays are usually at fixed locations and the users move sporadically and randomly, UAVs can move at high speed in 3D space with partially or fully controllable mobility. On one hand, the high mobility of UAVs  generally results in more frequent handovers and time-varying wireless backhaul links with  ground BSs/users. On the other hand, it also  leads to an important new design approach  of communication-aware UAV mobility control, such that the UAV's  position,  altitude,   speed, heading direction, etc., can be dynamically changed to better meet its communication objectives with the ground BSs/users.  For example, in UAV-assisted wireless communication, UAV-BSs/relays can design their trajectories (i.e., locations and speeds over time) either off-line or in real time to adapt to the locations and communication channels of their served ground users. Similarly,  for cellular-connected UAVs, they can also adjust their trajectories based on the locations of the ground BSs to find the best route to fulfill their mission requirements and in the meanwhile ensure a set of BSs along its trajectory to satisfy its communication needs. Furthermore, UAV 3D placement/trajectory design can be jointly considered with communication scheduling and resource allocation for further performance improvement.

\subsubsection{SWAP Constraints} Different from  terrestrial communication systems where the ground BSs/users usually have a stable power supply from the grid or rechargeable  battery,  the SWAP constraints of UAVs pose critical limits on their endurance and communication capabilities. For example, in the case of UAV-assisted wireless communications, customized BSs/relays, generally of  smaller size and lighter weight as well as with more compact antenna and  power-efficient hardware as compared to their terrestrial counterparts, need to be designed to  cater for the limited payload and size of UAVs. Furthermore, besides the conventional communication transceiver energy consumption, UAVs need to spend  the additional propulsion energy  to remain aloft and move freely over the air \cite{904}, \cite{980}, which is usually much more significant than the communication energy (e.g. in the order of kilowatt versus watt) for commercial UAVs. Thus, the energy-efficient design of UAV communication is more involved than  that for the conventional terrestrial systems considering the communication energy only \cite{800}, \cite{801}.

\begin{table*}\caption{Opportunities and challenges of cellular communication with UAVs.} \label{Table:OppurAndChallenge}
\centering
\begin{tabular}{|l|l|l|}
\hline
\bf Characteristic                                                   & \bf Opportunities                                                                                                                                                     & \bf Challenges                                                                                                                                                \\ \hline
\it High altitude                                                    & \begin{tabular}[c]{@{}l@{}}Wide ground coverage as aerial\\ BS/relay\end{tabular}                                                                                 & \begin{tabular}[c]{@{}l@{}}Require 3D cellular coverage for \\ aerial user\end{tabular}                                                                   \\ \hline
\begin{tabular}[c]{@{}l@{}}\it High LoS proba-\\ \it bility\end{tabular} & \begin{tabular}[c]{@{}l@{}}Strong and reliable communication\\ link; high macro-diversity; slow\\ communication scheduling and\\ resource allocation\end{tabular} & \begin{tabular}[c]{@{}l@{}}Severe aerial-terrestrial interference;\\ susceptible to terrestrial\\ jamming/eavesdropping\end{tabular}                      \\ \hline
\it High 3D mobility                                                 & \begin{tabular}[c]{@{}l@{}}Traffic-adaptive movement;\\ QoS-aware trajectory design\end{tabular}                                                                  & \begin{tabular}[c]{@{}l@{}}Handover management; wireless\\ backhaul\end{tabular}                                                                          \\ \hline
\it SWAP constraint                                                  & --                                                                                                                                                                 & \begin{tabular}[c]{@{}l@{}}Limited payload and endurance;\\ energy-efficient design; compact and\\ lightweight BS/relay and antenna\\ design\end{tabular} \\ \hline
\end{tabular}
\end{table*}

\subsection{Prior Work  and Our Contribution}
The  exciting new opportunities in a broad range of UAV applications have spawned extensive research recently. In particular, several magazine \cite{649,949,913,967,952,950} and survey \cite{khawaja2018survey,621,khuwaja2018survey,bekmezci2013flying,618,gupta2016survey,hayat2016survey,
shakhatreh2018unmanned,motlagh2016low,shakeri2018design,mozaffari2018tutorial,fotouhi2018survey} papers on wireless communications and networks with UAVs have appeared. Among them, the survey papers  \cite{khawaja2018survey,621,khuwaja2018survey} focus on air-ground channel models and experimental measurement results  of UAV communications.  The survey papers   \cite{bekmezci2013flying},  \cite{618} and \cite{gupta2016survey} mainly address  ad hoc networks for UAV communications by focusing on UAV-UAV communications.
Prior work  \cite{hayat2016survey} gives a survey on UAV-aided civil applications, while the survey paper \cite{shakhatreh2018unmanned} discusses other applications of UAVs and some promising technologies for them.   In \cite{motlagh2016low}, the UAV-enabled IoT services are overviewed with a particular focus on data collection, delivery, and processing, while in \cite{shakeri2018design} the challenges in designing and implementing  multi-UAV networks for a wide range of cyber-physical applications are reviewed. The recent works \cite{mozaffari2018tutorial} and \cite{fotouhi2018survey} provide  contemporary  surveys of UAV applications in cellular networks, focusing on academic literatures and industry activities, respectively.

Compared with the above survey papers, this paper aims to provide a more comprehensive survey and tutorial on UAV communications, with an emphasis on the two promising paradigms of cellular-connected UAVs and UAV-assisted wireless communications. Besides providing a state-of-the-art  literature survey from both academic and industrial research perspectives, this paper provides more technically in-depth results and discussions to facilitate and inspire  future research in this area. In particular, this  tutorial features a unified and general mathematical framework for UAV trajectory and communication co-design as well as a comprehensive overview on the various techniques to deal with the crucial air-ground interference issue in cellular communications with UAVs.

The rest of this paper is organized as follows. Section~\ref{sec:basics} introduces the fundamentals of UAV communications, including channel model, antenna model, UAV energy consumption model, and the mathematical framework for designing UAV trajectory and communication jointly. Section~\ref{sec:UAVAssisted} considers  UAV-assisted wireless communications, where the basic system models,  performance analysis, UAV placement/trajectory and communication co-design, as well as energy-efficient UAV  communications are discussed. We also highlight the promising new direction of learning-based UAV trajectory and  communication design at the end of this section.  In Section~\ref{sec:cellularConnected}, we address the other paradigm of cellular-connected UAVs. We start with a historical feasibility study on supporting aerial users in cellular networks by introducing some major field trials from 2G to 4G, as well as the latest standardization efforts by 3GPP. We then give an overview on some representative works evaluating the performance of the cellular network with newly added UAV users to draw useful insights.  Last, we present promising techniques to efficiently  embrace aerial users in the cellular network including air-ground interference mitigation and QoS-aware UAV trajectory planning.  In Section~\ref{sec:otherTopic}, we discuss  other related topics to provide promising directions for future research and investigation. Finally,  we conclude this paper in Section~\ref{sec:conclusions}.

{\it Notations: } In this paper, scalars and vectors are denoted by italic letters and boldface lower-case letters, respectively. $\mathbb{R}^{M\times 1 }$ and $\mathbb{C}^{M\times 1}$ denote the space of $M$-dimensional real- and complex-valued vectors, respectively. For a real number $a$, $\lceil a \rceil$ denotes the smallest integer greater than or equal to $a$. $j$ is the imaginary unit with $j^2=-1$. For a vector $\mathbf a$, $\mathbf a^T$, $\mathbf a^H$, $\|\mathbf a\|$, and  $[\mathbf a]_n$ denote its transpose, complex conjugate transpose, Euclidean norm, and the $n$th component, respectively.  The notation $\exp(\cdot)$ denotes the exponential function. For a twice differentiable time-dependent vector-function $\mathbf x(t)$, $\dot{\mathbf x}(t)$ and $\ddot{\mathbf x}(t)$ denote the first- and second-order derivatives with respect to time $t$, respectively. For a real-valued function $f(\mathbf q)$ with respect to a vector $\mathbf q$, $\nabla f(\mathbf q)$ denotes its gradient. For a random variable $X$, $\mathbb{E}[X]$ represents its statistical expectation, while  $\Pr(E)$ denotes the probability of an event $E$. Furthermore, $\mathcal {N}(\mu, \sigma^2)$ represents the Gaussian distribution with mean $\mu$ and variance $\sigma^2$.

\section{UAV Communication Fundamentals}\label{sec:basics}
In this section, we present some basic mathematical models pertinent to UAV communications, which are useful for research in  both frameworks of UAVs serving as aerial users or communication platforms. They include the channel model, antenna model, UAV energy consumption model, performance metrics, as well as mathematic formulation for performance optimization via joint UAV communication and trajectory design.

\subsection{Channel Model}\label{sec:channelModel}
UAV communications mainly involve three  types of links, namely the ground BS (GBS)-UAV link, the UAV-ground terminal (GT) link, and the UAV-UAV link. As the communication between UAVs with moderate distance typically occurs in clear airspace when the earth curvature is irrelevant, the UAV-UAV channel is usually characterized by the simple free-space path loss model \cite{617}, \cite{1023}.\footnote{For the special case of UAV-UAV links in a UAV swarm consisting of many UAVs in short-distances with each other, there may exist multipath due to the signal reflection/scattering among the UAVs.} Therefore, we focus on the channel models for GBS-UAV and UAV-GT links in this subsection, for cellular-connected UAVs and UAV-assisted terrestrial communications, respectively. In principle, the existing channel models for the extensively studied terrestrial communication systems can be applied to UAV communications. However, as UAV systems involve transmitters and/or receivers with altitude much higher than those in conventional terrestrial systems, customized mathematical models have been developed to more accurately characterize the unique propagation environment for UAV communications at different altitude. Significant efforts have been devoted to the channel measurements and modelling for UAV communications, where some recent surveys on them can be found in e.g., \cite{khawaja2018survey,621,khuwaja2018survey}.  Different from these existing surveys focusing   on channel measurement campaigns with detailed description of the measurement setup and data processing methods, we provide here a tutorial overview on the mathematical  UAV channel models to facilitate performance  analysis and evaluation  for UAV communication systems, as will be further  illustrated in more details  in Section~\ref{sec:UAVAssisted} and Section~\ref{sec:cellularConnected}.


%


We start with the general wireless channel model for baseband communication in a frequency non-selective channel,  where the complex-valued channel coefficient between a transmitter and a receiver can be expressed  as \cite{56}
\begin{align}
g=\sqrt{\beta (d)}\tilde{g}, \label{eq:g}
\end{align}
where $\beta(d)$ accounts for the large-scale channel attenuation including distance-dependent path loss and shadowing, with $d$ denoting the distance between the transmitter and the receiver, and $\tilde{g}$ is generally a complex random variable with $\mathbb{E}[|\tilde g|^2]=1$ accounting for the small-scale fading due to multi-path propagation. One classical model for $\beta(d)$ is the log-distance path loss (PL) model, where $\beta(d) [dB]= -\PL(d)[dB]$ with
\begin{align}
\PL(d) [dB] = 10  \alpha \log_{10}(d)+ X_0 [dB] +X_{\sigma} [dB], \label{eq:PL}
\end{align}
where $\alpha$ is the path loss exponent that usually has the value between $2$ and $6$, $X_0$ is the path loss at a reference distance of 1 m, $X_{\sigma}\sim \mathcal {N}(0, \sigma^2_X)$ accounts for the shadowing effect which is modelled as a normal (Gaussian) random variable with zero mean and a certain variance $\sigma^2_X$.

For  UAV communications, the choice of appropriate models for the  large-scale and small-scale channel parameters needs to take into account their unique propagation conditions. Firstly, different from terrestrial communication systems where Rayleigh fading is commonly used for small-scale fading, the more general Rician or Nakagami-m small-scale fading model is more appropriate for UAV-ground communications since the LoS channel component is usually present. While the above small-scale fading channel models have been well understood in existing literature, the modelling for the large-scale channel component in UAV-ground communications is generally more sophisticated due to the high altitude of UAVs and resultant 3D propagation space. Various customized models have been proposed, which can be generally classified into three categories, namely {\it free-space channel model}, {\it models based on altitude/angle-dependent parameters}, and {\it probabilistic LoS channel model}.

\subsubsection{Free-Space Channel Model}
For the ideal scenario in the absence of signal obstruction or reflection, we have the free-space propagation channel model where the effects of shadowing and small-scale fading vanish. In this case, we have $|\tilde g|=1$ and the channel power  in \eqref{eq:g} can be simplified  as
\begin{align}\label{eq:freeSpace}
\beta(d)=\left(\frac{\lambda}{4\pi d} \right)^2 = \tilde{\beta}_0d^{-2},
\end{align}
where $\lambda$ is the carrier  wavelength, $\tilde{\beta}_0\triangleq \left(\frac{\lambda}{4\pi}\right)^2$ is the channel power at the reference distance of $1$ m. With the above free-space path loss model, the channel power  is completely determined by the transmitter-receiver distance (or locations of the UAV and its communicating  GBS/GT), which is easily predictable if their locations  are known. As a result, free-space channel model has been widely assumed in early works on offline UAV trajectory optimization in communication systems \cite{641,904,919}.

In practice, free-space path loss model gives a reasonable approximation in rural area where there is little blockage or scattering, and/or when the altitude of UAV is sufficiently high so that a clear LoS link between the UAV and the ground node is almost surely guaranteed. However, for low-altitude UAV operating in urban environment where the building height is non-negligible as compared to UAV altitude, free-space propagation model is  oversimplified. In this case, more refined channel models are needed to reflect the change of propagation environment as the UAV altitude varies. Two approaches have been widely adopted to achieve this goal: using channel modelling parameters that are dependent on UAV altitude or elevation angle, or using a probabilistic LoS channel model by modelling the LoS and NLoS scenarios randomly but  governed by a certain probability distribution, as discussed in the following.

\begin{figure*}
\centering
\includegraphics[width=0.7\linewidth]{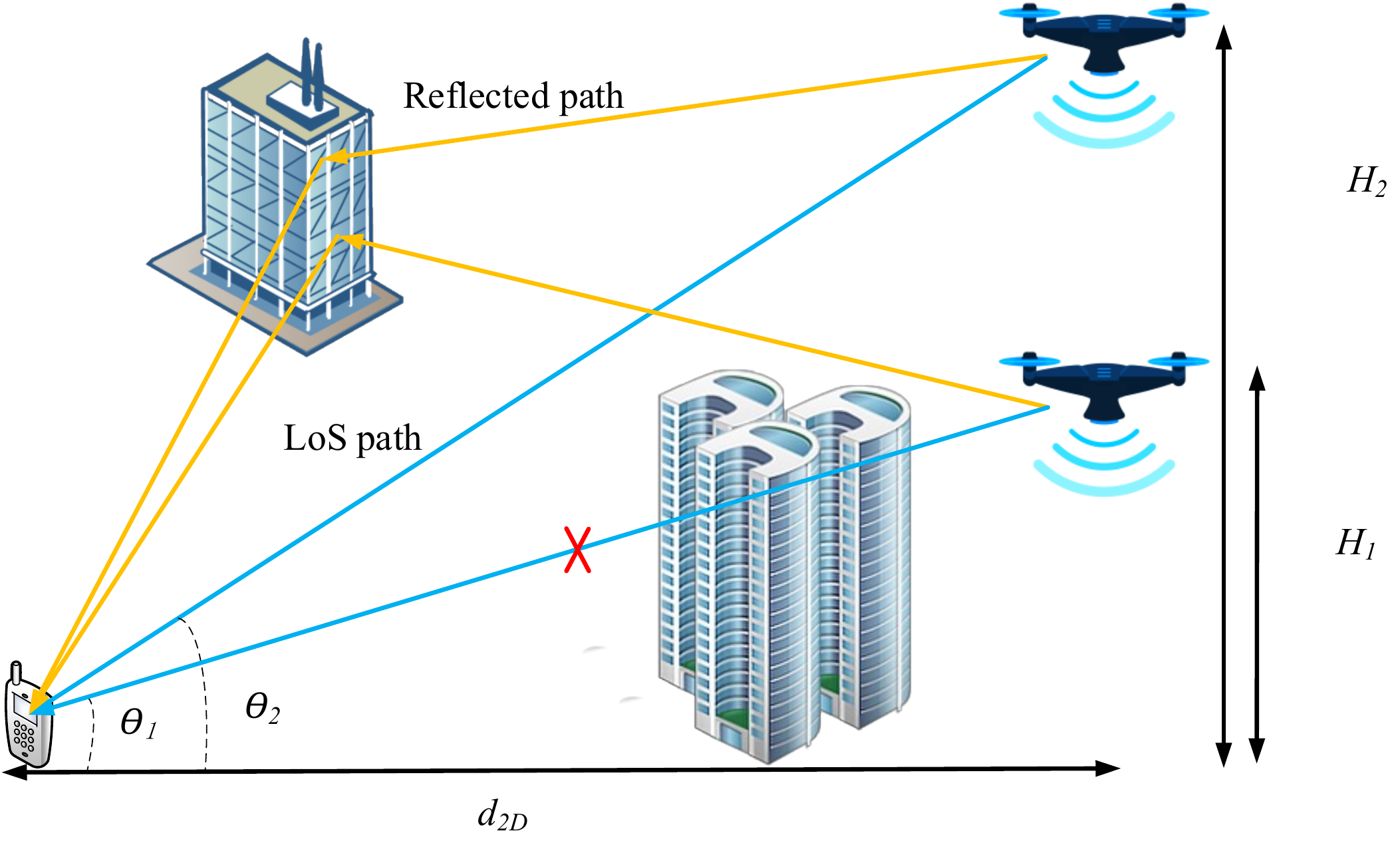}
\caption{An illustration of the UAV-ground propagation  in urban environment.}\label{F:Propagation}
\end{figure*}

\subsubsection{Altitude/Angle-Dependent Channel Parameters}
As illustrated in Fig.~\ref{F:Propagation}, in urban environment, as UAV moves higher, the effect of signal obstruction and scattering reduces. To explicitly model this, one approach is to use altitude- or angle-dependent channel parameters for the generic channel model in \eqref{eq:g}. Such parameters may include the path loss exponent $\alpha$ \cite{996}, \cite{954}, the Rician $K$ factor \cite{954}, the variance of the random shadowing $\sigma^2_X$ \cite{996}, or the excessive path loss relative to conventional terrestrial channels \cite{997}.

{\bf Altitude-dependent channel parameters:} In \cite{996},  the path loss exponent $\alpha$ for GBS-UAV link is modelled as a monotonically decreasing function of the UAV altitude $H_U$ as
\begin{align}
\alpha(H_U)= \max\left(p_1 -p_2 \log_{10}(H_U), 2\right),
\end{align}
where $p_1, p_2>0$ are modelling parameters that can be obtained via curve fitting based on channel measurement results. The above model explicitly reflects the fact that as the UAV moves higher, there are in general less obstacles and scattering, and hence smaller path loss exponent holds. As $H_U$ gets sufficiently large, we have the free space propagation model with $\alpha=2$. Similar altitude-dependent expressions have been suggested for $X_0$ and $\sigma_X^2$ in \eqref{eq:PL}. Note that while the above models were proposed in \cite{996} for GBS-UAV links with UAVs being aerial users of cellular BSs, it can be in principle applied to UAV-GT channels, but with different parameters to reflect the fact that GBS-UAV links are usually subject to less obstacles than UAV-GT links, due to the elevated GBS site.

{\bf Elevation angle-dependent channel parameters:} While the altitude-dependent channel model reveals the varying propagation environment for different UAV altitudes, it fails to model the fact that even with the same UAV altitude, the propagation environment may change if the UAV moves closer/further to/from the ground node \cite{997}. To address this issue, another approach is to model the channel modelling parameters as functions of the elevation angle $\theta$ (shown in Fig.~\ref{F:Propagation}), which depends   on both the UAV altitude and the horizontal (or 2D) distance with the corresponding ground node. For instance, in  \cite{954}, by considering the UAV-GT communications and assuming Rician fading channels, the Rician factor and the path loss exponent are respectively modelled as non-decreasing and non-increasing functions of $\theta$,  which implies that as $\theta$ increases, i.e., either the UAV flies higher or closer to the ground node, the LoS component becomes more dominating.

{\bf Depression angle-dependent excess path loss model:} For GBS-UAV communication, the elevation angle (termed as depression angle in \cite{997}) can be both positive (when UAV is higher than GBS) or negative (when UAV is lower than GBS). Under this setup, the authors in \cite{997} conducted both terrestrial and aerial experimental measurements in a typical suburban environment, by mounting the same handset on a car and on a UAV, respectively. By comparing the received signal power for these two measurement scenarios with roughly the same horizontal distance with the GBS, the authors proposed a path loss model for GBS-UAV channels by adding an {\it excess path loss}\footnote{Note that we follow the terminology used in \cite{997}, though the term ``excess'' could be misleading as it is possible that $\eta(\theta)$ is a negative value for small $\theta$.} on top of the conventional terrestrial path loss, where the excess path loss component is a function of the depression angle $\theta$, i.e.,
\begin{align}
\mathrm{PL}_{U}(d,\theta)=\PL_{\mathrm{ter}}(d)+\eta(\theta) + X_U(\theta),
\end{align}
where $\PL_{\mathrm{ter}}(d)$ is the conventional terrestrial path loss between the GBS and the point beneath the UAV that can be obtained based on \eqref{eq:PL}, $\eta(\theta)$ is the excess aerial path loss, and $X_U(\theta)\sim \mathcal{N}\big(0, \sigma_U^2(\theta)\big)$ represents the excess shadowing component. Furthermore, both $\eta(\theta)$ and $\sigma_U^2(\theta)$ are modelled as functions of $\theta$ as
\begin{align}
\eta(\theta)&=A(\theta-\theta_0)\exp\left(-\frac{\theta-\theta_0}{B}\right)+\eta_0,\\
\sigma^2_U(\theta)&=a\theta +\sigma_0,
\end{align}
where $A,B,\theta_0, a, \sigma_0$ are modelling parameters that can be obtained based on curve fitting using measurement data. It was suggested in \cite{997} that $A<0$ and thus $\eta(\theta)$ firstly decreases and then increases with $\theta$. This is due to the following two effects as $\theta$ increases: on one hand,  the obstruction and scattering are reduced as the UAV moves higher, while on the other hand, increased link distance and reduced GBS antenna gain are incurred.

\subsubsection{Probabilistic LoS Channel Model}\label{sec:ProbLoSModel}
In urban environment, the LoS link between UAV and ground nodes may be occasionally blocked by ground obstacles such as buildings. 
To distinguish the different propagation environment between LoS and NLoS scenarios, another common approach is to separately model the LoS and NLoS propagations  by taking into account their occurrence probabilities \cite{657,1043,971,642}, referred to as the {\it probabilistic LoS channel model}. Such probabilities are based on the statistical modelling of the urban environment, such as the density and height of buildings. For given transmitter and receiver positions, the probability that there is an LoS link between them is given by that of no buildings being  above the ray joining the transmitter and receiver \cite{1010}. Different expressions for LoS probability and the corresponding channel models have been proposed for UAV-ground communications. In the following, we discuss two well-known models, namely {\it elevation angle-dependent probabilistic LoS model} and the {\it 3GPP GBS-UAV channel model}.

{\bf Elevation angle-dependent probabilistic LoS  model:}  With this model, the large-scale channel coefficient $\beta(d)$ in \eqref{eq:g} is modelled as \cite{642,916,980}
\begin{align}
\beta(d)=\begin{cases}
\beta_0d^{-\alpha}, & \text{LoS environment} \\
\kappa \beta_0 d^{-\alpha}, & \text{NLoS environment},
\end{cases}
\end{align}
where $\beta_0$ is the path loss at the reference distance of 1 m under LoS condition, and $\kappa<1$ is the additional attenuation factor due to the NLoS propagation\footnote{A simplification has been made here by assuming that the shadowing parameter $\kappa$ is homogeneous in NLoS conditions, whereas in practice $\kappa$ is random and has a log-normal distribution.}.
Furthermore, the probability of having LoS environment is modelled as a logistic function of the elevation angle $\theta$ as \cite{642}
\begin{align}
P_{\LoS}(\theta)=\frac{1}{1+a \exp(-b(\theta-a))}, \label{eq:PrLoS}
\end{align}
where $a$ and $b$ are modelling parameters. The probability of NLoS environment is thus given by $P_{\NLoS}(\theta)=1-P_{\LoS}(\theta)$. Equation \eqref{eq:PrLoS} shows that the probability of having a LoS link increases as the elevation angle increases, and it approaches to $1$ as $\theta$ gets sufficiently large.

\begin{figure}
\centering
\includegraphics[width=1\linewidth]{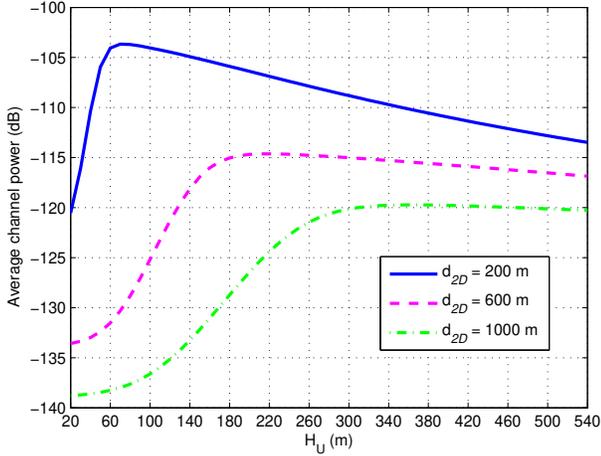}
\caption{Expected channel power versus  UAV altitude in the elevation-angle dependent probabilistic LoS channel model.}\label{F:ChannelVsAltitude}
\end{figure}

With such a model, the expected channel power, with the expectation taken over both the randomness of the surrounding buildings and small-scale fading, can be expressed as
\begin{align}
\bar g (d_{\mathrm{2D}}, H_U)&\triangleq \mathbb E[|g|^2]\\
&=P_{\LoS}(\theta) \beta_0d^{-\alpha}+(1-P_{\LoS}(\theta) )\kappa\beta_0d^{-\alpha}\\
& =\hat P_{\LoS}(\theta)\beta_0d^{-\alpha}, \label{eq:gbar}
\end{align}
where $d_{\mathrm{2D}}$ and $H_U$ are respectively the 2D distance and UAV altitude as illustrated in Fig.~\ref{F:Propagation}, $\hat P_{\LoS}(\theta)\triangleq P_{\LoS}(\theta) + (1-P_{\LoS}(\theta))\kappa$ can be interpreted as a regularized LoS probability by taking into account the effect of NLoS occurrence with the additional attenuation factor $\kappa$ \cite{980}. 
 A typical plot of $\bar g (d_{\mathrm{2D}}, H_U)$ versus $H_U$ for different $d_{\mathrm {2D}}$ values is shown in Fig.~\ref{F:ChannelVsAltitude}. It is observed that with given $d_{\mathrm{2D}}$, the expected channel power firstly increases with $H_U$, due to the enhanced chance of LoS connection, and then decreases as $H_U$ exceeds a certain threshold, at which the benefit of the increased LoS probability cannot compensate the increased path loss resulting from the longer link distance. Such a tradeoff on the UAV altitude has been extensively exploited for the UAV-mounted BS/relay placement optimization, as will be discussed in Section \ref{sec:UAVPlacement}.

{\bf 3GPP GBS-UAV channel model:} In early 2017, the 3GPP technical specification group (TSG) approved a new study item  on enhanced support for aerial vehicles via LTE networks. Detailed channel modelling between  GBSs and aerial vehicles with altitude varying from 1.5 m to 300 m has been suggested \cite{1012}, which includes the comprehensive modelling of LoS probability, path loss, shadowing, and small-scale fading. The suggested channel models are presented for three typical 3GPP deployment scenarios, namely Rural Macro (RMa), UMa, and UMi.

For all the three deployment scenarios, the LoS probability is specified by two parameters: the 2D distance $d_{\twoD}$ between the GBS and the UAV, as well as the UAV altitude $H_U$. If $H_U$ is below a certain threshold $H_1$, the model of LoS probability for conventional terrestrial users can be directly used for GBS-UAV channels. On the other hand, if $H_U$ is greater than a threshold $H_2$, 3GPP suggested a $100\%$ LoS probability. Of particular interest is the regime of $H_1\leq H_U \leq H_2$, where LoS probability is suggested as a function of $d_{\twoD}$ and $H_U$. For all the three scenarios, the LoS probability specified in \cite{1012}  can be uniformly  written as
\begin{align}
P_{\LoS}=
\begin{cases}
P_{\LoS, \mathrm{ter}}, & 1.5~ \text{m} \leq H_U \leq H_1, \\
P_{\LoS,\mathrm{U}}(d_{\twoD}, H_U), & H_1\leq H_U \leq H_2, \\
1, & H_2 \leq H_U \leq 300~  \text{m},
\end{cases}
\end{align}
where $P_{\LoS, \mathrm{ter}}$ is the LoS probability for conventional terrestrial GBS-UE channels specified in Table 7.4.2 of \cite{1017}, and $P_{\LoS,\mathrm{U}}(d_{\twoD}, H_U)$ is given by
\begin{align}
P_{\LoS,\mathrm{U}}(d_{\twoD}, H_U)=
\begin{cases}
1, & d_{\twoD} \leq d_1, \\
\frac{d_1}{d_{\twoD}} + \exp\left(\frac{-d_{\twoD}}{p_1}\right)\left(1-\frac{d_1}{d_{\twoD}}\right), & d_{2D}>d_1,
\end{cases}
\end{align}
with $p_1$ and $d_1$ given by logarithmic increasing functions of $H_U$ as specified in \cite{1012}. Note that for the three typical deployment scenarios, different values for $H_1$, $H_2$, $p_1$ and $d_1$ have been suggested. For example, $H_2=40$ m is suggested for RMa whereas $H_2=100$ m for UMa.

Based on the LoS and NLoS environment for the three deployment scenarios, the detailed path loss model and shadowing standard deviation are respectively specified in Table B-2 and Table B-3 of \cite{1012}. For moderate UAV altitude with $H_1\leq H_U\leq H_2$, the path loss exponent and shadowing standard deviation are given as decreasing functions of $H_U$, reflecting the fact of reduced obstruction and scattering as UAV moves higher. On the other hand, three different methods are suggested to model the small-scale fading, with modified values for multi-path angular spread, Rician factor, delay spread, etc \cite{1012}. Therefore, different from the other models discussed above, 3GPP model is in fact a combination of both approaches of altitude-dependent channel parameters and the probabilistic LoS channel model to characterize the different propagation environment with varying UAV altitude.

\subsubsection{Comparison of Different Models}
The choice of channel models for the study of UAV communications depends on the communication scenarios and the  purpose of the study, since they offer different tradeoffs between analytical tractability and modelling accuracy. For instance, the free-space channel model has been extensively used for the offline communication-oriented UAV trajectory design due to its simplicity and good approximation in rural environment or when the UAV altitude is sufficiently high. For urban environment, the models based on altitude/angle-dependent channel parameters and  LoS probabilities have been extensively used for theoretical analysis for UAV BS/relay placement and coverage performance optimization. On the other hand, the 3GPP model gives a very comprehensive modelling for various aspects of GBS-UAV channels, but it is more suitable for numerical simulations rather than theoretical analysis due to its complicated expressions. A qualitative comparison of the above  different UAV channel models is summarized in Table~\ref{Table:ChannelModel}.

\begin{table*}\caption{Comparison of main UAV-ground channel models.} \label{Table:ChannelModel}
\centering
\footnotesize
\begin{tabular}{|l|l|l|l|}
\hline
\bf Channel model                                                                                             & \bf Description                                                                                                                                                                                                                   & \begin{tabular}[c]{@{}l@{}}\bf Proposed application\\ \bf scenarios\end{tabular}                                                                       & \bf Pros and Cons                                                                                                                                                                                                                                \\ \hline
\begin{tabular}[c]{@{}l@{}}\it Free-space channel\\ \it model \cite{641,904}\end{tabular}                            & \begin{tabular}[c]{@{}l@{}}Channel power inversely proportional\\ to distance square, no shadowing or\\ small-scale fading\end{tabular}                                                                                       & \begin{tabular}[c]{@{}l@{}}GBS-UAV and UAV-GT\\ channels in rural area and/or\\ with very high UAV altitude\end{tabular}                       & \begin{tabular}[c]{@{}l@{}}Simple, useful for offline UAV\\ trajectory design; oversimplified\\ in urban environment\end{tabular}                                                                                                            \\ \hline
\begin{tabular}[c]{@{}l@{}}\it Altitude-dependent\\ \it channel parameters\\ \it \cite{996}\end{tabular}                & \begin{tabular}[c]{@{}l@{}}Channel modelling parameters such\\ as path loss exponent and shadowing\\ variance are functions of UAV\\ altitude\end{tabular}                                                                    & \begin{tabular}[c]{@{}l@{}}GBS-UAV in urban/suburban \\ environment\end{tabular}                                                             & \begin{tabular}[c]{@{}l@{}}Useful for theoretical analysis;\\ fails to model the change of\\ propagation environment when\\ UAV moves horizontally\end{tabular}                                                                              \\ \hline
\begin{tabular}[c]{@{}l@{}}\it Elevation angle-\\ \it dependent channel\\ \it parameters \cite{954}\end{tabular}        & \begin{tabular}[c]{@{}l@{}}Rician factor and path loss exponent\\ are functions of elevation angle\end{tabular}                                                                                                               & \begin{tabular}[c]{@{}l@{}}UAV-GT in urban/suburban\\ environment\end{tabular}                                                                 & \begin{tabular}[c]{@{}l@{}}Useful for theoretical analysis;\\ further experimental verification\\ required\end{tabular}                                                                                                                      \\ \hline
\begin{tabular}[c]{@{}l@{}}\it Depression angle-\\ \it dependent excess\\ \it path loss model \cite{997}\end{tabular}   & \begin{tabular}[c]{@{}l@{}}Excessive path loss depends on\\ depression (elevation) angle\end{tabular}                                                                                                                         & \begin{tabular}[c]{@{}l@{}}GBS-UAV channel in suburban\\ environment\end{tabular}                                                              & \begin{tabular}[c]{@{}l@{}}Small-scale fading model not\\ specified\end{tabular}                                                                                                                                                             \\ \hline
\begin{tabular}[c]{@{}l@{}}\it Elevation angle-\\ \it dependent\\ \it probabilistic LoS\\ \it model \cite{642} \end{tabular} & \begin{tabular}[c]{@{}l@{}}Separately model LoS and NLoS\\ propagations; LoS probability\\ increases with elevation angle\end{tabular}                                                                                        & \begin{tabular}[c]{@{}l@{}}UAV-GT channel in urban\\ environment with statistical\\ information of building\\ height/distribution\end{tabular} & \begin{tabular}[c]{@{}l@{}}Useful for theoretical analysis; \\ simplified shadowing; further \\experimental verification required\end{tabular}                                                                                                                      \\ \hline
\begin{tabular}[c]{@{}l@{}}\it 3GPP GBS-UAV\\ \it channel model \cite{1012}\end{tabular}                              & \begin{tabular}[c]{@{}l@{}}Separately model LoS and NLoS\\ propagations; LoS probability and\\ channel modelling parameters are\\ both functions of UAV altitude and\\ horizontal distance between GBS and\\ UAV\end{tabular} & \begin{tabular}[c]{@{}l@{}}GBS-UAV channel for UMa,\\ UMi and RMa scenarios\end{tabular}                                                       & \begin{tabular}[c]{@{}l@{}}Comprehensive models for path\\ loss, shadowing and small-scale\\ fading; useful for numerical\\ simulations but too complicated\\ for theoretical analysis or offline\\ UAV trajectory optimization\end{tabular} \\ \hline
\end{tabular}
\end{table*}

\subsubsection{Other Models and Directions of Future Work}
Besides the channel models discussed above, there are other models also proposed for UAV communications. For example, 3D geometry-based stochastic model for multiple-input multiple-output (MIMO)  UAV channels has been proposed in \cite{962}. For UAV communications above water, the classic two-ray model has been suggested \cite{1014,1016}. Furthermore, extensive channel measurements have been conducted  \cite{621,1014,1015,1016} on the air-ground channels in  the L-band (around 970 MHz) and C-band (around 5 GHz) at rather high UAV altitude, long-range (up to dozens of kilometers), and high aircraft speed (e.g. more than 70 m/s). The measurements were conducted over different environments, including  above-water environment \cite{1014}, mountainous/hilly environment \cite{1015}, and suburban and near-urban environments \cite{1016}. 
 Based on the measurement results, a modified log-distance path loss model was proposed to account for the flight direction \cite{1015,1016}
%
%
%
%
\begin{align}
\PL(d)=\PL_{\mathrm{ter}}(d)+\xi F,
\end{align}
where $\PL_{\mathrm{ter}}(d)$ is the classic log-distance path loss model as given in \eqref{eq:PL}, $\xi=-1$ if the aircraft travels towards the ground station and $\xi=1$ for travelling away from it, and $F$ is a small positive adjustment factor for direction of travel. It was explained in \cite{1015,1016} that such a correction factor is to account for the slightly different orientations of the aircraft in the two travel directions. For wide-band frequency-selective channel models, a tapped delay line (TDL) model has been developed in \cite{1015}, which includes the LoS component, a potential ground reflection and other intermittent taps.

It is worth mentioning that channel measurements and modelling for UAV communications are still  active and ongoing research. The incorporation of various other issues would be very useful for the accurate performance analysis and practical design of UAV communication systems in the future, such as the MIMO and massive MIMO channel modelling, the channel variations induced by UAV mobility and/or blade rotation, the millimeter wave (mmWave) UAV channel modelling \cite{1089}, and the wideband channel modelling in scattering environment.

Another important issue is channel estimation for UAV-ground communications. While the problem of acquiring the instantaneous channel state information (CSI) has been extensively studied for terrestrial communications, it deserves new investigations for UAV communications by exploiting the unique UAV-ground channel characteristics. For example, efficient channel estimation scheme could be designed when it is known a priori that the deterministic LoS component dominates, as typically the case for GBS-UAV channels in rurual/subrural environment, by tracking the Doppler frequency offset induced by the UAV movement. 
 As the performance of channel estimation schemes typically depends on the underlying channel models, more research endeavor is needed for devising efficient channel estimation schemes for the specific  UAV channel models  discussed above, especially for MIMO or massive MIMO based UAV communications.

\subsection{Antenna Model}\label{sec:AntModel}
Besides channel modelling,  antenna modelling at the transmitter/receiver is also crucial to the wireless communication link performance. Conventional terrestrial communication systems mostly assume that the transmitter-receiver distance is much larger than their antennas' height difference. As a result, signals are assumed to mainly propagate horizontally and antenna modelling mostly concerns the 2D antenna gain along the horizontal direction. However, 2D antenna modelling is generally insufficient for UAV communications, which involve aerial users or BSs with large-varying altitude. Instead, 3D antenna modelling is often needed to take into account both the azimuth and elevation angles for UAV-ground communications.

The simplest antenna modelling leads to the isotropic model, where the antenna radiates (or receives) equal power in all directions and the corresponding radiation pattern is a sphere in 3D. 
 Isotropic antenna is a hypothetical antenna modelling that is  mainly used for theoretical analysis as a baseline case. In practice, equal radiation in 2D only (say, in the horizontal dimension) can be easily realized (by e.g., dipole antennas), leading to the omnidirectional antenna. Isotropic or omnidirectional antenna modelling gives a reasonable approximation for scenarios when the antenna gains are approximately equal for the directions of interest. However, in modern wireless communication systems, {\it directional antennas with fixed radiation pattern} and advanced active antenna arrays for  {\it MIMO communications} are widely used.

\subsubsection{Directional Antenna with Fixed Radiation Pattern}\label{sec:fixedPattern}
For directional antenna with fixed radiation pattern, the antenna gain is completely specified by the deterministic function $G(\theta, \phi)$ with respect to the elevation and azimuth angles $\theta$ and $\phi$, respectively. There are two common approaches to realize directional antenna with fixed pattern. The first one is via carefully designing the antenna shape, such as the parabolic antennas and horn antennas. The other approach, as more commonly seen in modern wireless communications, uses antenna arrays consisting of multiple antenna elements, whose relative phase shifts are designed to achieve constructive signal superposition in desired directions. With the phase shift pre-determined and fixed, the array antenna works like a single antenna with pre-determined antenna gain in terms of $G(\theta, \phi)$.

\begin{figure*}
\centering
\begin{subfigure}{0.45\linewidth}
\centering
\includegraphics[width=80mm]{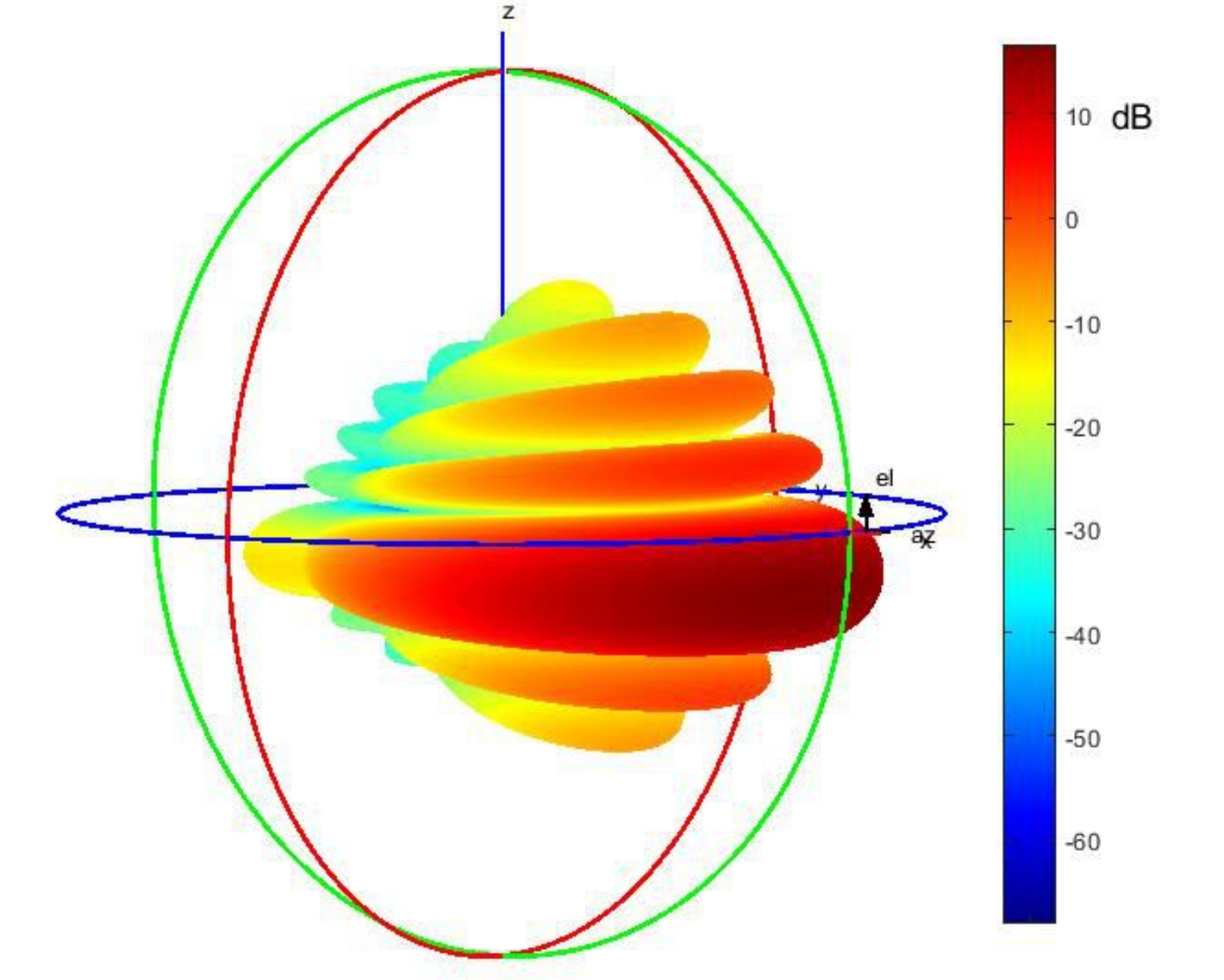}
\caption{3D plot}
\end{subfigure}
\hspace{0.02\textwidth}
\begin{subfigure}{0.45\linewidth}
\includegraphics[width=50mm]{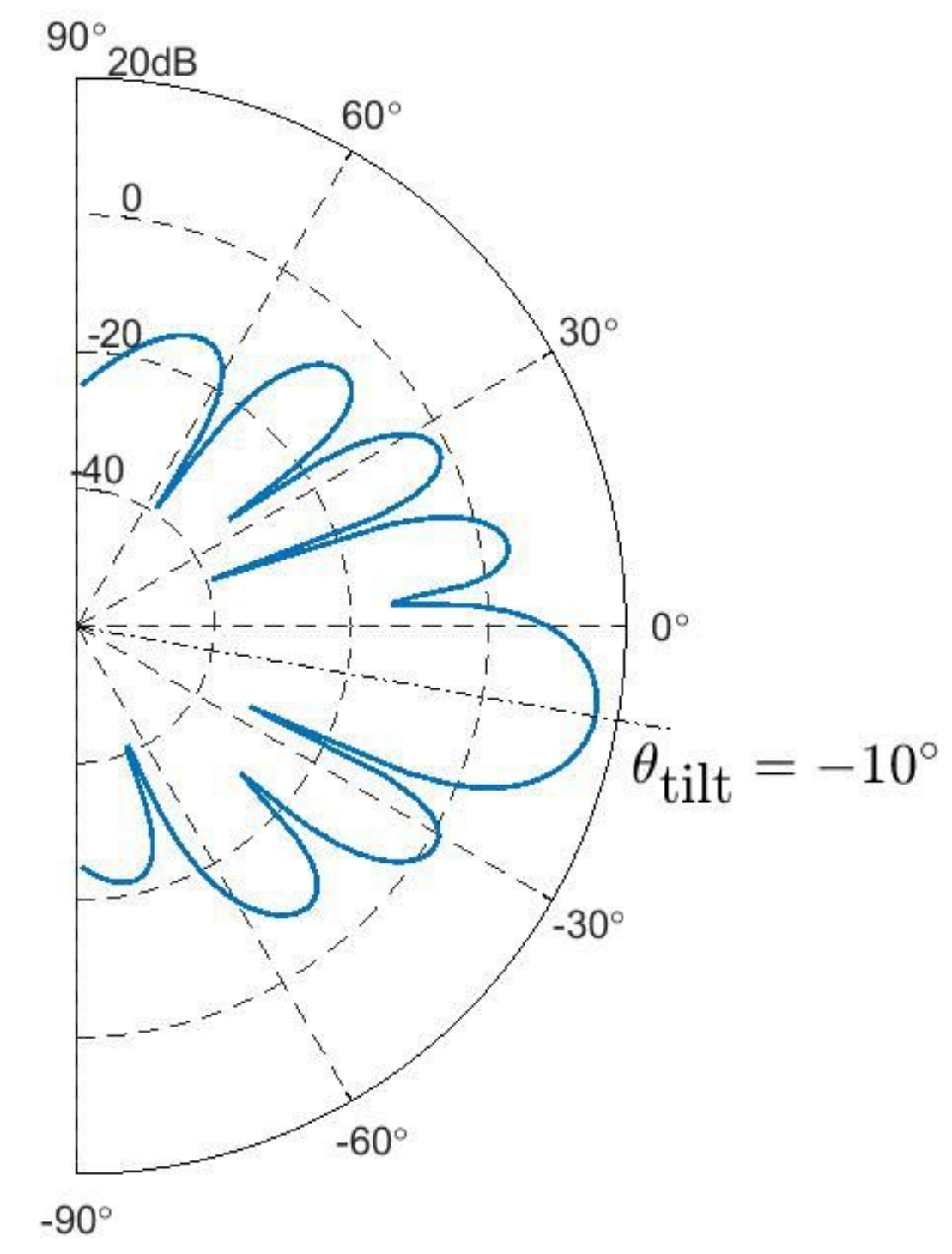}
\caption{2D plot for vertical pattern intercepted at azimuth angle $0^\circ$.}
\end{subfigure}
\caption{A typical antenna pattern of existing cellular BSs with ``fixed-pattern'' array configuration.}\label{F:BSAntPattern}
\end{figure*}

{\bf Cellular BS 3D directional antenna model:} Most existing cellular BSs are equipped with directional antennas with fixed radiation pattern, where sectorization technique is applied horizontally with e.g. three sectors for each BS site. Along the vertical dimension, the signal is usually downtilted towards the ground to cover the ground users and suppress the inter-cell interference.  For cellular BSs with fixed radiation pattern, i.e., without the full-dimensional MIMO (FD-MIMO) configuration, 3GPP suggested the array configuration  with $M$-element uniform linear array (ULA) placed vertically \cite{1012}, \cite{1024}. Each array element itself is directional, which is specified by its half-power beamwidths $\Theta_{\text{3dB}}$ and $\Phi_{\text{3dB}}$ along the vertical and horizontal dimensions, respectively. It is usually set that $\Theta_{\text{3dB}}=\Phi_{\text{3dB}}=65^{\circ}$. It is also possible that the antenna element is only directional along the vertical dimension but omni-directional horizontally  (Table 7.1-1 of \cite{1024}). To achieve antenna downtilt radiation pattern with downtilt angle $\theta_{\text{tilt}}$, where $\theta_{\text{tilt}}$ is defined relative to the horizontal plane of the BS site,  a {\it fixed} phase shift is applied  for each vertical antenna element, where the complex coefficient of the $m$th element is given by $w_m=\frac{1}{\sqrt M}\exp\left( -j \frac{2\pi}{\lambda}(m-1)d_V \sin \theta_{\text{tilt}}\right)$, where $d_V$ is the separation of adjacent antenna elements.
 It can be shown that with such phase shifts, the maximum antenna gain is achieved along the vertical direction $\theta_{\text{tilt}}$. As an illustration, Fig.~\ref{F:BSAntPattern} shows the 3D and 2D synthesized radiation pattern for an $8$-element ULA with adjacent elements separated by half-wavelength, i.e., $d_V=\lambda/2$, and $\theta_{\text{tilt}}=-10^\circ$. It can be observed that the main lobe is directing towards the elevation angle of $-10^\circ$, as desired. In addition, there are several side lobes with generally decreasing lobe gains as elevation angle increases. As will be discussed in Section~\ref{sec:cellularConnected},  these side-lobes make it possible to support UAV communications even using existing BSs with downtilt antennas.

The synthesized BS antenna gain based on the specified array configuration is quite useful for numerical simulations that require 3D BS antenna modelling, as will be illustrated in Section~\ref{sec:performanceEvalu}. However, it is difficult to be used for theoretical analysis due to the lack of closed-form expressions. To overcome this issue, one approach is to adopt the approximated two-lobe antenna model  consisting of one main lobe and one side lobe only, and all directions in each lobe have an identical antenna gain \cite{1025}. For cellular BSs serving aerial users where the vertical antenna gain is of particular interest, the two-lobe model can be expressed as
\begin{align}\label{eq:twoLobe}
G(\theta, \phi)=\begin{cases}
G, \ \theta \in [\theta_{\text{tilt}}-\frac{\Theta}{2}, \theta_{\text{tilt}}+\frac{\Theta}{2}], \\
g, \ \text{ otherwise},
\end{cases}
\end{align}
where $\Theta$ is the beamwidth of the main lobe, $G$ and $g$ are the antenna gains of the main lobe and side lobe, respectively. Note that in the above model, omnidirectional radiation is assumed in the horizontal domain \cite{983}. Such a simplified two-lobe antenna model gives a reasonable approximation for the performance analysis in conventional terrestrial systems \cite{1025}. However, it may not be sufficient for cellular UAV communications. The reason is that unlike terrestrial users which are usually served by the antenna main lobe of its closet BS, aerial users with altitude far exceeding the BS antenna height are typically served by the side lobe of a more distant BS. As a result, it is necessary to distinguish the strongest side lobe with other side lobes, since they will contribute to either desired signal or interference. Thus, more accurate antenna gain approximation than the two-lobe model is needed for improved performance analysis for cellular UAV communications \cite{1085}.

{\bf UAV directional antenna model:} In principle, similar techniques discussed above can be applied to model or synthesize the 3D directional antenna gains for UAVs. However, as the UAV orientation and its antenna boresight (i.e., the axis of maximum gain) may continuously change as it flies, additional care must be taken to define the signal direction with respect to the antenna boresight. On the other hand, for the convenience of mathematical representation and theoretical analysis, the directional antenna at UAVs is usually modelled with the main beam illuminating directly beneath the UAV and it is symmetric around the boresight \cite{940}. With the simple two-lobe approximation, the UAV directional antenna gain can be expressed as
\begin{align}\label{eq:twoLobeUAV}
G(r)=\begin{cases}
G, \ r\leq H_U \tan(\Psi), \\
g, \ \text{otherwise},
\end{cases}
\end{align}
where $r$ is the distance between the ground location of interest and the UAV's horizontal projection on the ground, and $\Psi$ is the half-beamwidth in radians (rad). In particular, the antenna gain of the main lobe can be approximated as $G\approx \frac{2.285}{\Psi^2}$ \cite{940}. Such antenna modelling has been used for both scenarios when UAV is used as aerial BS \cite{940,803,galkin2018stochastic} or aerial user \cite{983}.

\subsubsection{UAV MIMO Communications}
Different from directional antennas with fixed gain patterns, the antenna array for MIMO communications consists of elements each with a dynamically controllable complex weight coefficient. In this case, the antenna array can no longer be treated as a single antenna with fixed gain pattern as a function of the direction. Instead, the channel coefficients between different pairs of transmitting and receiving antennas are represented as a matrix, based on which transmit and receive spatial precoding/combining (also  generally known as beamforming) can be applied. This leads to the advanced MIMO communications, which have been extensively studied for terrestrial communications during the past two decades.

For UAV communications, the MIMO antenna modelling in general needs to take into account both the azimuth and elevation angles. With $M$ transmitting and $N$ receiving antennas, the MIMO channel can be modelled as
\begin{align}
\mathbf H = \sqrt{\beta(d)} \sum_{l=1}^L \mathbf a(\theta_l^R, \phi_l^R) \mathbf b^H(\theta_l^T,\phi_l^T),
\end{align}
where $L$ is the total number of multi-path, $\beta(d)$ is the large-scale channel coefficient as discussed in Section~\ref{sec:channelModel}, $\mathbf a(\cdot)\in \mathbb{C}^{N\times 1}$ and $\mathbf b(\cdot)\in \mathbb{C}^{M\times 1}$ are the array response vectors at the receiver and transmitter, respectively, $\theta_l^R$ and $\phi_l^R$ are respectively the elevation and azimuth  angles of arrival (AoAs) of the $l$th path, $\theta_l^T$ and $\phi_l^T$ are respectively the elevation and azimuth  angles of departure (AoDs) of the $l$th path.

To support MIMO UAV communications (as well as that of conventional users in high buildings), 3GPP has suggested the use of uniform rectangular arrays (URAs) at the cellular BSs \cite{1012}, \cite{1024},  with  antenna elements placed along both the vertical and horizontal dimensions. For instance, for UMa deployment scenario, one suggested BS antenna configuration is $(M_1,M_2,P)=(8,4,2)$ \cite{1012}, where $M_1$ is the number of antenna elements with the same polarization in each vertical column,  $M_2$ is the number of columns, and $P$ specifies the number of polarization dimensions, with $P=2$ for cross polarization and $P=1$ for co-polarization \cite{1088}. As 2D active arrays are used, signals in  both azimuth and elevation angles can be resolved, thus enabling 3D beamforming or FD-MIMO. As will be discussed in Section~\ref{sec:InterfMitigation}, 3D beamforming is a promising technique for dealing with the strong air-ground interference in cellular-connected UAV communications.



Conventional antenna array for MIMO communications requires one radio frequency (RF) chain for each antenna element. As the number of antennas increases as in massive MIMO and mmWave communications, the required cost and complexity  become prohibitive, in terms of hardware implementation, signal processing, and energy consumption \cite{851}. To overcome this issue, there have been significant research efforts on developing cost-aware MIMO transceiver architectures, such as analog beamforming \cite{573}, hybrid anlog/digital precoding \cite{576,578}, and lens antenna array communications \cite{823,1026}. In particular, for communication environment with limited channel paths, lens MIMO communication is able to achieve comparable performance with the fully digital MIMO communication, but with significantly reduced RF chain cost and signal processing complexity \cite{851}. This is particularly  appealing for UAV communications with the inherent multi-path sparsity due to  the high UAV altitude, as well as the imperative needs for  energy saving and  cost/complexity reduction for UAVs. Therefore, UAV MIMO communication with low-cost as well as compact and energy-efficient  transceivers is an importation problem that deserves further investigation.

\begin{figure*}
\centering
\begin{subfigure}{0.42\textwidth}
\centering
\includegraphics[width=\linewidth]{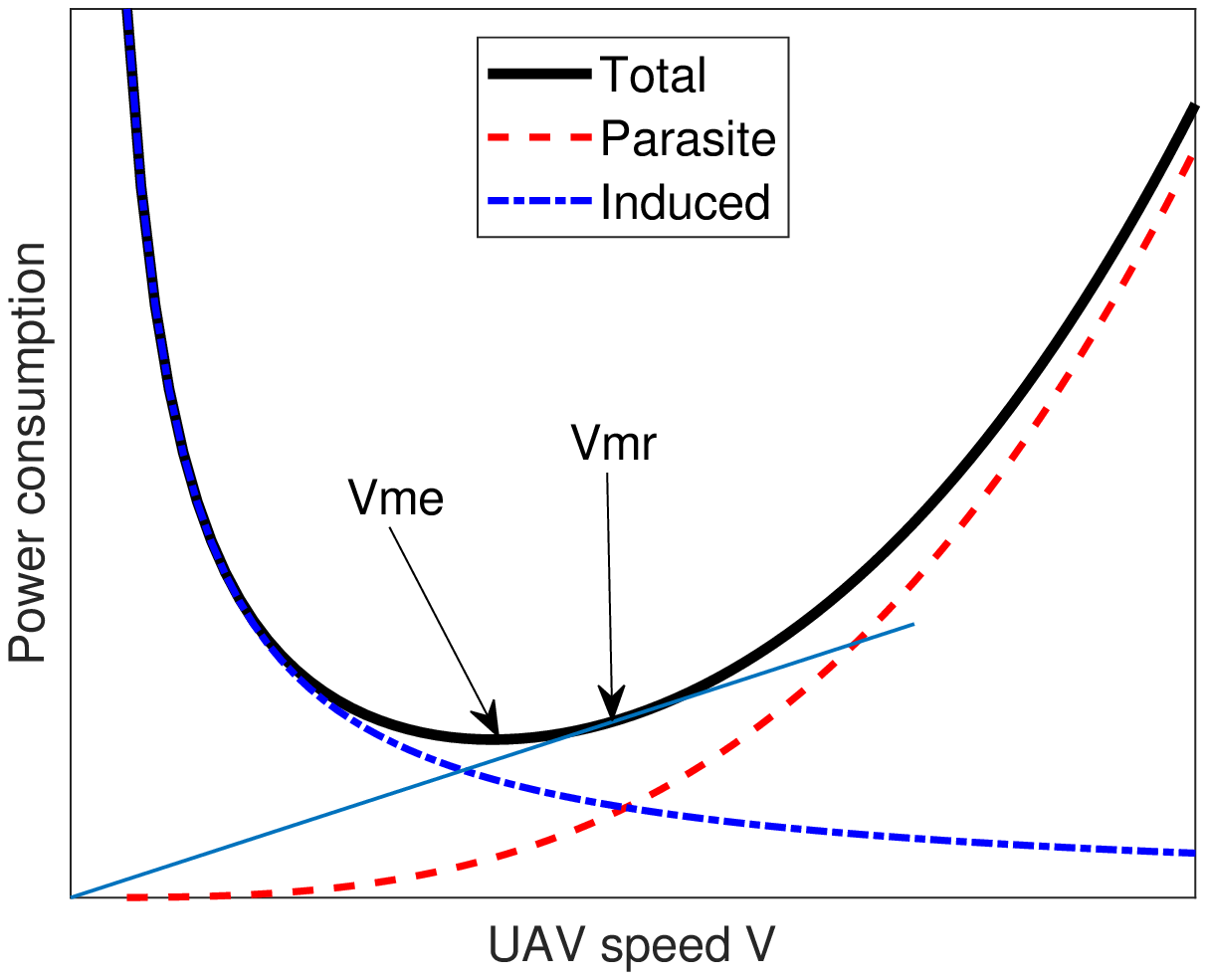}
\caption{Fixed-wing}
\end{subfigure}
\hspace{0.1\textwidth}
\begin{subfigure}{0.42\textwidth}
\includegraphics[width=\linewidth]{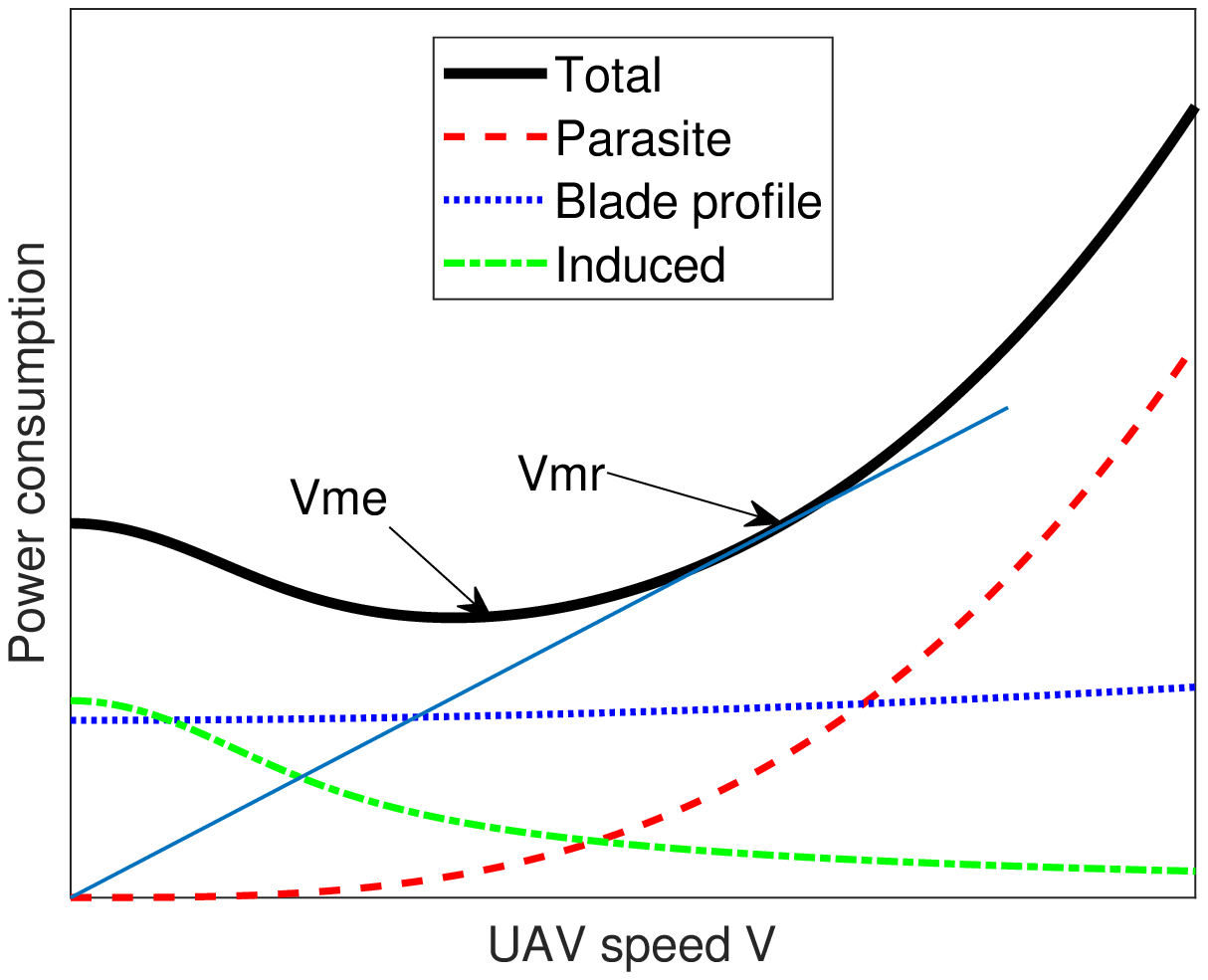}
\caption{Rotary-wing}
\end{subfigure}
\caption{Typical plot for UAV propulsion power consumption versus speed.}\label{F:PowervsSpeed}
\end{figure*}
\subsection{UAV Energy Consumption Model}\label{sec:energyModel}
One critical issue of UAV communications is the limited onboard energy of UAVs, which renders energy-efficient UAV communication particularly important. To this end, proper modelling for UAV energy consumption is crucial. Notice that besides the conventional communication-related energy consumption due to e.g., signal processing,  circuits, and power amplification, UAVs are subject to the additional propulsion energy consumption to remain aloft and move freely. Depending on the size and payload of UAVs, the propulsion power consumption may be much more significant than communication-related power expenditure. For scenarios where the communication-related energy is non-negligible, the existing models for communication energy consumption in the extensively studied terrestrial communication systems can be used for UAV communications. In contrast, the UAV propulsion energy consumption is unique for UAV communication, whereas its mathematical modelling had  received very little attention in the past.

Early works considering UAV energy consumption mainly targeted for various other applications rather than wireless communication, where empirical or heuristic energy consumption models were usually used. For example, in \cite{791}, an empirical energy consumption model was applied for the energy-aware UAV path planning for aerial imaging. To that end, experimental measurements were conducted  to study the energy consumption of a specific quadrotor UAV with different speeds. However, there is no mathematical model on UAV energy consumption suggested in \cite{791}, which makes the result difficult to be generalized for other UAVs. In \cite{790} and \cite{998}, the UAV energy (fuel) cost was modelled as the L$_1$ norm of the control force or acceleration vector, whereas in \cite{792}, it was modelled to be  proportional to the square of the UAV speed. However, no rigorous mathematical derivation was provided for such heuristic models. In fact, although  the power consumption of mobile robots moving on the ground can be modelled as a polynomial and monotonically increasing function with respect to its moving speed \cite{969}, such results are not applicable for UAVs due to their fundamentally different maneuvering mechanisms.

To fill such  gap, rigorous mathematical derivations were performed recently in \cite{904} and \cite{980} to obtain the theoretical closed-form propulsion energy consumption models for fixed-wing and rotary-wing UAVs, respectively, which are elaborated as follows.

{\bf Fixed-wing UAV energy model:} For a fixed-wing UAV in straight-and-level flight with constant speed $V$ in m/s, the propulsion power consumption can be expressed in a closed-form as \cite{904}
\begin{align}
P(V)= \underbrace{c_1 V^3}_{\text{parasite}}+\underbrace{\frac{c_2}{V}}_{\text{induced}},\label{eq:PVFixedWing}
\end{align}
where $c_1$ and $c_2$ are two parameters related to the aircraft's weight, wing area, air density, etc.

{\bf Rotary-wing UAV energy model:} On the other hand, for a rotary-wing UAV in straight-and-level flight  with speed $V$, the propulsion power consumption can be expressed as \cite{980}
\begin{align}
P(V)=& \underbrace{P_0 \left(1 + \frac{3 V^2}{U_{\mathrm{tip}}^2} \right)}_{\text{blade profile}}+ \underbrace{P_i  \left( \sqrt{1 + \frac{V^4}{4 v_0^4}}-\frac{V^2}{2v_0^2}\right)^{1/2}}_{\text{induced}} \nonumber\\
&+ \underbrace{\frac{1}{2} d_0 \rho s A V^3}_{\text{parasite}},\label{eq:PVRotaryWing}
\end{align}
where $P_0$ and $P_i$ represent the {\it blade profile power} and {\it induced power} in hovering status that depend on the aircraft weight, air density $\rho$, rotor disc area $A$, etc., $U_{\mathrm{tip}}$ denotes the tip speed of the rotor blade, $v_0$ is known as the mean rotor induced velocity in hovering, $d_0$ and $s$ are the fuselage drag ratio and rotor solidity, respectively.

The typical power versus speed  curves according to \eqref{eq:PVFixedWing} and \eqref{eq:PVRotaryWing} are plotted in Fig.~\ref{F:PowervsSpeed}(a) and Fig.~\ref{F:PowervsSpeed}(b), respectively. Several  observations can be made:
\begin{itemize}
\item First, for the extreme case with $V=0$, the required power consumption for fixed-wing UAV is infinity, whereas that for rotary-wing UAVs is given by a finite value $P_0+P_i$. This corroborates  the well-known facts that fixed-wing UAVs must maintain a minimum forward speed to remain airborne, while rotary-wing UAVs can hover with zero speed at fixed locations.
\item Secondly, for both types of UAVs, the power consumption consists of at least two components: the {\it parasite power} that is needed to overcome the parasite drag caused by the moving of the aircraft in the air, and the {\it induced power} for overcoming the induced drag resulted from the lift force to maintain  the aircraft airborne. For both UAV types, the parasite power increases in cubic with the aircraft speed $V$, while the induced power decreases as $V$ increases, with a more complicated expression for rotary-wing UAVs than fixed-wing UAVs.
\item Thirdly, compared to that for fixed-wing UAVs, the power consumption of rotary-wing UAVs has one additional term: the {\it blade profile power}, which is needed to overcome the profile drag due to the rotation of blades.
\end{itemize}
A comparison of the energy consumption models for fixed-wing versus rotary-wing UAVs is summarized in Table~\ref{Table:EnergyComparison}.

\begin{table*}\caption{Comparison of energy consumption models for fixed-wing versus rotary-wing UAVs.} \label{Table:EnergyComparison}
\begin{tabular}{|p{0.35\textwidth}|p{0.25\textwidth}|p{0.3\textwidth}|}
\hline
\bf  & \bf Fixed-Wing &  \bf Rotary-Wing \\
 \hline
  \it Convexity with respect to speed $V$ &  Convex & Non-convex \\
\hline
\it Components & Induced and parasite & Induced, parasite, and blade profile \\
\hline
 \it Power at $V=0$ & Infinity & Finite\\
\hline
\end{tabular}
\end{table*}


For both UAV types, two particular UAV speeds that are of high practical interest are the {\it maximum-endurance (ME) speed} and the {\it maximum-range (MR) speed}, which are denoted as $V_{\mathrm{me}}$ and $V_{\mathrm{mr}}$, respectively.

 {\bf ME speed:} By definition, the ME speed $V_{\mathrm{me}}$  is the optimal UAV speed that maximizes the UAV endurance for any given onboard energy, which can be obtained as
 \begin{align}
 V_{\me}&=\mathrm{arg}~\underset{V\geq 0}{\min}\ P(V).
 \end{align}
 For fixed-wing UAV, $V_{\me}$ can be obtained based on \eqref{eq:PVFixedWing} to be $V_{\me}=\left(\frac{c_2}{3c_1}\right)^{1/4}$, whereas it can be obtained numerically for rotary-wing UAVs. Note that even for rotary-wing UAVs, hovering is not the most power-conserving status since  $V_{\me}\neq 0$ in general. This may seem counter-intuitive at the first glance, but it is fundamentally due to the fact that the induced power, which is the dominant power consumption component  at low UAV speed, reduces as $V$ increases.

{\bf MR speed:} On the other hand, the MR speed $V_{\mr}$ is the optimal UAV speed that maximizes the total traveling distance with any given onboard energy, which can be obtained as
\begin{align}
V_{\mr}&=\mathrm{arg}~\underset{V\geq 0}{\min}\ E_0(V)\triangleq \frac{P(V)}{V}.
\end{align}
Note that $E_0(V)$ in Joule/meter (J/m) represents the UAV energy consumption per unit travelling distance. For fixed-wing UAVs, $V_{\mr}$ can be obtained in closed-form as $V_{\mr}=\left(\frac{c_2}{c_1}\right)^{1/4}=3^{1/4}V_{\me}$, while  it can be  obtained numerically for rotary-wing UAVs. Alternatively, for both UAV types, $V_{\mr}$ can be obtained graphically based on the  power-speed curve $P(V)$, by drawing a tangential line from the origin to the power curve that corresponds to the minimum slope (and hence power/speed ratio), as illustrated in Fig.~\ref{F:PowervsSpeed}(b). Last, it can be  shown that $V_{\mr}>V_{\me}$ for both UAV types.

{\bf Extensions and  directions of future Work:}
Note that \eqref{eq:PVFixedWing} and \eqref{eq:PVRotaryWing} only give the instantaneous power consumption for UAVs in straight-and-level flight with constant speed $V$. For UAVs flying in 3D airspace with arbitrary  trajectory $\mathbf q(t) \in \mathbb{R}^{3\times 1}$, $0\leq t \leq T$,  with $T$ denoting the time horizon of interest, the energy consumption in general depends on both the 3D velocity vector $\mathbf v(t)=\dot {\mathbf q}(t)$ and acceleration vector  $\mathbf a(t)=\ddot {\mathbf q}(t)$. In \cite{904}, for arbitrary 2D trajectory with level flight (i.e., constant altitude), a closed-form expression of energy consumption was derived for fixed-wing UAVs. The result has a nice interpretation based on the work-energy principle. 
 Based on \eqref{eq:PVRotaryWing}, similar expression can be derived for rotary-wing UAVs given  arbitrary 2D trajectory with level flight. However, for arbitrary 3D UAV trajectory $\mathbf q(t)$ with UAV climbing or descending over time, to the authors' best knowledge, no closed-form expression has been rigorously derived for the UAV energy consumption as a function of $\mathbf q(t)$. One heuristic closed-form approximation might be
\begin{align}
E(\mathbf q(t))&\approx \int_0^T P\left(\|\mathbf v(t)\|\right)dt + \underbrace{\frac{1}{2}m\left(\|\mathbf v(T)-\mathbf v(0)\|^2\right)}_{\Delta_K} \nonumber\\
&+ \underbrace{mg\left([\mathbf q(T)]_3 - [\mathbf q(0)]_3 \right)}_{\Delta_P}, \label{eq:Eqt}
\end{align}
where $P(\cdot)$ is given by \eqref{eq:PVFixedWing} or \eqref{eq:PVRotaryWing} with $\|\mathbf v(t)\|$ being the instantaneous UAV speed, $m$ is the aircraft mass, and $g$ is the gravatational acceleration. Note that the second and third terms in \eqref{eq:Eqt} represent the change of kinetic energy and potential energy, respectively. It is worth remarking that proper care should be taken while using \eqref{eq:Eqt}, since it ignores the effect of UAV acceleration/deceleration on the additional external forces (or work) that must be provided by the engine. More research endeavors are  thus needed to rigorously derive the UAV energy consumption with arbitrary 3D trajectory and evaluate  the accuracy of the approximation in \eqref{eq:Eqt}. In addition, the derivations in  \cite{904} and \cite{980} assumed a zero wind speed. The energy consumption model by taking into account the effect of wind is a challenging problem that deserves further investigation. Furthermore, it will be worthwhile to practically validate the theoretical energy consumption models by flight experiment and measurement.

\subsection{UAV Communication  Performance Metric}\label{sec:PerformanceMetrics}
For UAV communications, similar performance metrics as for  conventional terrestrial communications can be used, such as link signal-to-interference-plus-noise ratio (SINR), outage/coverage  probability,  communication throughput, delay, spectral efficiency,  energy efficiency, etc. In addition, in certain  scenarios, new  performance metrics such as UAV  mission completion time \cite{957,953,1082} and energy consumption \cite{904,980} are of practical interest. In the following, we model the above performance metrics in the context of UAV-ground communications.


\subsubsection{SINR}
\begin{figure*}
\centering
\includegraphics[width=0.9\linewidth]{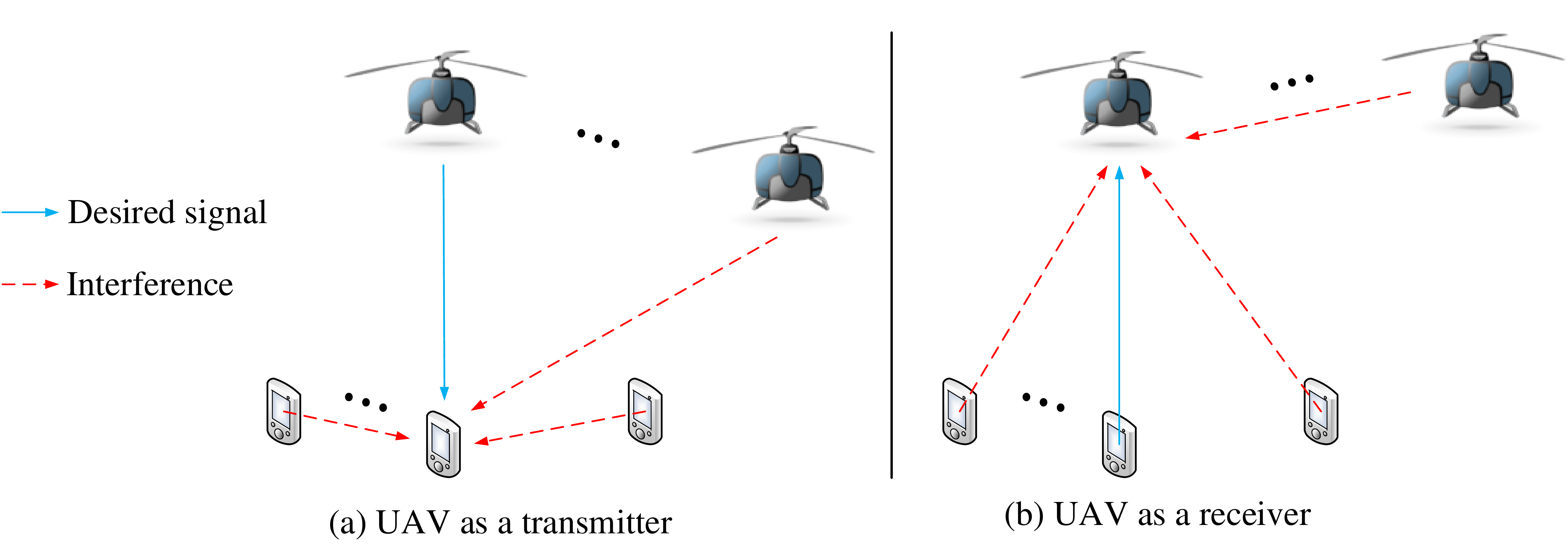}
\caption{An illustration of the possible interferences when the UAV acts as a transmitter or a receiver.}\label{F:interference}
\end{figure*}

Consider a generic UAV communication system with  $K$ co-channel UAVs communicating with their respective ground nodes (GBSs or GTs). Each UAV can be either a transmitter or a receiver. Let $\mathcal Q=\{\mathbf q_k\}_{k=1}^K$ denote the 3D locations of all the $K$ UAVs at a given time instant, and $\mathcal Q_k^-$ denote all other  UAV locations excluding that of UAV $k$. For the communication link between each UAV $k$ and its associated ground node, the interference scenarios are shown in Fig~\ref{F:interference}, for the cases that  UAV $k$ is a transmitter or a receiver. When UAV $k$ is transmitting information, the SINR at its corresponding ground receiver can be expressed as (see Fig.~\ref{F:interference}(a))
\begin{align}
\gamma_k(\mathcal Q)=\frac{S(\mathbf q_k)}{I_{\mathrm{ter}}+I_{\mathrm{aer}}(\mathcal Q_k^-)+\sigma^2}, \label{eq:gammak1}
\end{align}
where $S(\mathbf q_k)$ is the desired received  signal power that changes with  the location of UAV $k$, $I_{\mathrm{ter}}$ is the aggregate  interference from other transmitting  ground nodes, and $I_{\mathrm{aer}}(\mathcal Q_k^-)$ is the aggregate  interference from other transmitting  UAVs which changes with their locations, and $\sigma^2$ is the receiver noise power.  On the other hand, when UAV $k$ is receiving information, its  SINR can be similarly written as
\begin{align}
\gamma_k(\mathcal Q)=\frac{S(\mathbf q_k)}{I_{\mathrm{ter}}(\mathbf q_k)+I_{\mathrm{aer}}(\mathcal Q)+\sigma^2},\label{eq:gammak2}
\end{align}
where the difference with \eqref{eq:gammak1} lies in  that for this case both the terrestrial and aerial interference powers change with $\mathbf q_k$. Such a difference has the following important implication: For the air-ground link with a UAV transmitter, changing the UAV location has an effect on its own link SINR only through the desired signal power; while in the case with a UAV receiver, it affects the link SINR in a more complicated manner, through both the desired signal and undesired interference powers. This observation is useful for the design of interference-aware UAV trajectory in practice.

In both \eqref{eq:gammak1} and \eqref{eq:gammak2}, the desired signal power $S(\mathbf q_k)$ can be further written as
\begin{align}
S (\mathbf q_k) = P_tG_t(\mathbf q_k)G_r(\mathbf q_k)\beta(\mathbf q_k)|\tilde g|^2, \label{eq:Sqk}
\end{align}
where  $P_t$ is the transmission power, $G_t$ and $G_r$ are the transmit and receive antenna gains, respectively, $\beta$ is the large-scale channel power  including path loss and shadowing, and $\tilde g$ is a random variable accounting for the small-scale fading. Note that in \eqref{eq:Sqk}, $S(\mathbf q_k)$ explicitly depends on the UAV location $\mathbf q_k$ via the following three aspects: the transmit antenna gain, the receive antenna gain, and the large-scale channel power. Specifically, for directional transmission with either fixed antenna pattern or flexible beamforming, the relative position between UAV $k$ and its associated GBS/GT determines  the AoDs and AoAs of the signal propagation, which thus affects the transmit and receive antenna gains. 
On the other hand, the dependence of the large-scale channel power  $\beta(\mathbf q_k)$ on the UAV location $\mathbf q_k$ is evident based on our discussions in   Section~\ref{sec:channelModel}. 

Similarly, the dependence of the interference from the terrestrial and other aerial users on the UAVs' locations can be drawn for the above two cases, respectively.

\subsubsection{Outage Probability}
The  SINR in \eqref{eq:gammak1} and \eqref{eq:gammak2} generally varies in both space and time and thus can be modelled as a random variable. For a target SINR  threshold $\Gamma$, the outage probability for the link of an arbitrary UAV $k$ can be expressed as\footnote{Note that $1-\Prb_{\out,k}(\mathcal Q)$ is usually referred to as the non-outage or coverage probability.}
\begin{align}
\Prb_{\out,k}(\mathcal Q)&=\Pr\left(\gamma_k (\mathcal Q)< \Gamma \right).\label{eq:Prbout1}
\end{align}
Note that for the given UAV locations $\mathcal Q$, the above outage probability needs to take into account the randomness in both time (e.g., due to small-scale fading) as well as space (say, due to the LoS/NLoS probabilities). 

\subsubsection{Communication Throughput}
Assuming the capacity-achieving Gaussian signaling and Gaussian distributed interference and noise, the achievable rate for the link of UAV $k$ is given by
$R_k(\mathcal Q)=\log_2\left(1+\gamma_k(\mathcal Q)\right)$ in bits per second per Hertz (bps/Hz) with each given channel realization.
The average achievable communication throughput over the random channel realizations is thus given by
\begin{align}
\hat R_k(\mathcal Q) &= \mathbb E \left[ \log_2\left(1+\gamma_k(\mathcal Q)\right)\right].\label{eq:firstLine}
\end{align}

For the case of flying UAVs with the $K$ UAVs following certain trajectories $\mathcal Q(t)=\{\mathbf q_k(t)\}_{k=1}^K$, $0\leq t \leq T$, the average communication throughput of UAV $k$ can be written as
\begin{align}
\bar R_k \left(\mathcal Q(t)\right)&=\mathbb E \left[ \int_0^T R_k(\mathcal Q(t))dt \right] =\int_0^T \hat R_k(\mathcal Q(t)) dt. \label{eq:barRsum}
\end{align}

\subsubsection{Energy Efficiency}\label{sec:EnergyEfficiency}
Energy efficiency is measured by the number of information bits that can be reliably  communicated  per unit energy consumed, thus measured in  bits/Joule \cite{800,zhang2016fundamental,wu2016overview,qing15_wpcn_twc}. Of particular interest for UAV communications is the energy efficiency taking into account the unique UAV's propulsion energy consumption. For UAV $k$, the link energy efficiency can be defined as
\begin{align}\label{eq:EE}
\EE_k \big(\mathcal Q(t) \big) = \frac{\bar R_k\left(\mathcal Q(t)\right)}{E\big(\mathbf q_k(t)\big)+E_{\com}},
\end{align}
where the numerator is the average  communication throughput of UAV $k$ given  in \eqref{eq:barRsum} that in general depends on its own trajectory as well as those of all other co-channel UAVs due to their interference, while the denominator includes both its propulsion energy consumption $E\big(\mathbf q_k(t)\big)$ given in e.g. \eqref{eq:Eqt}  that depends only  on its own   trajectory, as well as communication energy consumption, denoted by $E_{\com}$. Besides the above per-link energy efficiency, there are also  other definitions of energy efficiency, such as the network  energy efficiency, which is given by  the sum communication throughput  of all UAVs'  links normalized by their  total (propulsion and communication) energy consumption.


\subsubsection{Special Case (Orthogonal Communication with Isotropic Antennas)}
For the purpose of illustration, we consider the special case with orthogonal communications over all  the UAV and terrestrial links, and where all UAVs and ground nodes are equipped with isotropic antennas, under which  the performance metrics discussed above can be greatly simplified. Specifically, with orthogonal communications, all UAV links are interference-free and therefore they can be considered separately. Furthermore, with isotropic transmit and receive antennas, we have $G_t(\mathbf q_k)=G_r(\mathbf q_k)=1$, $\forall \mathbf q_k$.  Then the communication throughput of each UAV $k$'s link in \eqref{eq:barRsum} can be simplified  as
\begin{align}
\bar R_k\left(\{\mathbf q_k(t)\} \right)& =\mathbb{E} \left[\int_0^T \log_2\left(1+ \frac{P|g_k(t)|^2}{\sigma^2} \right) dt \right], \label{eq:barRk}
\end{align}
where $P$ is the transmit power and $g_k(t)=\sqrt{\beta_k(t)}\tilde{g}_k(t)$ is the instantaneous channel between UAV $k$ and its associated ground node as in \eqref{eq:g}. The expression in \eqref{eq:barRk} is difficult to be directly used for the performance analysis and UAV trajectory design, because obtaining its closed-form expression as an explicit function of UAV trajectory $\mathbf q_k(t)$ is challenging. If the probabilistic LoS channel model is adopted, by applying Jensen's inequality and a homogeneous approximation of the LoS probability, we have  \cite{980}
\begin{align}
\bar R_k\left(\{\mathbf q_k(t)\} \right)& \leq  \int_0^{T} \log_2\left(1+ \frac{P\mathbb{E}\left[|g_k(t)|^2\right]}{\sigma^2 }\right)dt\\
& =\int_0^{T}  \log_2\left(1+ \frac{\tilde {\gamma}_0\hat P_{k,\LoS}(t)}{\|\mathbf q_k(t)-\mathbf w_k\|^\alpha}\right)dt \label{eq:RkApprox1}\\
& \approx  \int_0^{T}  \log_2\left(1+ \frac{\gamma_{k}}{\|\mathbf q_k(t)-\mathbf w_k\|^\alpha}\right)dt, \label{eq:RkApprox2}
\end{align}
where $\mathbf w_k\in \mathbb{R}^{3\times 1}$ denotes the location of the ground node associated with UAV $k$, $\tilde{\gamma}_0\triangleq P\beta_0/\sigma^2$, and $\hat P_{k,\LoS}(t)=P_{k,\LoS}(t) + (1-P_{k,\LoS}(t))\kappa$ is the  regularized LoS probability as defined in \eqref{eq:gbar}. Note that in \eqref{eq:RkApprox2}, a  homogeneous approximation of the LoS probability is made by letting $\hat P_{k,\text{LoS}}(t)\approx \bar P_{k,\text{LoS}}$, $\forall t$, and $\gamma_k\triangleq \tilde{\gamma}_0\bar P_{k,\text{LoS}}$. This provides a simple closed-form  approximation of the expected communication throughput as a function of trajectory $\mathbf q_k(t)$ in \eqref{eq:RkApprox2}, which can be readily used for performance analysis or trajectory optimization. As revealed in \cite{980}, such approximation gives rather satisfactory accuracy for suburban or rural environment with sufficiently  large modelling parameter $b$ in \eqref{eq:PrLoS}, and it becomes  exact for the case with LoS link only, as commonly assumed in prior work on UAV trajectory optimization \cite{641,904,919,wu2017GCjoint}. The more accurate approximation of the expected throughput over general UAV-ground channels and the corresponding UAV trajectory optimization is  nontrivial, which requires further investigation (see, e.g. \cite{1086}).

%

\subsection{Mathematical Formulation for UAV Communication and Trajectory Co-Design}\label{sec:ConstrUAVTraj}
For performance optimization of UAV communication systems, besides the traditional communication design such as multi-user transmission scheduling and resource allocation, we also need to consider the new design DoF by exploiting the UAV's high mobility. Without loss of generality, let $\mathcal R(t)$ represent all relevant variables related to communication design over time $t$, such as transmit  power, bandwidth, time allocation, beamforming, etc., and $\mathcal Q(t)$ denote  the trajectories of all UAVs over time $t$. Then a generic mathematical problem for UAV performance optimization can be formulated as
\begin{align}
\mathrm{(P1):} \quad \underset{\mathcal Q(t), \mathcal R(t)}{\max} \ &  U\left(\mathcal Q(t), \mathcal R(t) \right)\notag \\
\text{s.t.}\ & f_i\left(\mathcal Q(t) \right)\geq 0, \ i=1,\cdots, I_1,\\
& g_i \left( \mathcal R(t)\right) \geq 0, \ i=1,\cdots, I_2,\\
& h_i \left(\mathcal Q(t), \mathcal R(t)\right) \geq 0,\  i=1,\cdots, I_3.
\end{align}
Note that $ U\left(\cdot, \cdot \right)$ represents the utility function to be maximized, which could correspond to any of the performance metrics exemplified in the previous subsection,  $f_i\left(\cdot \right)$'s represent the constraints solely on the UAV trajectories, $g_i\left(\cdot\right)$'s denote the constraints solely on the communication design variables, and finally,  $h_i(\cdot, \cdot)$'s specify  the coupled constraints (if any) involving both UAV trajectories and communication variables. One typical example of such coupled constraints is the  interference constraint \cite{1087}, which limits the transmit power and trajectory of each UAV such that its interference power at any of the other UAV links' receivers needs to be below a certain threshold. Note that even without any coupled constraints on $h_i(\cdot, \cdot)$'s, the objective utility function of the above generic optimization problem $\mathrm{(P1)}$ in general has coupled trajectory and communication variables as shown in the previous subsection, which thus calls for a new UAV trajectory and communication co-design approach. Also note  that while $\mathrm{(P1)}$ is given in the general form in terms of UAV trajectory optimization, it includes the UAV placement optimization as a special case, for which we have $\mathcal Q(t)=\mathcal Q$, $\forall t$.


 While the constraints on communication design have been extensively studied in wireless communication, those on UAV mobility are relatively new.   In practice, the UAV trajectory constraints could be  due to the  aircraft mechanical limits, mission requirements, and/or flying regulations imposed by government authorities. For the purpose of illustration, we list down some typical UAV trajectory constraints for a single UAV with trajectory denoted by $\mathbf q(t)$ as follows.
\begin{itemize}
\item Minimum/maximum altitude: In the operational rules released by FAA for small UAVs \cite{936}, it is required that the aircraft should not fly more than 400 feet (122 m) above the ground level. Thus, we generally express  the maximum and minimum UAV altitude constraints as
    \begin{align}
    H_{\min}\leq [\mathbf q(t)]_3\leq H_{\max}, \forall t.\label{eq:HeightContr}
    \end{align}
\item Initial/final locations: In many scenarios, the UAV's initial and/or final locations for the time horizon of interest $[0, T]$ are predetermined when  e.g., the UAV can only be launched or landed  at certain given locations, or its mission specifies  the initial and final locations (e.g., for package delivery). Mathematically, we have
    \begin{align}
    \mathbf q(0)=\mathbf q_I, \ \mathbf q(T)=\mathbf q_F,\label{eq:StuatusConstr}
    \end{align}
    where $\mathbf q_I, \mathbf q_F\in \mathbb{R}^{3\times 1}$ are the given initial/final locations. 
\item Maximum/minimum UAV speed:
\begin{align}
V_{\min}\leq \|\mathbf v(t)\|\leq V_{\max}, \forall t, \label{eq:speedConstr}
\end{align}
where $\mathbf v(t)\triangleq \dot{\mathbf q}(t)$ denotes the the UAV velocity. Note that we usually have $V_{\min}=0$ for rotary-wing UAVs, whereas $V_{\min}>0$ for fixed-wing UAVs.
\item Maximum acceleration constraint:
\begin{align}
\|\mathbf a(t)\|\leq a_{\max}, \forall t, \label{eq:accConstr}
\end{align}
where $\mathbf a(t)\triangleq \ddot{\mathbf q}(t)$ denotes the UAV acceleration. Note that as shown in \cite{904}, for fixed-wing UAVs with banked level turn, the maximum acceleration constraint \eqref{eq:accConstr} implies a constraint on the UAV's maximum turning angle.
\item Obstacle avoidance: To ensure that the UAV avoids a given  obstacle with known location $\mathbf r\in \mathbb R^{3\times 1}$, we could impose the constraint
\begin{align}
\|\mathbf q(t)-\mathbf r\|\geq D_{1}, \forall t, \label{eq:obstacleAvoidance}
\end{align}
where $D_1$ is the safety distance with the  obstacle.
\item Collision avoidance: For a  multi-UAV system, the collision avoidance constraint among the UAVs  can be expressed as
\begin{align}
\|\mathbf q_k(t)-\mathbf q_{k'}(t)\| \geq D_2, \forall k>k', \ \forall t,
\end{align}
where $k$ and $k'$ represent the UAV indices.
\item No-fly zone: The mathematical constraints of a given no-fly zone depend on its shape. For example, if the no-fly zone is of a ball shape, constraints in the form of \eqref{eq:obstacleAvoidance} can be imposed. On the other hand, if it is a cubic volume, the following constraints need to be satisfied
    \begin{align}
    \bigcup_{i=1}^6 \mathbf a_i^T\mathbf q(t)\geq b_i,\ \forall t,
    \end{align}
    where $\{\mathbf a_i,  b_i\}_{i=1}^6$ specifies the 6 hyperplanes corresponding to the faces of the cubic volume, and for two conditions $C_1$ and $C_2$, $C_1 \bigcup C_2$ denotes that either $C_1$ or $C_2$ needs to be satisfied.
\end{itemize}

The optimization problem $\mathrm{(P1)}$ for UAV trajectory and communication co-design  is in general difficult to be solved for two main  reasons. Firstly, the formulated problem is usually non-convex with respect to communication and  trajectory variables. In fact, even by fixing one of the two types of variables, the problem may be  still non-convex over the other.  Secondly,  the optimization problem involves continuous time $t$, which results in infinite variables and thus is difficult to be directly optimized. In the following two sections, several useful techniques will be introduced to address such challenges under different UAV communication setups. 

\section{UAV-Assisted Wireless Communications}\label{sec:UAVAssisted}
\subsection{Section Overview and    Organization}
In this section, we focus on the first framework of UAV-assisted wireless communications, where UAVs are employed as aerial communication platforms to provide wireless access for terrestrial users from the sky. Under this framework, three typical use cases have been envisioned \cite{649}:

%
%
%
%
%
%

{\it UAV-aided ubiquitous coverage}, where UAVs are used as aerial BSs  to achieve seamless coverage for a given  geographical area. In this case, UAVs possess  the essential functionalities of traditional terrestrial BSs, but operate  from a much higher altitude and with more flexible 3D deployment and movement. Applications of this use case include UAV-enabled wireless coverage in remote areas, temporary traffic offloading in cellular hot spots \cite{795}, fast communication service recovery for disaster relief \cite{sharma2016intelligent}.

{\it UAV-aided relaying}, where UAVs are employed as aerial  relays to establish or strengthen the wireless connectivity between far-apart  terrestrial users or user groups. Typical applications include UAV-enabled cellular coverage extension, wireless backhaul, big data transfer, emergency response, and military operations \cite{1027}.

{\it UAV-aided information dissemination and data collection}, where UAVs are employed as aerial access points (APs) to disseminate (or collect) information to (from) ground nodes. Typical applications include UAV-aided wireless sensor network and IoT communications.

\begin{figure*}
\centering
\includegraphics[width=0.765\linewidth]{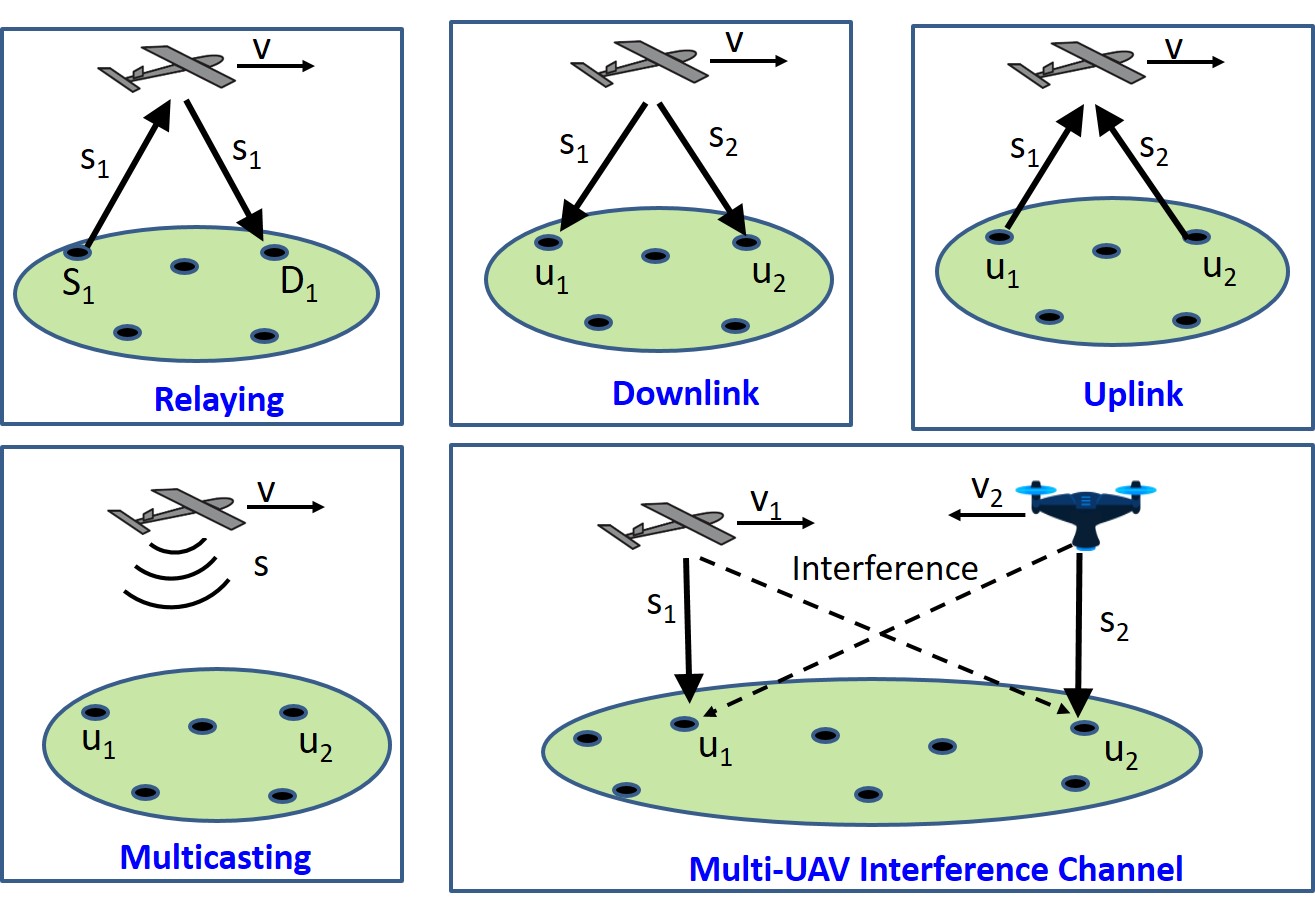}
\caption{Some basic models  for UAV-assisted communications.}\label{F:CommunicationModes}
\end{figure*}

Similar to the conventional terrestrial communications, UAV-assisted communications may have various basic models as illustrated in Fig.~\ref{F:CommunicationModes}. These include: (i) {\it UAV-enabled relaying}, where the UAV assists the communication from source node to destination node; (ii) {\it UAV-enabled downlink}, where the UAV sends independent information to multiple ground nodes; (iii) {\it UAV-enabled uplink}, where the UAV receives independent information from multiple ground nodes; (iv) {\it UAV-enabled multicasting}, where the UAV transmits common information to multiple ground nodes; (v) {\it Multi-UAV interference channel},  where there are multiple UAVs each communicating with its respective ground node subjected to the co-channel interference from the others. In general, a UAV-assisted communication system  may involve one or more of the above communication models \cite{1056}, possibly under the co-existence with other terrestrial BSs/APs/relays.

%

Depending on the UAV mobility,  research on UAV-assisted wireless communications in the  literature can be loosely classified into two categories. In the first category, UAVs are used as (quasi-)stationary aerial communication platforms which remain static for a very long period of time once deployed. Under such a setup, extensive research effort  has been devoted to UAV placement optimization and performance analysis by taking into account the unique characteristics of UAV-ground  channels. In the second category, UAVs are employed as flying platforms to serve the terrestrial users. In this case, the high  UAV mobility offers further performance enhancement over stationary UAV platforms by exploiting the new DoF of UAV trajectory design. In general, UAV trajectory optimization needs to be jointly considered with multiuser communication scheduling and  resource allocation, as formulated in (P1).  Note that while (quasi-)static UAVs may be  easier for practical implementation as they  can be tethered with  ground vehicles for stability control and reliable energy supply,  flying UAVs are more flexible for deployment and dynamic movement  to best suite the communication needs. Therefore, the practical choice of static or flying UAVs depends on the application requirement.

The remaining part of this section is organized as follows. In Section~\ref{sec:performanceAnalysis}, we review the state-of-the-art results on performance analysis of  UAV-assisted communications, for static and flying UAV platforms, respectively.  Section~\ref{sec:UAVPlacement} focuses on (quasi-)static UAV platforms, where the important problems of 2D/3D UAV placement are discussed. By exploiting the highly controllable UAV mobility, Section~\ref{sec:coDesign} introduces another important line of research for trajectory and communication  co-design for flying UAVs. Considering its importance and unique characterization in UAV communications, energy-efficient UAV communication is addressed dedicatedly in Section~\ref{sec:EnergyEfficientCommun}, which is an extension of the UAV trajectory and communication co-design discussed Section~\ref{sec:coDesign}. In Section~\ref{sec:learning}, we discuss some recent results  on designing  UAV trajectory and communication by leveraging machine learning techniques.

\subsection{Performance Analysis}\label{sec:performanceAnalysis}
For any  UAV-assisted communication system deployed or to be deployed, one important issue is to validate/evaluate its performance after/before the deployment. This can be achieved by conducting experimental field test \cite{1032}, and computer-based  simulations \cite{1028,1029,974} or theoretical analysis \cite{1031,1030,954,916,1038,1036,966,1035,1033,955,1002}, respectively. In particular, theoretical performance analysis not only predicts the expected performance of the  UAV system to be deployed, but also helps reduce the extensive simulations time. Furthermore, it can also offer useful insights and guidelines to design the UAV system and optimize its performance. Therefore, performance analysis for UAV-assisted wireless communications  has received significant research attention recently. While most works on performance analysis considered similar performance metrics such as the coverage/outage probability  given  in \eqref{eq:Prbout1} or the expected communication throughput given  in \eqref{eq:firstLine}, they differ in terms of the spatial  modelling of the aerial/ground nodes involved, the considered system  setup, as well as the UAV channel and antenna models assumed. In the following, we present some representative works on performance analysis for static and flying UAV platforms, respectively, by further addressing the two different scenarios where the locations/trajectories of UAVs are modelled deterministically or stochastically.


\subsubsection{Static UAV Platform}
\textcolor{white}{xx}

{\bf Deterministic modelling of UAV location:}
In this case, the number as well as locations of UAVs are deterministic and known a priori \cite{1031,1030,954,916,1038}, whereas their associated ground nodes could be modelled either deterministically or stochastically. 


For example, in \cite{954}, the authors considered one single UAV communicating with a ground node either directly or through a terrestrial relay. The relaying nodes are randomly distributed following a Poisson Point Process (PPP). By using Rician channel model for the small-scale fading, with elevation angle-dependent Rician $K$ factor and path loss exponent as discussed in Section~\ref{sec:channelModel}, the authors derived the outage probability as a function of the UAV altitude with three communication modes between the UAV and the associated ground node: direct air-to-ground communication, decode-and-forward (DF) relaying by a selected ground relay, and cooperative communication. It was found that the outage probability first decreases and then increases with the UAV altitude $H_U$. This is expected since at relatively small $H_U$, as $H_U$ increases, the benefits of reduced path loss exponent and increased Rician $K$ factor dominates the loss caused by the increased link distance. However, the reverse is true if $H_U$ exceeds a certain threshold. 

In \cite{916}, a UAV-enabled communication system with underlaid D2D links was studied. The UAV was assumed to hover at a given altitude serving multiple ground users in a given area, and the D2D users are spatially distributed following a PPP. With elevation-angle dependent probabilistic LoS channel model for the UAV-ground links as discussed in Section~\ref{sec:ProbLoSModel}, the outage probabilities of the downlink user served by the UAV and the D2D users were respectively derived. It was revealed that as the UAV altitude increases, the outage probability of D2D users firstly increases and then decreases, while the reverse is true for that of the downlink UAV user. This is expected due to the different roles that the UAV plays for the D2D users and the UAV user, namely as an interference  source versus the desired information  source.

{\bf Stochastic  modelling of UAV location:}
When multiple UAV BSs are used, one effective method is to model their 3D locations  stochastically according to a random point process, by which the powerful analytical tool of stochastic geometry can be applied to attain the  network-level performance analysis. Different from the deterministic UAV modelling which was typically applied for one UAV BS at a given location, the stochastic analysis of UAV network involving multiple UAV BSs needs to consider the UAV-to-ground interference, by analyzing the distance distributions of the desired and interfering links. While stochastic geometry has been extensively used for the tractable performance analysis of terrestrial communication systems, its application to UAV networks is usually more challenging. Apart from the more sophisticated UAV channel model as reviewed in Section~\ref{sec:channelModel}, the following factors also complicate the stochastic analysis of UAV networks. Firstly, as the UAV BSs can be freely deployed in 3D space, their stochastic spatial modelling in general requires 3D point process, as opposed to 2D point process for terrestrial BSs. Some initial attempts have been made along this direction with 3D PPP modelling for UAV BSs with given altitude range \cite{955,1002}. However, for analytical simplicity, most of the existing works are still based on 2D point process by assuming given UAV altitude  \cite{1036,966,1035,1033}. Secondly, while conventional terrestrial BSs are usually modelled as an infinite-size homogeneous PPP (HPPP), it is not quite suitable for UAV-enabled communications \cite{966}, especially for the current  deployment applications with typically small  number of UAV BSs. To reflect this fact, Binomial point process (BPP) has been applied for the performance analysis of finite-size UAV network \cite{1036,966}, where the number of UAV BSs is finite and known a priori.

For example, in \cite{966}, the authors derived the downlink coverage performance for a given finite number of UAV BSs deployed in a plane of fixed altitude, which are modelled as a uniform 2D BPP. 
 By assuming that each  ground user is always associated with its closest UAV BS and suffers co-channel interference from other UAVs, the closed-form expression for the coverage probability of a typical ground user was derived. It was revealed that the coverage probability degrades as the UAVs' altitude increases. The reason is that as the altitude increases, the distance differences between the communication link and the interfering links diminish, and hence the average signal-to-interference ratio (SIR) degrades. Note that such results were obtained based on the classic log-distance path loss model with the Nakagami-m small-scale fading, without taking into account the change of propagation environment as the UAV altitude varies. The impact of the UAV altitude was observed to be different, depending on the assumed UAV channel and/or directional antenna models, as reported in \cite{1035,1033}.

    Specifically, in  \cite{1033}, the UAV BSs were modelled as a 2D PPP with directional UAV antennas, which were assumed to have one main lobe and negligible side lobes as in \eqref{eq:twoLobe}. The maximum-power association rule was applied, where the user is associated to the UAV that provides the maximum power. Different from  \cite{966} as discussed above, \cite{1033} demonstrated that as the UAV altitude increases, the coverage probability firstly increases and then decreases. Similar observations have been obtained in \cite{1035}. 



\begin{table*}\caption{A summary of representative works on performance analysis of UAV-assisted wireless communications.}\label{Table:performanceAnalysis}
\centering
\footnotesize
\begin{tabular}{|l|l|l|l|l|l|}
\hline
\bf Reference & \begin{tabular}[c]{@{}l@{}}\bf Number\\ \bf of UAV\\ \bf BSs\end{tabular} & \begin{tabular}[c]{@{}l@{}}\bf Static or\\ \bf Flying\end{tabular} & \bf Setup                                                                                                                                                                   & \begin{tabular}[c]{@{}l@{}}\bf UAV channel\\ \bf Model\end{tabular}                                                                      & \bf Main Findings                                                                                                                                                         \\ \hline
{\cite{954}}   & One                                                           & Static                                                      & \begin{tabular}[c]{@{}l@{}}UAV BS serving ground\\ users with a terrestrial relay\end{tabular}                                                                            & \begin{tabular}[c]{@{}l@{}}Elevation-angle\\ dependent channel\\ parameters, Rician\\ fading\end{tabular}                        & \begin{tabular}[c]{@{}l@{}}Outage probability first \\ decreases and then increases\\ with UAV altitude\end{tabular}                                                  \\ \hline
{\cite{916}}   & One                                                           & Static                                                      & \begin{tabular}[c]{@{}l@{}}UAV BS with underlaid\\terrestrial  D2D links\end{tabular}                                                                                               & \begin{tabular}[c]{@{}l@{}}Elevation-angle\\ dependent\\ probabilistic LoS,\\ Rayleigh fading\end{tabular}                       & \begin{tabular}[c]{@{}l@{}}UAV altitude has different \\ effects on the D2D user and \\downlink UAV user\\ performances\end{tabular}                              \\ \hline
{\cite{966}}   & Multiple                                                      & Static                                                      & \begin{tabular}[c]{@{}l@{}}UAV BSs at the same\\ altitude modelled as a BPP;\\ each user associates with the\\ closest UAV BS\end{tabular}                              & \begin{tabular}[c]{@{}l@{}}Log-distance path\\ loss, Nakagami-m\\ fading\end{tabular}                                            & \begin{tabular}[c]{@{}l@{}}Coverage probability degrades \\as UAV altitude increases\end{tabular}                                                                   \\ \hline
{\cite{1035}}   & Multiple                                                      & Static                                                      & \begin{tabular}[c]{@{}l@{}}UAV BSs modelled as a PPP\\ with the same altitude;\\ directional UAV antenna;\\ each user associates with the\\ closest UAV BS\end{tabular} & \begin{tabular}[c]{@{}l@{}}Elevation-angle\\ dependent\\ probabilistic LoS\\ and shadowing, no\\ small-scale fading\end{tabular} & \begin{tabular}[c]{@{}l@{}}Coverage probability firstly\\ increases and then decreases\\ with UAV altitude\end{tabular}             \\ \hline
{\cite{1033}}   & Multiple                                                      & Static                                                      & \begin{tabular}[c]{@{}l@{}}UAV BSs modelled as a PPP\\ with a given altitude,\\ directional UAV antenna;\\ maximum-power based\\ association\end{tabular}               & \begin{tabular}[c]{@{}l@{}}Probabilistic LoS,\\ Nakagami-m\\ fading\end{tabular}                                                 & \begin{tabular}[c]{@{}l@{}}Coverage probability firstly\\ increases and then decreases\\ with UAV altitude\end{tabular}                                               \\ \hline
{\cite{1020}}   & One                                                           & Flying                                                      & \begin{tabular}[c]{@{}l@{}}UAV relay following a\\ circular trajectory periodically\end{tabular}                                                                      & \begin{tabular}[c]{@{}l@{}}Log-distance path\\ loss model, Rician\\ fading\end{tabular}                                          & \begin{tabular}[c]{@{}l@{}}With a periodic circular UAV\\ trajectory, variable-rate\\ communication outperforms\\ fixed-rate communication\end{tabular}                 \\ \hline
{\cite{887}}   & One                                                           & Flying                                                      & \begin{tabular}[c]{@{}l@{}}UAV BS following a line\\ trajectory periodically\end{tabular}                                                                               & \begin{tabular}[c]{@{}l@{}}Free space path\\ loss\end{tabular}                                                                   & \begin{tabular}[c]{@{}l@{}}A tradeoff between throughput\\ and access delay\end{tabular}                                                                              \\ \hline
{\cite{1022}}   & Multiple                                                      & Flying                                                      & \begin{tabular}[c]{@{}l@{}}UAV BSs at the same\\ altitude with stochastically\\ modelled movement\end{tabular}                                                           & \begin{tabular}[c]{@{}l@{}}Log-distance path\\ loss model,\\ Nakagami-m\\ fading\end{tabular}                                    & \begin{tabular}[c]{@{}l@{}}Stochastically flying UAV BSs \\achieve similar coverage\\ performance as static BSs, but \\with significantly  reduced AFD\end{tabular} \\ \hline
\end{tabular}
\end{table*}

\subsubsection{Flying UAV Platform}
For the performance evaluation of flying UAV platforms, some early results  on field experiments \cite{1037,1021} or computer  simulations \cite{1001} were reported. Recently, the theoretical performance analysis of flying UAV platforms in various setups has  received growing interest. Most of such works were based on the deterministic modelling of UAV trajectories \cite{916,1020,887,1007}, whereas there was also an  initial attempt to consider stochastically modelled random UAV trajectories \cite{1022}.



{\bf Deterministic  modelling of UAV trajectory:}
In \cite{1020}, a UAV-assisted relaying system was studied, where a fixed-wing UAV is employed to assist the communication between two ground nodes without the direct communication link. As fixed-wing UAV must maintain a forward speed to remain airborne, the UAV was assumed to fly along a circle at a constant height and thus its location changes periodically. By considering DF relaying and delay-sensitive applications such that the UAV forwards the information as soon as it receives and decodes it, the authors in \cite{1020} derived the link outage probability by assuming Rician fading channel models. It was found that with the periodic circular UAV trajectory, the variable-rate communication outperforms the fixed-rate communication.

In \cite{887}, the authors studied the UAV-assisted communication system with the UAV flying cyclically among the ground users, thus resulting in a cyclical variation pattern of each  UAV-user channel strength. By considering the basic setup where  all ground users are located in a line and served by the UAV alternately, a tradeoff between the average access delay and the network common  throughput was revealed. This  study was further extended in \cite{1007} for a hybrid wireless network consisting of a flying UAV BS and a conventional terrestrial BS, where the UAV flies cyclically along the cell edge to help offload the data traffic from the terrestrial BS. 



{\bf Stochastic  modelling of UAV trajectory:}
Different from the above works with deterministic UAV flying trajectories, the performance of flying UAV BSs was analyzed in \cite{1022} with stochastic UAV flying trajectories. To this end, the stochastic geometry analysis of \cite{966} was extended to the case of flying UAV BSs, where the UAVs are assumed to fly following stochastic trajectory processes, i.e., at any snapshot, the UAV BSs can be modelled as a BPP. Two types  of stochastic trajectory processes were considered, namely spiral and oval processes. The results demonstrated that compared to the static UAV BSs, the stochastically moving UAV BSs achieve comparable coverage performance but with  significantly reduced channel  average fade duration (AFD).

Table~\ref{Table:performanceAnalysis} summarizes the above  representative works on performance analysis for both static and flying UAV-assisted wireless communication systems.

\subsection{UAV Placement}\label{sec:UAVPlacement}

In this subsection, we focus on  (quasi-)static UAV communication platforms, where the locations of UAVs remain unchanged for the duration of interest. For such setups, one important design problem is to determine the UAV locations to achieve the best communication performance, which has received extensive research attention recently \cite{642,803,793,922,937,999,940,914,917,886,1083}. Different from the conventional 2D cell planning with terrestrial BSs of typically pre-determined BS heights, the altitude of UAV BS can be flexibly determined, thus leading to the new 3D BS placement problems. Besides, the unique characteristics of UAV-ground channels as discussed in Section~\ref{sec:channelModel} also need to be considered for the  UAV placement.

The optimal altitude of UAV communication platforms depends on the propagation environment and the antenna models. For the simple isotropic antennas with the free-space path loss model, it is not difficult to see that the UAV placed at the minimum possible altitude $H_{\min}$ leads to the smallest path loss and thus   the best communication channel with the GTs. On the other hand, for urban environment with signal blockage and multipath scattering, the optimization of UAV BS altitude becomes non-trivial. Specifically, as the UAV altitude increases, there are less obstacles and therefore the communication link is more likely to be dominated by the strong LoS component, e.g., with larger Rician $K$ factor and/or higher LoS probability. However, as the altitude further increases, the benefit of having stronger LoS link cannot compensate for the higher path loss incurred due to the increased link distance, as illustrated in Fig.~\ref{F:ChannelVsAltitude}.  Depending on how much information pertaining the user locations is available at the UAV BS, we consider  the following different scenarios for UAV placement optimization assuming: (i) No user location information (ULI); (ii) Perfect ULI; and (iii) Partial ULI.

\subsubsection{No ULI} When there is completely no ULI available, the UAV placement is usually optimized to maximize the geographic area covered by the UAV.

 One representative work along this line is \cite{642}, which focused on the 1D altitude optimization. By assuming the elevation angle-dependent probabilistic LoS channel model, the coverage radius $R_{\cov}$ of a UAV BS is defined as the maximum horizontal distance from the UAV projected location on the ground so that the expected path loss is below a given threshold, where the expectation is taken with respect to the LoS and NLoS occurrence probabilities. An implicit expression was derived between $R_{\cov}$ and the UAV altitude $H_U$ in \cite{642}, and it numerically showed that $R_{\cov}$ firstly increases and then decreases with $H_U$. Thus, an optimal UAV altitude exists that is in general between the minimum and maximum allowable altitude.

 In \cite{643}, by assuming that two UAV BSs are employed to serve a target rectangular area on the ground, the 3D locations of both UAVs were determined to maximize the fraction of the area covered by the UAV BSs. For the interference-free scenario, the two UAVs were placed so that they are separated as much as possible, while ensuring that neither UAV covers outside the target area. The above work was then extended to \cite{803}, where by using directional UAV antenna model similar as \eqref{eq:twoLobeUAV}, the 3D locations of a given number of UAV BSs were obtained to maximize the total coverage area by leveraging the circle packing problem.

\subsubsection{Perfect ULI}
On the other hand, when the ULI or even the instantaneous CSI of the served GTs is known, the UAV placement can be designed for various objectives, such as maximizing the number of covered users \cite{793,922,937}, maximizing the communication throughput \cite{999,940,914}, or minimizing the number of required UAVs \cite{917,886}.

UAV placement optimization for maximizing the number of covered users can be usually formulated as mixed integer nonlinear programming  \cite{793,922,937}, with the binary variables indicating whether the users are served by each  UAV or not. Such formulations were extended in \cite{999}, which took into account the limited backhaul capacity of the UAV BSs and the rate requirement for different users.

The UAV placement may also be designed to directly maximize the system throughput \cite{999,940,914}. By assuming the free-space path loss channel model and the  directional UAV antenna with dynamically adjustable beamwidth \eqref{eq:twoLobeUAV}, the authors in \cite{940} investigated the joint UAV altitude and beamwidth optimization problems for throughput maximization in three basic  multiuser communication models, namely {\it downlink multicasting} (MC) where UAV sends common information to all ground users, {\it downlink broadcasting} (BC) where UAV sends independent information to different users, and {\it uplink multiple access} (MAC) where each user sends independent information to the UAV. It was revealed that for the considered UAV directional antenna model, the UAV altitude should be set as the maximum possible value for downlink MC, but the minimum possible value for downlink BC, while it has no effect to the throughput performance of uplink MAC.

Another sensible design objective for UAV placement is to minimize the number of required UAVs while satisfying the communication requirement of ground users \cite{917,886}. In \cite{917}, by assuming that the user rate requirements are known, a heuristic algorithm based on particle swarm optimization was proposed to find the 3D locations of UAV BSs to minimize the number of UAV BSs. In \cite{886}, by assuming that the UAVs hover at a fixed altitude, an efficient spiral UAV placement algorithm was proposed to find the minimum number of UAV BSs and their 2D horizontal locations to ensure that all GTs are covered by at least one UAV. The main idea is to place the UAV BSs successively, starting from the area perimeter of those uncovered GTs and moving inwards along a spiral path toward the center of the area. Compared to the benchmark strip-based placement, the proposed spiral based algorithm better utilizes the location information of GTs and thus generally leads to less  number of required UAV BSs.

\subsubsection{Partial ULI}
In many practical scenarios, instead of perfect ULI, it is more feasible  to gain the partial information regarding the user locations, such as the statistic distribution of the users or some side information at each locations realization. In \cite{1075}, a traffic-aware adaptive UAV deployment scheme was proposed, where starting from the current location, the displacement direction and distance of the UAV was optimized. The proposed scheme requires very limited knowledge of the  GT locations at each  realization, namely only the number of GTs for each given sub-area, rather than their  exact ULI. Based on the simple majority-vote rule, the UAV adjusts its location towards the sub-area that has the largest  number of GTs, with the displacement distance optimized to maximize the average throughput or the successful transmission probability for all GTs in the network.

A summary of the above  representative works on UAV placement is given in Table~\ref{Table:UAVPlacement}.

\begin{table*}\caption{A summary of representative works on UAV placement.}\label{Table:UAVPlacement}
\centering
\footnotesize
\begin{tabular}{|l|l|l|l|l|}
\hline
\bf Reference          & \begin{tabular}[c]{@{}l@{}}\bf Number\\ \bf of UAV\\ \bf BSs\end{tabular} & \bf Design variable & \bf Design Objective                                                                                                                                                     & \bf Main techniques                                                                                                        \\ \hline
{\cite{642} }            & Single UAV                                                    & 1D altitude     & Maximize coverage area                                                                                                                                               & \begin{tabular}[c]{@{}l@{}}Implicit expression between\\ coverage radius and UAV \\ altitude\end{tabular}              \\ \hline
{\cite {643}}           & Two UAVs                                                      & 3D location     & \begin{tabular}[c]{@{}l@{}}Given a target rectangular area,\\ maximize the fraction of coverage area \\using two UAV BSs\end{tabular}                                & \begin{tabular}[c]{@{}l@{}}Maximum separation of the\\ two UAV BSs subject to\\ coverage area constraint\end{tabular}  \\ \hline
{\cite{793,922,937} } & Single UAV                                                    & 3D location     & \begin{tabular}[c]{@{}l@{}}Given user locations, maximize the\\ number of served users\end{tabular}                                                                  & \begin{tabular}[c]{@{}l@{}}Mixed-integer nonlinear\\ programming\end{tabular}                                          \\ \hline
{\cite{803} }            & \begin{tabular}[c]{@{}l@{}}Multiple\\ UAVs\end{tabular}       & 3D location     & Maximize the total coverage area                                                                                                                                     & Circle packing                                                                                                         \\ \hline
{\cite{999}}           & Single UAV                                                    & 3D location     & \begin{tabular}[c]{@{}l@{}}With UAV backhaul capacity constraint,\\ maximize the number of served users \\ or sum-rate\end{tabular}                                  & Branch-and-bound method                                                                                                \\ \hline
{\cite{914}}           & Single UAV                                                    & 2D location     & \begin{tabular}[c]{@{}l@{}}With UAV serving as relay, maximize\\ the throughput or minimize\\ communication power\end{tabular}                                       & \begin{tabular}[c]{@{}l@{}}Smart local search for LoS\\ propagation\end{tabular}                                       \\ \hline
{\cite{940}}            & Single UAV                                                    & 3D location     & \begin{tabular}[c]{@{}l@{}}Joint altitude and beamwidth\\ optimization for three basic multiuser\\ communication models\end{tabular}                                 & \begin{tabular}[c]{@{}l@{}}Closed-form throughput \\ expressions in terms of UAV\\ altitude and beamwidth\end{tabular} \\ \hline
{\cite{917}}           & \begin{tabular}[c]{@{}l@{}}Multiple\\ UAVs\end{tabular}       & 3D location     & \begin{tabular}[c]{@{}l@{}}Minimize the number of UAVs to satisfy\\ the user rate requirement\end{tabular}                                                           & Particle swarm optimization                                                                                            \\ \hline
{\cite{886}}           & \begin{tabular}[c]{@{}l@{}}Multiple\\ UAVs\end{tabular}       & 2D location     & \begin{tabular}[c]{@{}l@{}}Minimize the number of UAVs to ensure\\ that all GTs are covered\end{tabular}                                                             & Spiral BS placement                                                                                                    \\ \hline
{\cite{1075}}           & Single UAV                                                    & 2D location     & \begin{tabular}[c]{@{}l@{}}Optimize UAV displacement direction\\ and distance for maximizing average\\ throughput or success transmission\\ probability\end{tabular} & \begin{tabular}[c]{@{}l@{}}UAV displacement to the\\ sub-area with the most users\end{tabular}                         \\ \hline
\end{tabular}
\end{table*}

\subsection{Trajectory and Communication Co-Design}\label{sec:coDesign}

Compared to conventional terrestrial BSs or quasi-stationary UAV BSs,  flying UAV communication platforms offer an additional DoF via UAV trajectory optimization. Note that the concept of exploiting node mobility for boosting communication performance is not new, which has been  studied in MANET  \cite{651} or ground mobile robotics \cite{969}, \cite{970}. However, there are some important differences between such systems and the UAV communication system. Firstly, nodes moving on the ground are usually subject to many obstacles, which greatly limits their flexibility for path adaption. Therefore, most existing works on exploiting ground node mobility assumed either the random mobility model \cite{651} or deterministic mobility along pre-determined path \cite{970}. In contrast, UAVs moving in 3D airspace offers more design DoF in path/trajectory optimization for communication performance improvement. Secondly, due to the generally rich scattering environment, the wireless channels for ground robotic communications usually suffer from severe fading, which is difficult to be efficiently predicted at any location. In contrast, the UAV-ground communications often contain strong LoS link, making it easier for channel prediction and thus facilitating the offline trajectory optimization. Last but not least, robots and UAVs differ significantly in terms of energy consumption model, as discussed in Section~\ref{sec:energyModel}. The above differences are summarized in Table~\ref{Table:mobility}, which render the communication-aware  UAV trajectory optimization significantly different from that for the traditional terrestrial communications.

\begin{table*}\caption{Exploiting mobility in UAV versus terrestrial communication systems.} \label{Table:mobility}
\centering
\begin{tabular}{|l|l|l|}
\hline
                                                   & \bf Terrestrial System                                                                                                                                                                                         & \bf UAV System                                                                                                                                    \\ \hline
\it Mobility                                                        & \begin{tabular}[c]{@{}l@{}}{\small $\bullet$} Nodes usually move randomly\\  ~~\,(e.g., in a MANET)\\ {\small $\bullet$} Nodes move with predetermined\\  ~~\,path (e.g., mobile robotics)\\ {\small $\bullet$} Very restrictive path planning\end{tabular} & \begin{tabular}[c]{@{}l@{}}{\small $\bullet$} UAV mobility highly controllable/\\~~\,predictable\\ {\small $\bullet$} More flexible path adaptation in 3D\\  ~~\,space\end{tabular} \\ \hline
\begin{tabular}[c]{@{}l@{}}\it Communication\\ \it channel\end{tabular} & \begin{tabular}[c]{@{}l@{}}{\small $\bullet$} Severe shadowing and multipath\\   ~~\,fading\\ {\small $\bullet$} Difficult to predict offline\end{tabular}                                                                                        & \begin{tabular}[c]{@{}l@{}}{\small $\bullet$} Less shadowing and fading\\ {\small $\bullet$} More predictable\end{tabular}                                                      \\ \hline
\it Energy consumption                                              & \begin{tabular}[c]{@{}l@{}}{\small $\bullet$} Polynomial and  increasing \\   ~~\,function of speed\end{tabular}                                                                                                                       & {\small $\bullet$} More complicated (see Section II-C)                                                                                                             \\ \hline
\end{tabular}
\end{table*}

For UAV-assisted communications, the UAV trajectory optimization is in general closely coupled with  communication scheduling and resource allocation, for which a generic optimization problem has been formulated in Section~\ref{sec:ConstrUAVTraj}. Note that problem $\mathrm{(P1)}$ is difficult to be efficiently and optimally  solved due to its non-convexity in general. In the following, we present several useful techniques to address  problem $\mathrm{(P1)}$ for UAV-assisted communications. In particular, we first present the classic travelling salesman problem (TSP) and pickup-and-deliver problem (PDP) as two  useful techniques for initial UAV path planning, and then introduce the more general optimization framework with block coordinate descent (BCD) and successive convex approximation (SCA) techniques.

\subsubsection{TSP and PDP for Initial Path Planning}
In general, UAV trajectory optimization involves two aspects: {\it path planning} to determine the flying route, and {\it speed optimization} that essentially determines how much time should be spent on each location along the route. While path planning has been extensively studied for UAV systems, early works mainly focused on UAV navigation applications, rather than targeting for optimizing the communication performance \cite{620,790,920,921}. For such scenarios, mixed-integer linear program (MILP) has been shown to be an effective approach \cite{790,1006,998,1004}. Recently, there have been a handful of works on UAV  path planning for  communication purposes by partially optimizing some of the trajectory parameters. For example, in \cite{658} and \cite{659}, by assuming that the UAV flies with a constant speed, the UAV's heading (or flying direction) was optimized for UAV-based wireless relaying and uplink communications, respectively. In \cite{788}, a UAV-based mobile relay node was considered for forwarding  independent data to different user groups. The data downloading volume as well as the relay trajectory in terms of the visiting sequence to the different user groups were optimized by  a genetic algorithm. In \cite{619} and \cite{785}, the deployment/movement of UAVs was optimized to improve the network connectivity of a UAV-assisted ad-hoc network. More recently, the use of more  powerful optimization techniques for communication-aware UAV trajectory design has received growing interest, discussed as follows.

\begin{figure}
\centering
\includegraphics[width=0.85\linewidth]{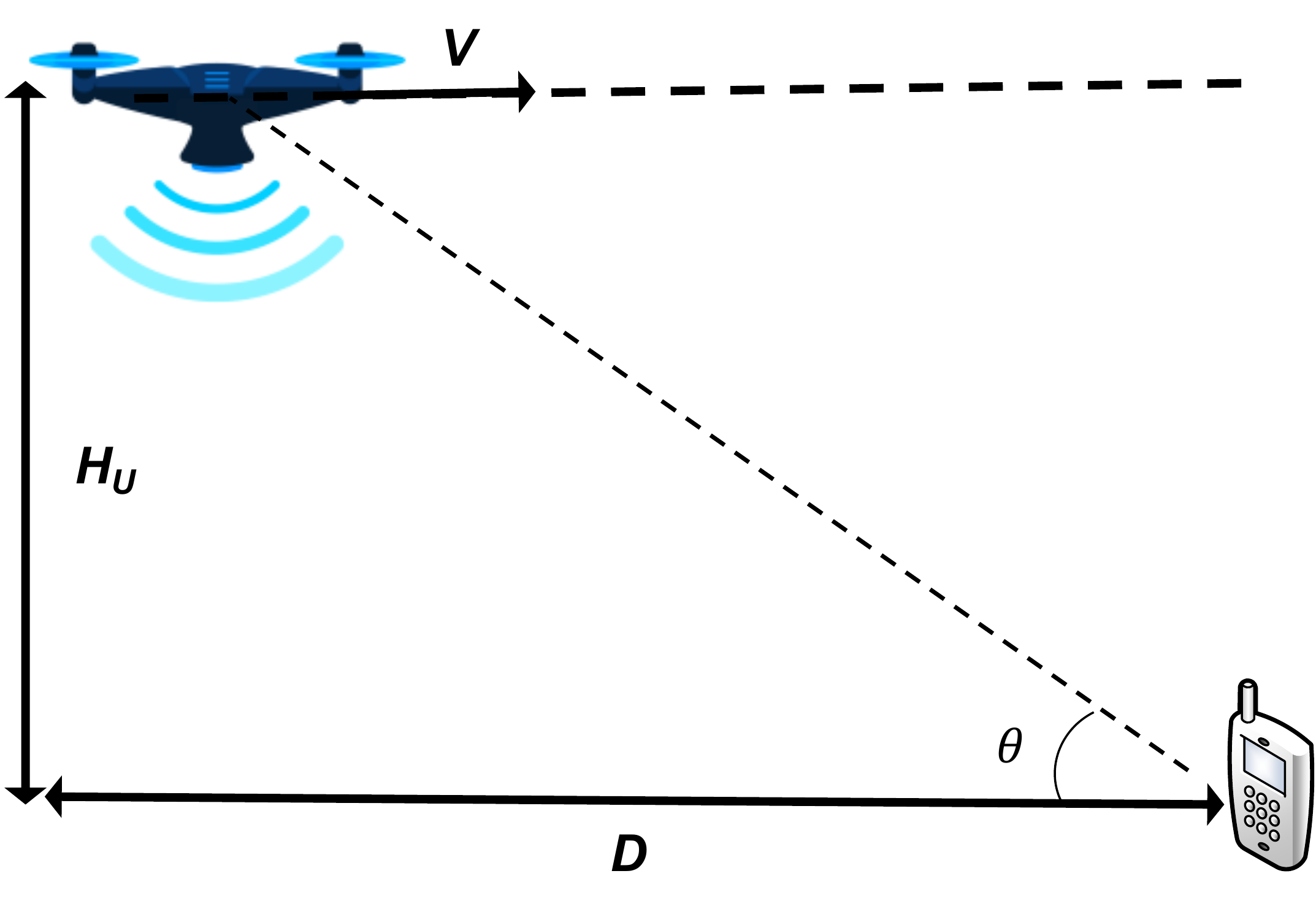}
\caption{A point-to-point link with the UAV flying towards the GT.}\label{F:StraightPath}
\end{figure}

\begin{figure*}
\centering
\begin{subfigure}{0.3\textwidth}
\centering
\includegraphics[width=\linewidth]{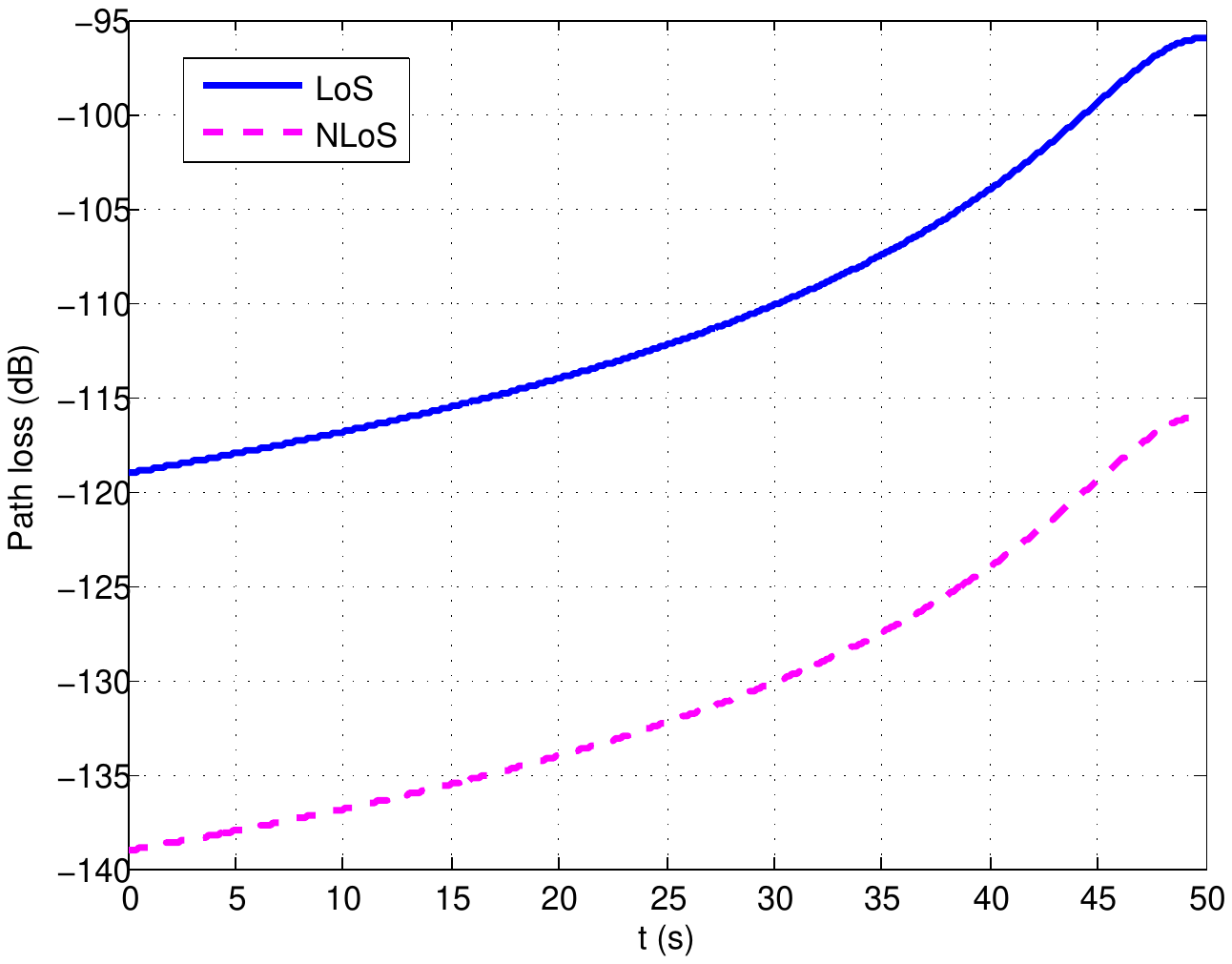}
\caption{}
\end{subfigure}
\hspace{0.01\textwidth}
\begin{subfigure}{0.3\textwidth}
\includegraphics[width=\linewidth]{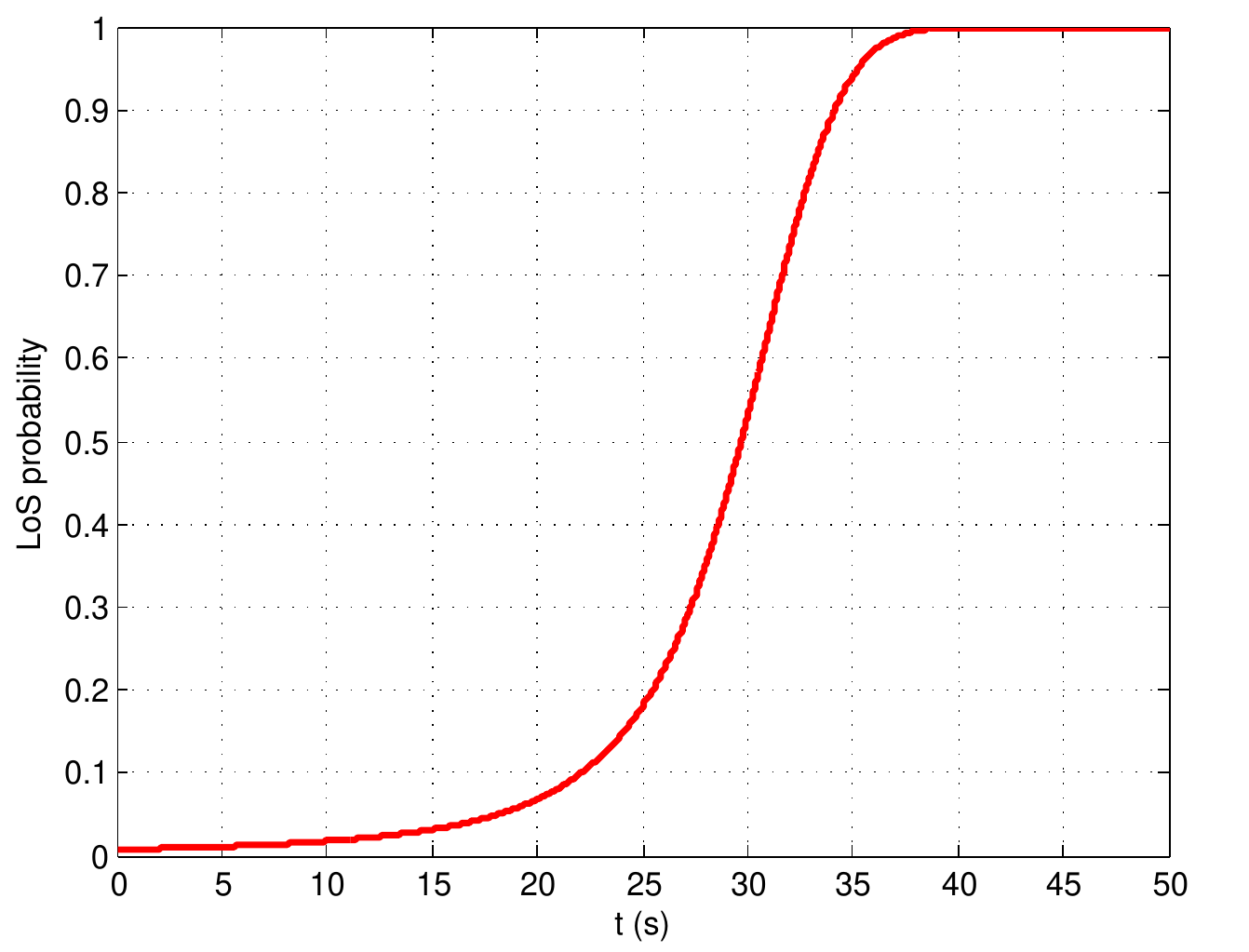}
\caption{}
\end{subfigure}
\hspace{0.01\textwidth}
\begin{subfigure}{0.3\textwidth}
\includegraphics[width=\linewidth]{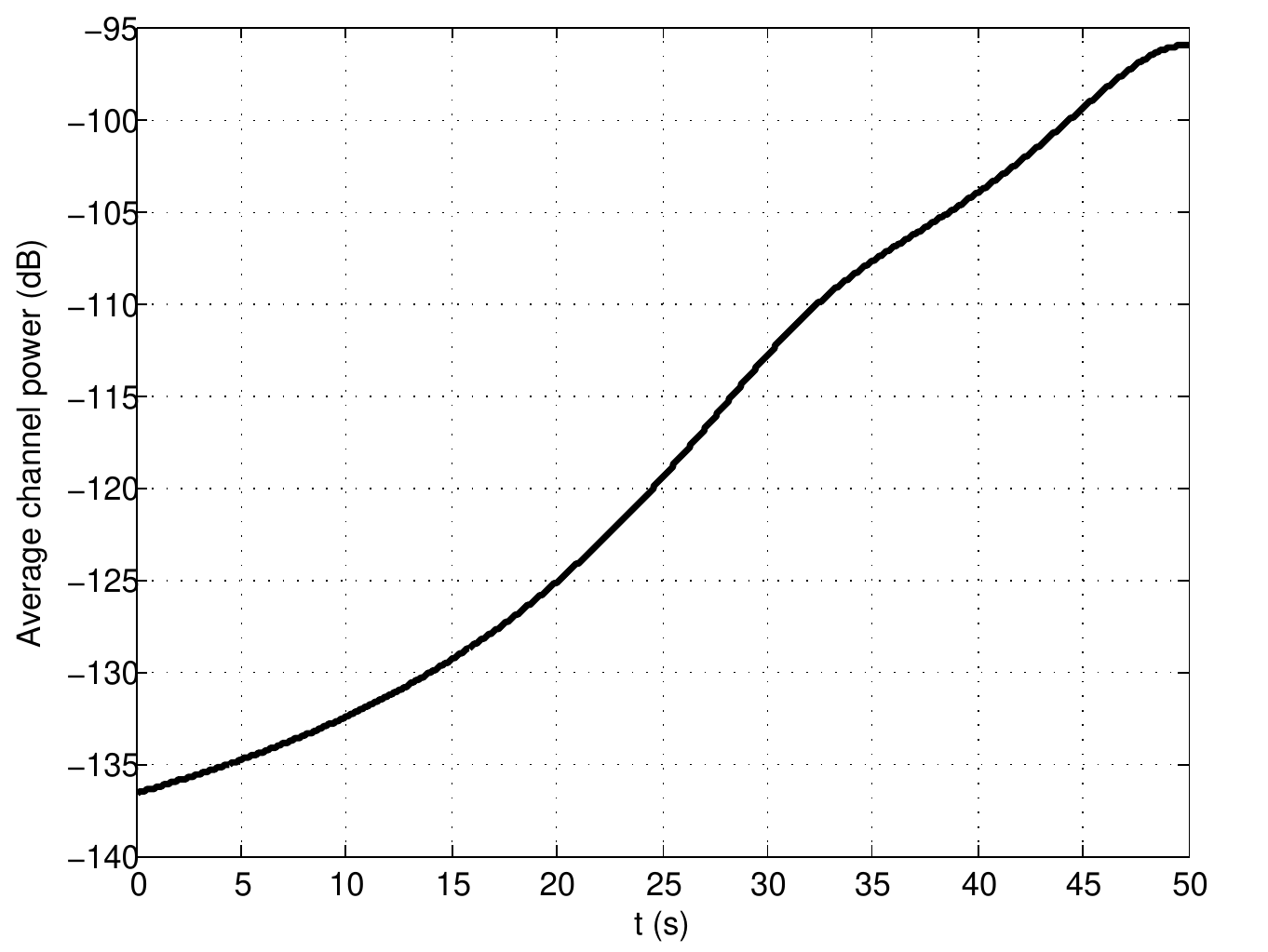}
\caption{}
\end{subfigure}
\caption{Variation of channel quality as UAV flies towards the GT: (a) Channel path loss for LoS and NLoS conditions; (b) LoS probability; (c) Average channel power.}\label{F:ChannelQualityVsTime}
\end{figure*}

Intuitively, for enhancing the communication link quality, the UAV should move closer to its communicating  GT. This not only reduces the link distance, but also increases the likelihood of establishing a LoS communication link with it, especially in dense urban environment.  As a toy example to illustrate this fact,  we consider a basic point-to-point communication setup with a UAV at fixed altitude $H_U$ communicating with a static ground node, with their initial horizontal distance denoted by $D$, as shown in Fig.~\ref{F:StraightPath}. Fig.~\ref{F:ChannelQualityVsTime} plots the channel path loss, the LoS probability, and the {\it average} channel power versus time as the UAV flies towards the ground node with a constant speed $V$. Note that the channel path loss  in Fig.~\ref{F:ChannelQualityVsTime}(a) is based on the classical log-distance path loss model in \eqref{eq:PL} by averaging over the shadowing, whereas the average  channel power is obtained by averaging over the occurrence of LoS and NLoS realizations, with the elevation angle-dependent LoS probability model given in \eqref{eq:PrLoS}. The following parameters are used: $D=1000$ m, $H_U=100$ m, $V=20$ m/s, $\alpha=2.3$, $X_0[dB]=50$ dB, $a=10$, $b=0.6$, and $\kappa=0.01$. It is observed that as the UAV moves closer to the ground node, the channel path loss  in Fig.~\ref{F:ChannelQualityVsTime}(a) is significantly improved  by about  23 dB for both LoS and NLoS cases, and there is an overall gain of about 40 dB shown  in Fig.~\ref{F:ChannelQualityVsTime}(c)  for the average  channel power, due to the additional benefit of enhanced LoS probability as shown in Fig.~\ref{F:ChannelQualityVsTime}(b). This demonstrates the promising benefit of UAV trajectory design to enhance the channel quality, especially for delay-tolerant applications so that there is sufficient time for the UAV to move towards its served GTs. 
Motivated by this, in the following, we  discuss two useful techniques for UAV path planning, following  the principle of bringing the UAV to each of its served GTs as closer as possible. Such techniques are useful to find an initial UAV flying path, which  can be used for trajectory initialization for the more refined UAV trajectory and communication joint optimization to be discussed in Section~\ref{sec:jointOptimization}.

\begin{figure*}
\centering
\begin{subfigure}{0.32\textwidth}
\centering
\includegraphics[width=\linewidth]{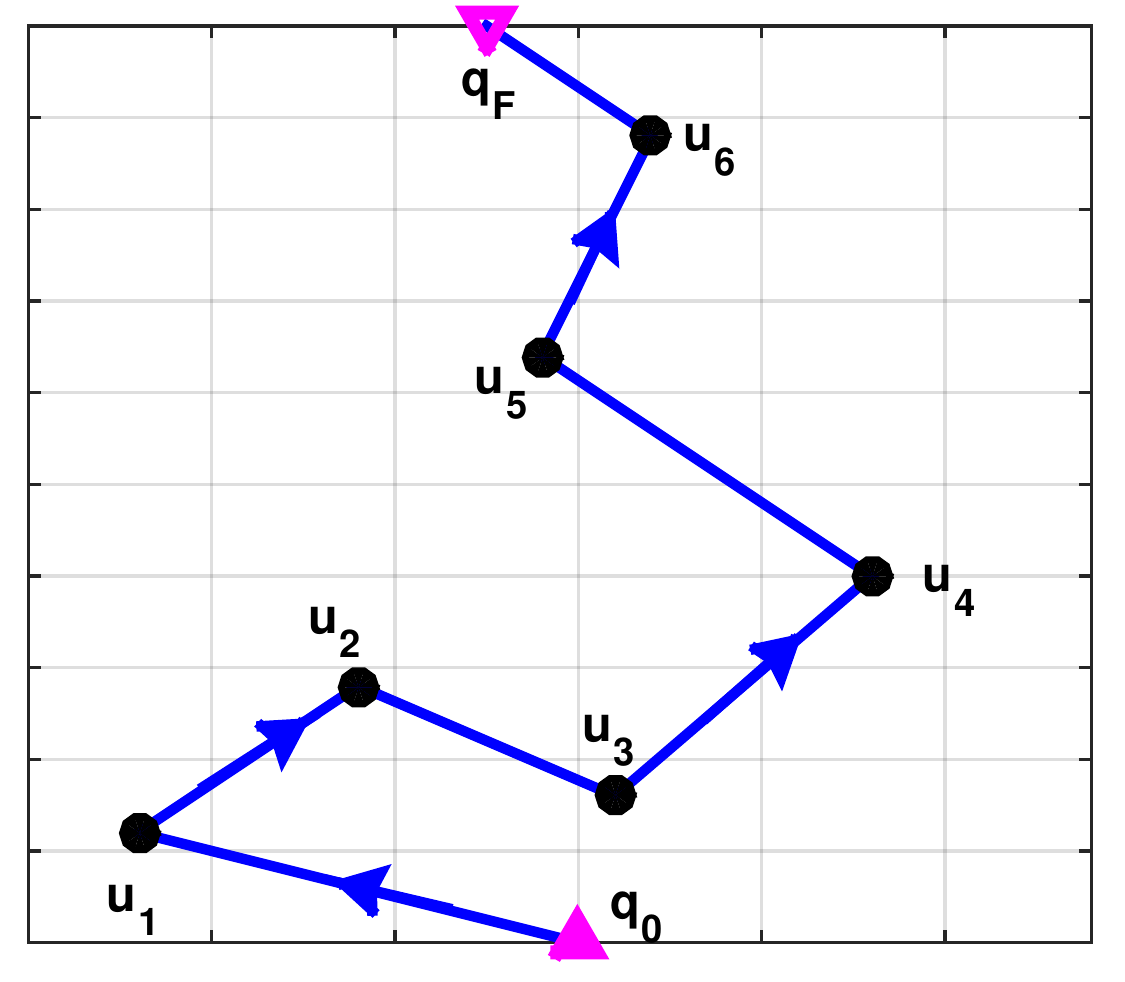}
\caption{Travelling salesman problem.}
\end{subfigure}
\hspace{0.01\textwidth}
\begin{subfigure}{0.3\textwidth}
\includegraphics[width=\linewidth]{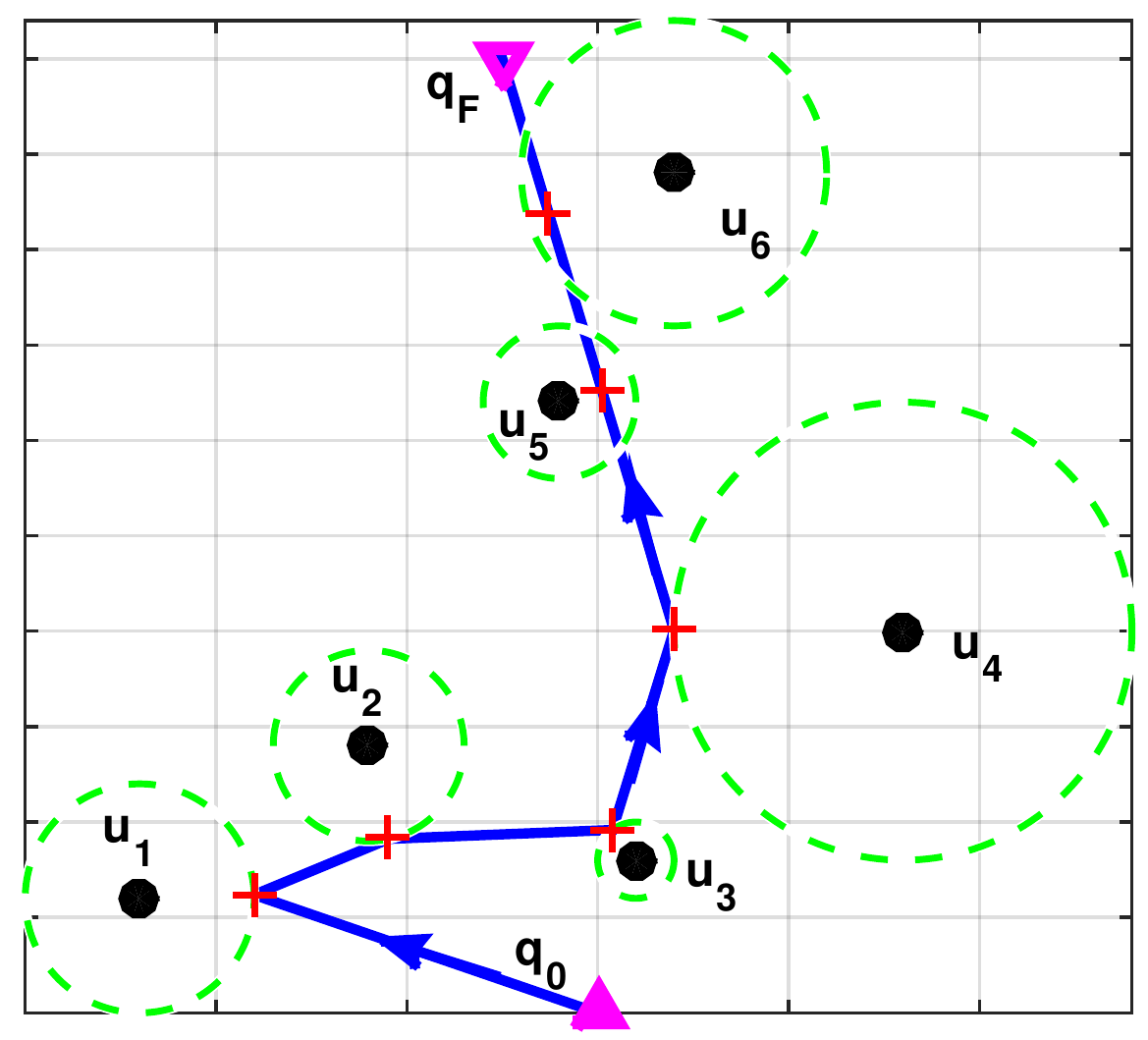}
\caption{Travelling salesman problem with neighbourhood.}
\end{subfigure}
\hspace{0.01\textwidth}
\begin{subfigure}{0.3\textwidth}
\includegraphics[width=\linewidth]{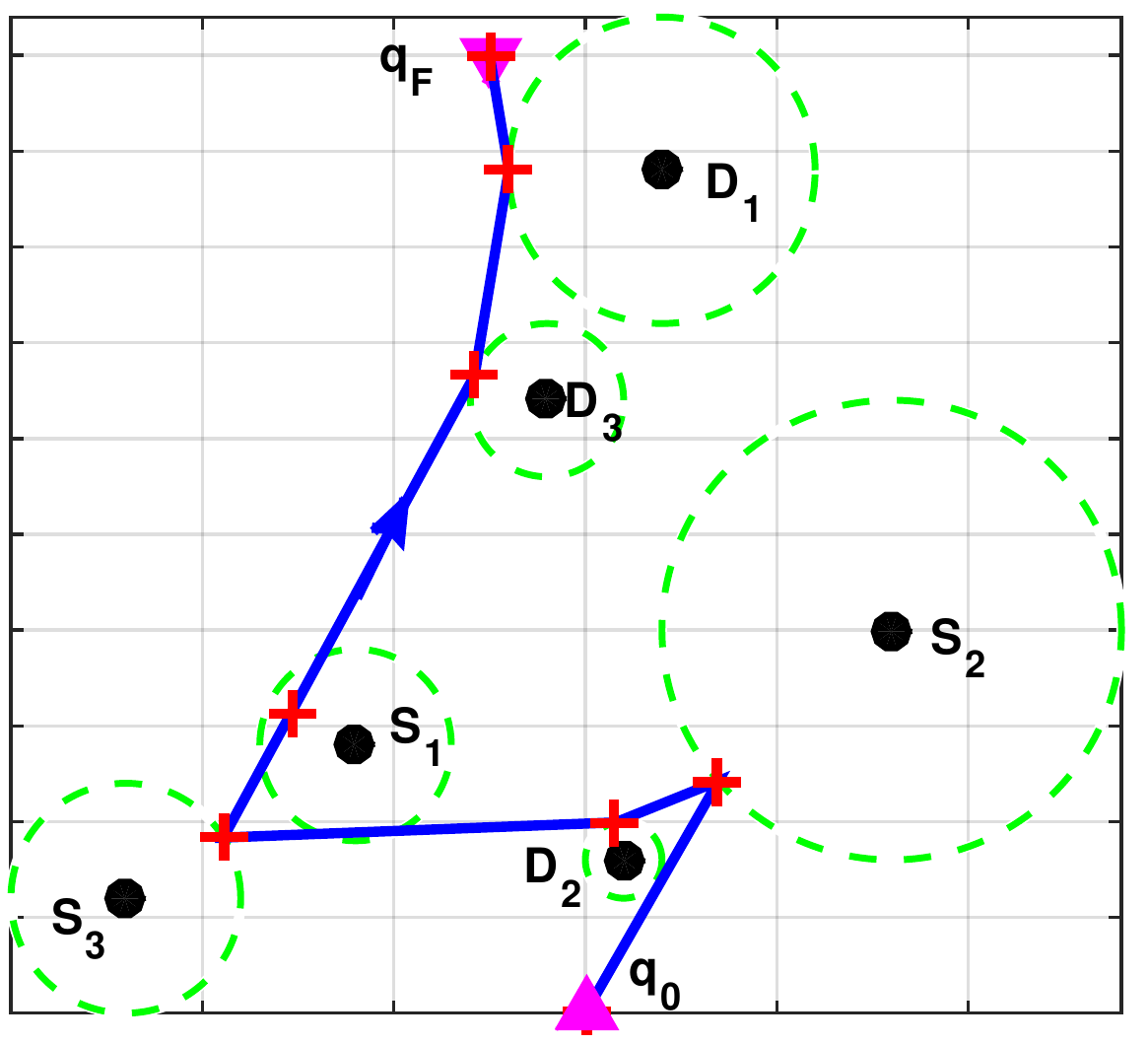}
\caption{Pickup and delivery problem with 3 source-destination pairs.}
\end{subfigure}
\caption{TSP/TSPN and PDPN for UAV initial path planning. $\mathbf q_0$ and $\mathbf q_F$ denote the pre-determined initial and final locations, respectively.}\label{F:TSPAndPDP}
\end{figure*}

{\bf Travelling salesman problem (TSP):} To best exploit the UAV mobility for multiuser communications, the UAV in general needs to fly sequentially towards multiple GTs served by it. Intuitively, the sooner the UAV reaches each of the GTs, the more time will be left  for the UAV to enjoy the best communication links with them. In this regard, a closely related problem is the celebrated {\it TSP} \cite{909,TSP,908,907}, which can be applied to determine the UAV flying path as well as the serving order of the GTs, as illustrated in Fig.~\ref{F:TSPAndPDP}(a). The standard TSP is described as follows: given a set of cities and the distances between each pair of the cities, a traveler wishes to start and end at the same city and visit each city exactly once. The problem then aims to find the route (or the sequence of visited cities) such that the total traveling distance is minimized. TSP is known to be NP-hard, but various efficient algorithms have been proposed to find high-quality solutions \cite{TSP,908,907}, e.g., via solving binary integer problems. Note that the standard TSP algorithms deal with the scenario that the traveller/UAV needs to return to the initial city/location where it starts the tour. However, for UAV communications, the UAV needs not necessarily return to the initial location, and its initial/final location might be pre-specified, as in \cite{641,904}. In this case, variations of TSP algorithms  can be applied by adding dummy cities/GTs whose distances with the existing cities/GTs are properly set \cite{957}.

TSP is feasible only when the given UAV operation duration $T$ is sufficiently large so that  the UAV can reach all GTs. Besides, in certain scenarios, it is simply unnecessary for the UAV to reach exactly on top of each GT (e.g., when only few data needs to be collected from some GTs).  In this case, another closely related problem is the  TSP with neighborhood (TSPN), as illustrated in Fig.~\ref{F:TSPAndPDP}(b). TSPN is a generalization of TSP in the sense that the traveller does not have to visit each city/GT exactly, but needs to reach a given  neighborhood region around the city/GT. TSPN is also NP hard, with various algorithms proposed to obtain approximate solutions \cite{927,928,957}. In fact, in the context of UAV communications, the resultant problem is even more general than TSPN, as the size (radius) of each neighborhood area can also be a design  variable depending on the communication requirement. One useful method for addressing such problems is as follows: Firstly, solve the TSP based on the $K$ cities/GTs to obtain the visiting order, by  ignoring the neighborhood regions. Then with the obtained order,  use convex optimization techniques to obtain the optimal visiting locations inside the neighborhood regions. This method was firstly proposed in \cite{957}, and was later applied in various other setups \cite{1056,980}. In fact, the above process for alternately updating the visiting order and the visiting locations can be repeated until convergence is reached. Another variation of TSP is the selective TSP \cite{1070}, also known as the orienteering problem \cite{1071}, where instead of visiting all nodes (or neighborhood regions), the goal is to determine a path and a subset of the nodes (or neighborhood regions) for visiting to maximize a certain utility, such as the number of nodes (or neighborhood regions) visited within a finite duration. This technique was applied in \cite{1069} for trajectory design for UAV-enabled distributed estimation via maximizing the number of sensors visited by the UAV within a given time horizon.

{\bf Pickup-and-delivery problem (PDP):} For UAV-enabled mobile relaying, we usually have the additional {\it information-causality constraint} \cite{641,1056}, i.e., the UAV needs to firstly receive data from a source node before forwarding to its corresponding destination node. In this case, a useful approach  for determining the UAV flying path is by solving the {\it PDP}. PDP can be regarded as another generalization of TSP, with the additional precedence constraints, i.e., for each pair of source-destination nodes, the UAV needs to firstly visit the source node before the destination node to meet the above information-causality constraint. PDP is also NP hard, but various algorithms have been proposed to yield high-quality approximate solutions. Furthermore, in the general scenario where the given UAV operation duration $T$ is insufficient to visit all the GTs, the extended PDP with neighborhood (PDPN) can be applied to obtain the visiting order of the GTs, as illustrated in Fig.~\ref{F:TSPAndPDP}(c). 


\subsubsection{Joint Trajectory and Communication Optimization}\label{sec:jointOptimization}
While TSP and PDP are useful techniques to determine the initial UAV flying path or serving order of the GTs, they are in general suboptimal for the generic problem $\mathrm{(P1)}$. On one hand, the  UAV flying trajectory needs to take into account the communication performance more explicitly, which also depends on the communication user scheduling and resource allocation with any given UAV trajectory. On the other hand, in practical scenarios where  UAVs are subject to various mobility constraints such as those exemplified in Section~\ref{sec:ConstrUAVTraj}, the simple TSP and PDP solutions, which ignore such constraints, may lead to infeasible UAV path. To tackle such issues, it is inevitable to address the trajectory and communication joint optimization problem $\mathrm{(P1)}$. However, $\mathrm{(P1)}$ is difficult to be directly solved for two reasons.   Firstly, it involves objective/constraint functions with essentially an infinite number of variables due to the continuous time. Secondly, it is generally non-convex  with respect to communication  and UAV trajectory design variables. In fact, even by fixing one of these two types of  variables, the problem is usually still non-convex with respect to the other. In the following,  we first introduce two trajectory discretization techniques to convert $\mathrm{(P1)}$ into more tractable forms with a finite number of optimization variables, and then elaborate the BCD and SCA techniques to deal with the non-convexity.

{\bf Trajectory discretization:} To transform the  optimization problem $\mathrm{(P1)}$ into a more tractable form with a finite number of variables, it is necessary to discretize the UAV trajectory as well as other related variables. The basic idea of trajectory discretization is to approximate the continuous UAV trajectory by a piece-wise linear trajectory, which is represented by a finite number of line segments and the duration that the UAV needs to spend on each line segment. In order to ensure sufficient discretization accuracy, the length of each line segment should not exceed a certain threshold, say $\Delta_{\max}$, whose value could be pre-specified based on practical requirements. For example, within each line segment, the distance between the UAV and all ground nodes of interest should be approximately unchanged in order to maintain constant average channel gains to facilitate the communication design and performance characterization. In this case, one may choose $\Delta_{\max}$ such that $\Delta_{\max}\ll H_{\min}$, with $H_{\min}$ denoting the minimum UAV altitude. For any given $\Delta_{\max}$, two trajectory discretization approaches have been proposed in the literature, namely {\it time discretization}  \cite{641}, \cite{904} and {\it path discretization} \cite{980}.

\begin{figure}
\centering
\includegraphics[scale=0.5]{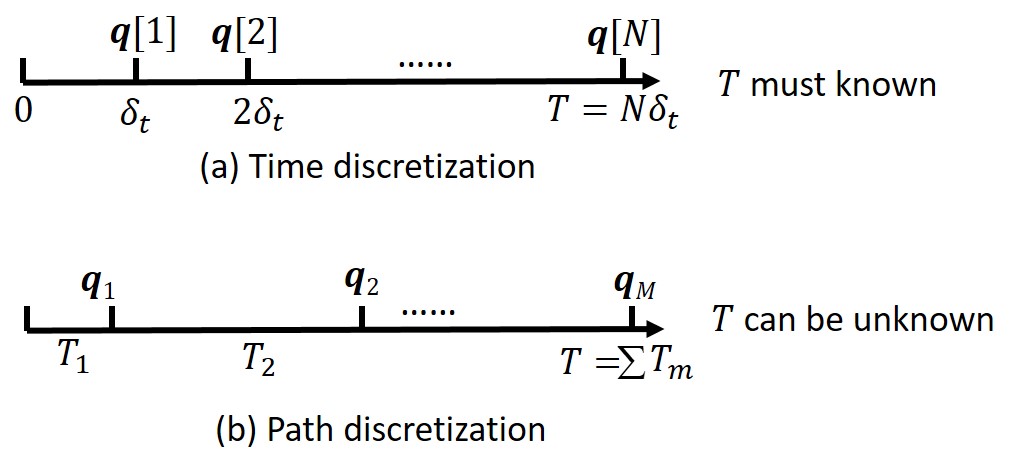}
\caption{Time versus path discretization.}\label{F:discretization}
\end{figure}

{\it Time Discretization:} As illustrated in Fig.~\ref{F:discretization}(a), with time discretization, the given time horizon $[0, T]$ is divided into $N$ equal time slots with sufficiently small slot length $\delta_t$ \cite{641}, \cite{904}, where $T=N\delta_t$. Let $V_{\max}$ denote the UAV's maximum flying speed. Then it is necessary to ensure that each segment length does not exceed $\Delta_{\max}$ even with the maximum flying speed,  for which $\delta_t$ should be chosen as $\delta_t\leq \Delta_{\max}/V_{\max}$. Thus, the minimum number of  segments required  with time discretization is
$N=\lceil TV_{\max}/\Delta_{\max} \rceil$.
 As a result, the continuous UAV trajectory $\mathbf q(t)$, $0\leq t \leq T$, can be approximated by the $N$-length sequence $\{\mathbf q[n]\}_{n=1}^N$, which need to satisfy the maximum UAV speed and acceleration constraints. 
With time discretization, the UAV movement can be approximated in a linear state-space model with respect to the UAV location, velocity, and acceleration (note that $\delta_t$ is chosen and fixed), which is given by \cite{904}
\begin{align}
\vbd[n+1]&= \vbd[n]+\mathbf a[n] \delta_t, \forall n, \label{eq:vn} \\
\q[n+1]&= \q[n]+\vbd[n] \delta_t+\frac{1}{2}\mathbf a[n]\delta_t^2, \forall n, \label{eq:qn}
\end{align}
where $\mathbf v[n]$ and $\mathbf a[n]$ are respectively the velocity and acceleration vectors in 3D, which are assumed to be constant  within each time slot $n$. As such, the UAV trajectory constraints given in Section~\ref{sec:ConstrUAVTraj} can be approximated in discrete forms accordingly.

\begin{table*}\caption{Comparison between time and  path discretization.} \label{Table:discretization}
\centering
\begin{tabular}{|l|l|l|}
\hline
 & \bf Time discretization                                                                                                                                                       & \bf Path discretization                                                                                                                                                                   \\ \hline
\it Pros     & \begin{tabular}[c]{@{}l@{}}{\small $\bullet$} Equal time slot length\\ {\small $\bullet$} Linear state-space representation\\ {\small $\bullet$} Incorporate the maximum\\ ~~\,acceleration constraint easily\end{tabular}        & \begin{tabular}[c]{@{}l@{}}{\small $\bullet$} Fewer variables if UAV hovers or\\ ~~\,flies slowly most of the time\\ {\small $\bullet$} No need to know mission completion\\ ~~\,time $T$ a priori\end{tabular}                   \\ \hline
\it Cons     & \begin{tabular}[c]{@{}l@{}}{\small $\bullet$} Excessively large number of time\\ ~~\,slots when UAV hovers or moves\\~~\,slowly\\ {\small $\bullet$} Need to know mission completion\\ ~~\,time $T$ a priori\end{tabular} & \begin{tabular}[c]{@{}l@{}}{\small $\bullet$} Difficult to incorporate the maximum\\ ~~\,acceleration constraint\\ {\small $\bullet$} More variables if UAV flies with\\ ~~\,high/maximum speed most of the\\ ~~\,time\end{tabular} \\ \hline
\end{tabular}
\end{table*}
{\it Path Discretization \cite{980}:}  Another approach for discretized representation of UAV trajectory is to divide the  UAV path (instead of time) into $M$ consecutive line segments of generally unequal lengths as shown in Fig.~\ref{F:discretization}(b), which are represented by a sequence of segment start/end locations $\{\mathbf q_m\}$ along the path, together with the time sequence $\{T_m\}$ representing the duration that the UAV  spends on each line segment. Path discretization can be interpreted as the more general form of time discretization, with flexibly chosen unequal time slot lengths for different line segments. Specifically, instead of fixing the slot length to $\delta_t=\Delta_{\max}/V_{\max}$ that is bottlenecked by the maximum flying speed,  with path discretization,  the time slot length $T_m$  is  dynamically determined by the actual flying speed $V_m$ that is assumed to be constant over each  line segment. In this case, we  have $T_mV_m\leq \Delta_{\max}$, $\forall m$. Note that since $V_m\leq V_{\max}$, we have $T_m\geq \delta_t$, $\forall m$. In other words, given the same value for the maximum segment length $\Delta_{\max}$, path discretization entails longer time slot length in general. As a result, given the same trajectory to be discretized with the total operation duration $T=N\delta_t=\sum_{m=1}^M T_m$, we have $M\leq N$ in general, i.e., fewer line segments are needed by path discretization than time discretization, especially when the UAV flies with a speed lower than the maximum speed for a significant portion of the operation duration.

To  further illustrate the above fact,  we consider the scenario that the UAV needs to hover at a particular location for 1000 s. If time discretization approach is used (say with time interval of 1 s), then we need 1000 variables $\mathbf q[1],\cdots \mathbf q[1000]$ (all are equal) to represent this status, even though the UAV remains stationary. In contrast, with path discretization, only three variables are sufficient, namely $\mathbf q_1$ and $\mathbf q_2$ (with $\mathbf q_1=\mathbf q_2$) representing the hovering location and $T_1=1000$ s representing the hovering duration.  Another advantage of path discretization is that it does not require to specify the  mission completion time $T$ a priori. Instead, a coarse estimation of the total flying distance $\hat D$ is sufficient to determine the required number of segments $M$,  for which $M$ is chosen to be sufficiently large so that $M\Delta_{\max}\geq \hat{D}$. This is appealing since in many  practical trajectory design problems  such as that  for UAV energy consumption minimization \cite{980}, the UAV operation time itself is a variable and there is no monotonic relationship for efficiently searching its optimal value  (e.g., by the  bisection method), thus only the time-consuming  exhaustive search is applicable.  If time discretization is used, it would require solving prohibitively large number of optimization problems, each for a pre-assumed and fixed $N$, which is impractical.

On the other hand, note that time discretization also has its own merit. First, as  the time interval $\delta_t$ is fixed, time discretization leads to the simple linear state-space model as given in \eqref{eq:vn} and \eqref{eq:qn}, which can easily handle the UAV maximum acceleration constraint. In contrast, such linear  relationship is not preserved for path discretization with $\{T_m\}$ also being the optimization variables. Second, if ignoring the acceleration variable, time discretization requires only one variable for each line segment, namely the UAV locations $\{\mathbf q[n]\}$, as the UAV velocity for each line segment $n$ can be directly obtained as $\mathbf v[n]=(\mathbf q[n+1]-\mathbf q[n])/\delta_t$ with $\delta_t$ given. By contrast, path discretization needs two variables for each line segment (namely both the UAV end  location and time duration). Thus, if given the same number of line segments, i.e., $N=M$ (e.g., when the UAV always flies at its maximum speed during the operation), then path discretization needs to double the number of variables as compared to time discretization. The comparison of these two UAV trajectory discretization techniques is   summarized in Table~\ref{Table:discretization}.

By applying the above trajectory  discretization techniques, the optimization problem $\mathrm{(P1)}$ can be transformed into the following generic form with a finite number of variables:
\begin{align}
\mathrm{(P2):} ~ \underset{\{\mathcal Q[n]\}, \{\mathcal R[n]\}}{\max} \ &  U\left(\{\mathcal Q[n]\}, \{\mathcal R[n]\} \right)\notag \\
\text{s.t.}~~~~~\ & f_i\left(\{\mathcal Q[n]\} \right)\geq 0, \ i=1,\cdots, I_1, \label{eq:trajConstr1}\\
& g_i \left( \{\mathcal R[n]\}\right) \geq 0, \ i=1,\cdots, I_2,\\
& h_i \left(\{\mathcal Q[n]\}, \{\mathcal R[n]\}\right) \geq 0,\  i=1,\cdots, I_3\label{eq:trajConstr3}.
\end{align}
In the above, $\{\mathcal Q[n]\}$ and $\{\mathcal R[n]\}$ denote the discretized UAV trajectories and communication design variables, respectively.

{\bf BCD and SCA for resource and trajectory optimization:} Problem $\mathrm{(P2)}$ involves the joint optimization of UAV trajectory and communication resource allocation, which is usually non-convex and difficult to be solved optimally. To tackle this problem efficiently, one useful approach to obtain a generally locally optimal solution for it is by alternately updating one block of variables with the other block fixed, which is known as the BCD method \cite{1090,hong2016unified}. 
Note that for any given feasible UAV trajectory, problem $\mathrm{(P2)}$ reduces to the extensively studied communication resource allocation problem, for which the existing techniques developed under the terrestrial communication setup can be directly applied. However, for any fixed communication resource allocation, the UAV trajectory optimization problem is relatively new, which is thus discussed in detail as follows. In particular, we introduce an effective technique, namely SCA, which is useful for solving non-convex  UAV trajectory optimization problems. For the purpose of easy illustration, we consider the case of one UAV with discretized trajectory denoted as $\{\mathbf q[n]\}$. The corresponding sub-problem of $\mathrm{(P2)}$ for trajectory optimization with given communication resource allocation can be written as
\begin{align}
\mathrm{(P3):} \quad \underset{\{\mathbf q[n]\}}{\max} \ &  f_0\left(\{\mathbf q[n]\} \right)\notag \\
\text{s.t.} \ & f_i\left(\{\mathbf q[n]\} \right)\geq 0, \ i=1,\cdots, I,
\end{align}
where $f_0\left(\cdot \right)$ represents the utility to be maximized, and $f_i\left(\cdot \right)$'s are the corresponding constraints in \eqref{eq:trajConstr1} and \eqref{eq:trajConstr3} of $\mathrm{(P2)}$ which involve the UAV trajectory with $I=I_1+I_3$. Note that problem $\mathrm{(P3)}$ is non-convex if at least one of the functions $f_i(\cdot)$ is non-concave with respect to $\{\mathbf q[n]\}$, $i=0,1,\cdots, I$. This is usually the case, since most utility and constraint functions given in Section~\ref{sec:PerformanceMetrics} and Section~\ref{sec:ConstrUAVTraj} are non-concave over $\{\mathbf q[n]\}$, due to  which standard  convex optimization techniques cannot be directly applied to solve $\mathrm{(P3)}$. Fortunately, recent work has shown that SCA is a useful technique for transforming the non-convex optimization problem into solving a series of convex optimization problems, with guaranteed monotonic convergence to at least a Karush-Kuhn-Tucker (KKT) solution under some mild conditions \cite{1092,768}. Thus, we apply SCA to solve the UAV trajectory optimization problem $\mathrm{(P3)}$ in the following.

\begin{figure*}
\centering
\includegraphics[width=0.6\linewidth]{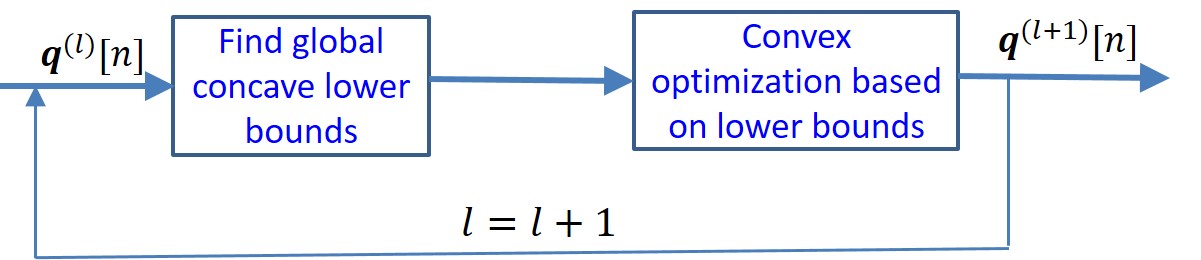}
\caption{Successive convex approximation for trajectory optimization.}\label{F:SCA}
\end{figure*}

\begin{figure*}
\centering
\includegraphics[width=0.5\linewidth]{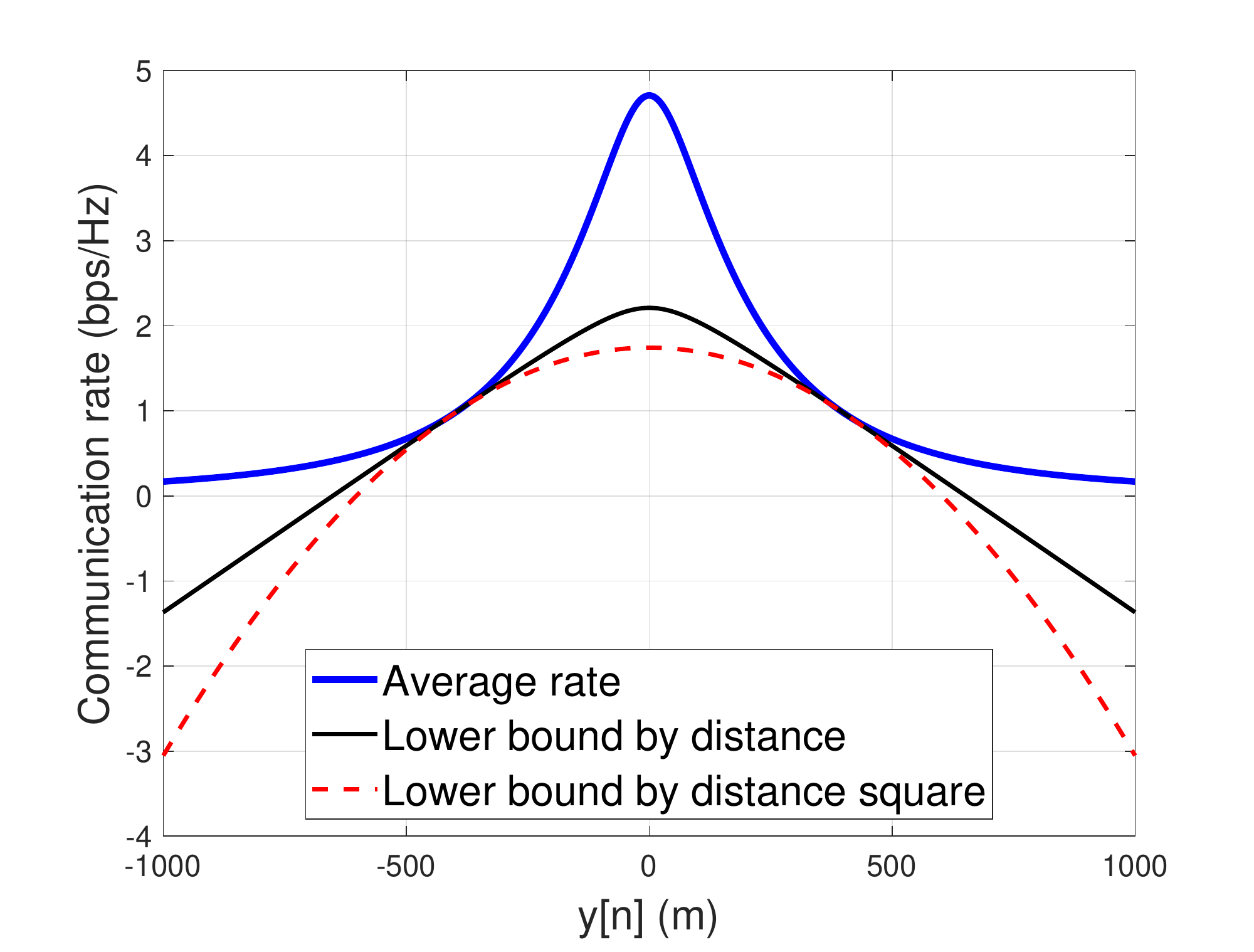}
\caption{An illustration of the global concave lower bounds for communication rate.}\label{F:RateLowerBound}
\end{figure*}
SCA is an iterative optimization  technique.  Specifically, at  each iteration  $l$, we need to firstly find a global {\it concave} lower bound for those non-concave functions $f_i\left(\{\mathbf q[n]\} \right)$ in $\mathrm{(P3)}$, such that
\begin{align}
f_i\left(\{\mathbf q[n]\} \right)\geq f_{i,\lb}^{(l)}\left(\{\mathbf q[n]\} \right), \forall \mathbf q[n]. \label{eq:LB2}
\end{align}
Then by replacing those non-concave functions $f_i\left(\{\mathbf q[n]\}\right)$ in $\mathrm{(P3)}$ with their corresponding concave lower bounds  $f_{i,\lb}^{(l)}\left(\{\mathbf q[n]\} \right)$, we have the following convex optimization problem
\begin{align}
\mathrm{(P4):} \quad \underset{\{\mathbf q[n]\}}{\max} \ &  f_{0,\lb}^{(l)}\left(\{\mathbf q[n]\} \right)\notag \\
\text{s.t.}\  &  f_{i,\lb}^{(l)}\left(\{\mathbf q[n]\} \right)\geq 0, \ i=1,\cdots, I. \label{eq:convexLB}
\end{align}
As $\mathrm{(P4)}$ is convex, its optimal solution, denoted as $\{\mathbf q^{(l)}[n]\}$, can be efficiently obtained based on standard convex optimization techniques or readily available software toolbox such as CVX \cite{227}. In addition, due to the global lower bound of \eqref{eq:LB2}, it can be verified that $\{\mathbf q^{(l)}[n]\}$ is also feasible to the non-convex problem $\mathrm{(P3)}$, and the corresponding optimal value  provides at least a lower bound to that of problem $\mathrm{(P3)}$. Furthermore, if the lower bound \eqref{eq:LB2} is tight at the local point $\{\mathbf q^{(l-1)}[n]\}$ at the $l$th iteration, i.e.,
\begin{align}
f_{i,\lb}^{(l)} \left( \{\mathbf q^{(l-1)}[n]\} \right) =  f_i \left( \{\mathbf q^{(l-1)}[n]\}  \right), \label{eq:eqValue}
\end{align}
then the sequence $f_0\left(\{\mathbf q^{(l)}[n]\}\right)$ monotonically increases and converges to a finite limit \cite{768}. With the additional condition that the gradient at the local point is also tight, i.e.,
\begin{align}
\nabla f_{i,\lb}^{(l)} \left( \{\mathbf q^{(l-1)}[n]\} \right) =  \nabla f_i \left( \{\mathbf q^{(l-1)}[n]\}  \right), \label{eq:eqGrad}
\end{align}
then under some mild constraint qualifications, $\{\mathbf q^{(l)}[n]\}$ converges to a solution fulfilling the KKT conditions of problem $\mathrm{(P3)}$ \cite{768}. Thus, by iteratively updating the local point $\{\mathbf q^{(l)}[n]\}$ and solving a sequence of convex optimization problems $\mathrm{(P4)}$, a KKT solution of the non-convex trajectory optimization problem $\mathrm{(P3)}$ can be obtained. The main idea of SCA for trajectory optimization is shown in Fig.~\ref{F:SCA}.

The remaining task is then to find the concave lower bounds for the involved UAV utility and constraint functions satisfying the above properties. Fortunately, such bounds can be found for the typical utility/constraints functions specified in Section~\ref{sec:PerformanceMetrics} and Section~\ref{sec:ConstrUAVTraj} \cite{641,904,919}, by using the fact that for convex differentiable functions, the first-order Taylor approximation provides a global lower bound \cite{202}. For example, at the $l$th iteration with the given local point $\{\mathbf q^{(l)}[n]\}$ and $\mathbf v^{(l)}[n]$, the following bounds are useful for the non-convex minimum speed constraint \eqref{eq:speedConstr} \cite{904}
\begin{align}
\|\mathbf v[n]\|^2 \geq \|\mathbf v^{(l)}[n]\|^2 + 2 \mathbf v^{(l)T}[n]\left(\mathbf v[n]- \mathbf v^{(l)}[n]\right), \ \forall \mathbf v[n].
\end{align}
Similar bounds can be obtained for most of other constraints. Besides, for the average communication rate in \eqref{eq:RkApprox2}, by defining the convex function $h(z)=\log_2\left(1+\frac{\gamma_k}{z^\alpha}\right), z\geq 0$ and letting $z=\|\mathbf q[n]-\mathbf w_k\|$, the following concave lower bound can be obtained
\begin{align}
&\log_2\left( 1+ \frac{\gamma_k}{\|\mathbf q[n] - \mathbf w_k\|^\alpha}\right)
\geq  A_k[n] \nonumber\\
&~~~~~~~~~-B_k[n] \left(\|\mathbf q[n] -\mathbf w_k \| -\|{\mathbf q}^{(l)}[n]-\mathbf w_k\| \right), \label{eq:RateLB1}
\end{align}
where
\begin{align}
A_k[n]&=\log_2\left( 1+ \frac{\gamma_k}{\|\mathbf {q}^{(l)}[n] - \mathbf w_k\|^\alpha}\right),\\
B_k[n]&=\frac{\gamma_k\alpha(\log_2 e)}{\|\mathbf q^{(l)}[n]-\mathbf w_k\|\left(\|\mathbf q^{(l)}[n]-\mathbf w_k\|^{\alpha} +\gamma_k\right)}.
\end{align}
Note that for the given local point ${\mathbf q}^{(l)}[n]$, all terms on the right hand side of \eqref{eq:RateLB1} are constants, except the term $\|\mathbf q[n]-\mathbf w_k\|$, which is the distance between the UAV and GT. Thus, we refer \eqref{eq:RateLB1} as {\it lower bound by distance}. In fact, depending on the chosen convex function for which the first-order Taylor approximation is applied, there may exist more than one global concave lower bounds satisfying \eqref{eq:eqValue} and \eqref{eq:eqGrad}. For example, for the average communication rate function \eqref{eq:RkApprox2}, by defining another convex function $h(z)=\log_2\left(1+\frac{\gamma_k}{z^{\alpha/2}}\right), z\geq 0$ and letting $z=\|\mathbf q[n]-\mathbf w_k\|^2$, an alternative lower bound in terms of $\|\mathbf q[n]-\mathbf w_k\|^2$ can be obtained, which we term as {\it lower bound by distance square} and has been extensively used in prior work on UAV trajectory optimization \cite{641,904,918,919}.

Fig.~\ref{F:RateLowerBound} gives a 1D illustration for the above concave lower bounds, where $\mathbf w_k=\mathbf 0$ and $\mathbf q[n]=[0, y[n], H]^T$. In other words, the UAV is assumed to fly along the $y$-axis with a constant altitude $H$ communicating with a GT located at the origin. The following parameters are used: $\alpha=2.3$, $H=100$ m, and $\gamma_k=60$ dB. The average rate (i.e., the left hand side of \eqref{eq:RateLB1}) versus $y[n]$ is plotted in Fig.~\ref{F:RateLowerBound}, together with the two lower bounds discussed above obtained at the local point $y^{(l)}[n]=400$ m. It is observed that the {\it lower bound by distance} is in fact tighter than that by distance square, though the latter has been extensively used in the literature. It is thus interesting to investigate whether this new tighter bound would lead to better performance of the converged trajectory in future work.

To summarize, UAV communications usually involve the joint optimization of UAV trajectory and communication resource allocation, as represented by the generic problem formulation $\mathrm{(P1)}$. For multiuser systems, the classic TSP and PDP algorithms  can be used to find the initial UAV path planning. On the other hand, time- and path-discretization techniques can be applied to convert the continuous-time  optimization problem approximately  into more tractable forms with a finite number of discrete variables. To deal with the problem  non-convexity, BCD can be used to alternately update the communication resource allocation and UAV trajectory. In particular, for the non-convex trajectory optimization subproblem, SCA is found to be effective to obtain a KKT suboptimal solution in general. Note that as the SCA-based UAV trajectory optimization requires iterative procedures, a feasible initial UAV trajectory needs to be specified. The TSP/PDP based path planning offers a good starting point to obtain the initial UAV trajectory for SCA. However, when the UAV trajectories are subject to various constraints shown in Section~\ref{sec:ConstrUAVTraj}, more general  methods need to be developed to determine a sound feasible initial path satisfying such constraints, which deserve further investigation.

The use of BCD and SCA  for joint UAV trajectory and communication resource allocation was firstly proposed in \cite{641} in the context of UAV-enabled mobile relaying. It was later successfully applied in various other setups, such as energy-efficient UAV communications \cite{904,980}, multi-UAV enabled downlink communication \cite{919,Yuxu2018}, UAV-enabled data collection \cite{918}, physical-layer security for UAV communications \cite{zhang2018securing,1073,Xiaobo18uav}, UAV-enabled mobile edge computing \cite{jeong2016mobile,cao2018mobile}, and UAV-enabled wireless power transfer \cite{956} and wireless powered communications \cite{1072}. Note that one drawback  of alternately updating UAV trajectory and communication resource allocation is the likelihood of trapping into undesirable local optimums, if the initialisation is not properly designed. 
Therefore, there have been recent efforts on investigating the simultaneous update of these two blocks of variables for certain setups via developing new concave lower bound functions \cite{980, 1041}. The use of alternating direction method of multipliers (ADMM) technique to reduce the computation complexity for multi-UAV trajectory design has also been reported in \cite{1041}. UAV placement and movement optimization has been studied in \cite{975} for multi-UAV uplink  coordinated multipoint (CoMP) communications, where each UAV forwards its received signals from all ground users to a central processor for joint decoding. While the above works mostly assumed either orthogonal multiuser communications or treating interference as noise, the capacity region of the UAV-enabled two-user broadcast channel has been characterized in \cite{wu2018uavRegion} and \cite{JR:wu2017_capacity}, which requires superposition coding and interference cancellation in general. Under this setup, it was revealed that the capacity-achieving UAV trajectory follows the simple hover-fly-hover (HFH) pattern, where the UAV successively hovers at a pair of initial and final locations.     

It is worth remarking that due to the practically finite UAV flying speed, exploiting UAV mobility for communication performance enhancement is most appropriate for delay-tolerant applications. In fact, for UAV platforms serving multiple users, there exists a new tradeoff between communication throughput and access delay, which was firstly studied in \cite{887} for a UAV flying with fixed trajectory, and was later extended in \cite{wu2018common,wu2017delayAPCC} via joint design of UAV trajectory and communication resource allocation in orthogonal frequency division multiple access (OFDMA) systems.

While the above works on communication-trajectory co-design mostly focused on 2D trajectory with fixed UAV altitude, more research efforts are needed for 3D trajectory-communication co-design to fully exploit the 3D UAV mobility, especially in dense urban environment \cite{1086}. To this end, more sophisticated channel models and performance metrics as discussed in Section~\ref{sec:channelModel}  and Section~\ref{sec:PerformanceMetrics} need to be used. Besides, the consideration of more practical antenna models, such as the directional antenna with fixed pattern or more advanced MIMO beamforming  as discussed in Section~\ref{sec:AntModel}, is expected to have a significant  impact on the joint optimization of communication resource and UAV trajectory, which is worthwhile for further investigation. Furthermore, for UAV-assisted communication in real-time applications, high-capacity wireless backhauling needs to be established between UAV and the core network on the ground. This brings a new design consideration to achieve the optimal balance between the wireless backhaul and radio access, via joint UAV position/movement and resource optimization, which deserves further studies.

\subsection{Energy-Efficient UAV Communication}\label{sec:EnergyEfficientCommun}
Energy-efficient wireless communication has been an active research avenue during the past decade. It was  driven not only by the     need to reduce the operation cost and green gas emission of the information and communications technology (ICT) industry, but also due to the importance to prolong the battery usage or lifespan of various types of communication devices. For UAV communications, the need for energy saving is even more imperative, due to the highly limited onboard energy and the additional propulsion energy consumption, besides the conventional communication energy expenditure.

Energy-efficient UAV communications were initially focused only on the saving of the communication-related energy consumption of either the ground nodes  \cite{796,918,1040} or the UAV \cite{797,798}. For example, in \cite{796}, adaptive link selection and transmission schemes were studied to minimize the energy consumption of ground nodes for a hybrid communication system with both aerial relay and direct terrestrial communications. In \cite{918}, the authors studied the UAV-enabled data collection to minimize the maximum energy consumption of all sensor nodes via jointly optimizing the UAV trajectory and the wake-up schedule of the sensor nodes. In \cite{797}, the UAV-enabled downlink communication was studied, where the locations of the UAVs and the cell boundaries are optimized to minimize the required transmit power of UAVs, while satisfying the user rate requirement.

Note that for UAV communication systems, the UAV propulsion energy consumption is usually much more significant as compared to the communication counterpart  and thus poses  the fundamental limit on the UAV endurance and communication performance. Therefore, there have been growing research efforts on energy-efficient UAV communications by rigorously taking into account the UAV's propulsion energy consumption \cite{904, 980, 972, eom2018uav}. This usually leads to significantly different design problems as compared to those for the conventional terrestrial systems considering the communication energy only, due to the new tradeoff between minimizing the UAV propulsion energy consumption versus maximizing the communication throughput, both dependent on the UAV trajectory, as discussed in Section~\ref{sec:energyModel} and Section~\ref{sec:PerformanceMetrics}. To illustrate such a tradeoff, consider the basic setup where a UAV needs to communicate with a ground node. From the throughput maximization perspective, the UAV should stay stationary at the nearest possible location from the ground node so as to maintain the best channel for communication. However, as shown in Fig.~\ref{F:PowervsSpeed}, hovering is power-inefficient for rotary-wing UAVs and even impossible for fixed-wing UAVs. Therefore, energy-efficient UAV communication in general requires a non-trivial UAV trajectory design, jointly with the communication resource allocation, to achieve an optimal balance between energy saving  and throughput enhancement. One commonly used design objective is the energy efficiency as defined  in Section~\ref{sec:EnergyEfficiency}.

\begin{figure}
\centering
\includegraphics[width=0.85\linewidth]{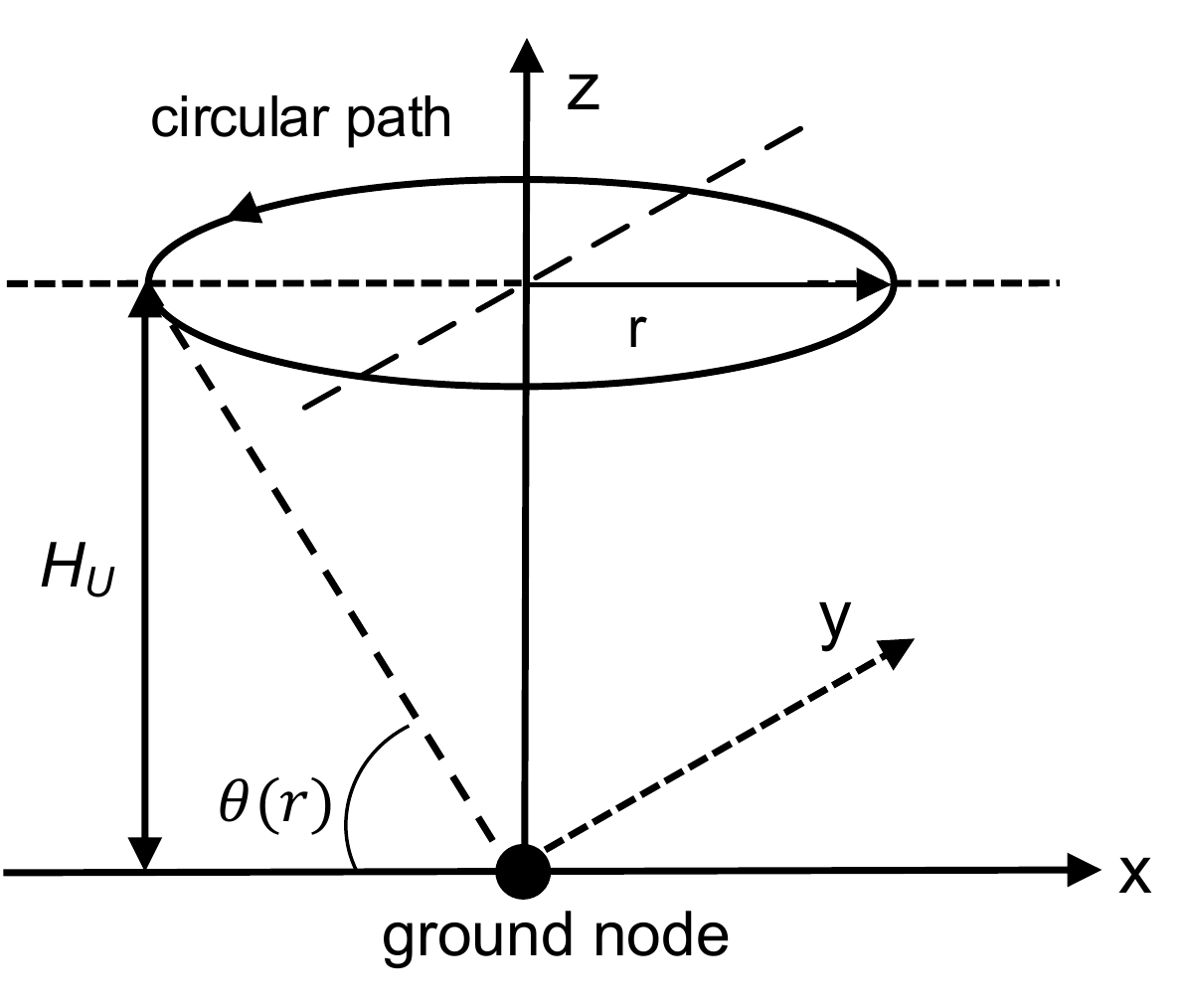}
\caption{A point-to-point link where a fixed-wing UAV flies following a circular path with radius $r$.}\label{F:CircularPath}
\end{figure}

\begin{figure}
\centering
\includegraphics[scale=0.45]{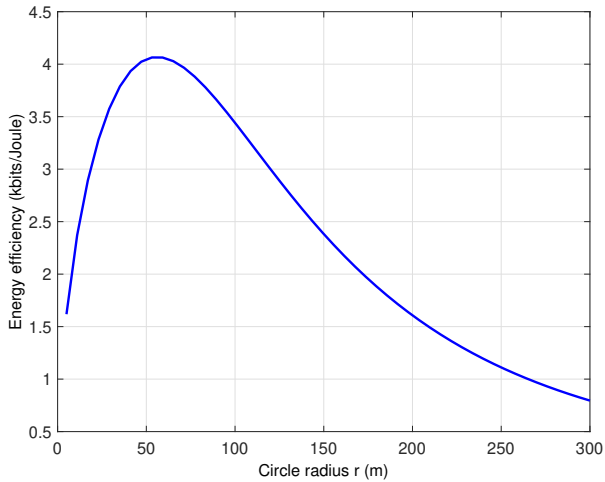}
\caption{A typical plot of energy efficiency versus circle radius $r$.}\label{F:EEVsRadius}
\end{figure}

 As a simple illustration for energy-efficient UAV communications, let's  consider the scenario that a fixed-wing UAV flies at a constant altitude $H_U$ while communicating with a ground node. Assume that the UAV follows the simple circular path on the horizontal plane with radius $r$ and the projection of the circle center  on the ground coincides with the ground node, as shown in Fig.~\ref{F:CircularPath}. The elevation angle is a function of $r$ given by $\theta(r)=\tan^{-1}(H_U/r)$. By using the elevation-angle dependent probabilistic LoS channel model and extending the result presented in \cite{904} based on the Jensen's inequality approximation of the expected communication throughput, the energy efficiency can be expressed as a closed-form expression of the radius $r$ as
\begin{align}
\EE (r) = \frac{\log_2 \left(1+ \frac{\hat P_{\LoS}(r)\gamma_0}{(H_U^2+r^2)^{\alpha/2}} \right) }{A\left(c_1+\frac{c_2}{g^2r^2}\right)^{1/4}+P_{\com}}, \label{eq:EEr}
\end{align}
where $c_1$ and $c_2$ are the constants for the fixed-wing UAV energy consumption model  as in \eqref{eq:PVFixedWing}, $A=\left(3^{-3/4}+3^{1/4} \right)c_2^{3/4}$, $\gamma_0=P_t\beta_0/\sigma^2$ is the received signal-to-noise ratio (SNR) at the reference distance of $1$ m with $P_t$ denoting the transmit power, $P_{\com}$ is the communication-related power consumption of the UAV, and $\hat P_{\LoS}(r)=P_{\LoS}\left( \theta(r)\right)+(1-P_{\LoS}\left( \theta(r)\right))\kappa$, which decreases with $r$ and can be interpreted as the regularized LoS probability, with $P_{\LoS}(\theta)$ given in \eqref{eq:PrLoS}. It is observed that as $r$ increases, both the terms involving $r$ in the denominator and  numerator in \eqref{eq:EEr} decrease. Thus, there must exist an optimal value $r^\star$ that maximizes $\EE (r)$, which is validated by Fig.~\ref{F:EEVsRadius} showing one typical plot of $\EE(r)$ against $r$. The same parameters as for Fig.~\ref{F:ChannelQualityVsTime} are used for the channel modelling, and the UAV energy consumption parameters are set as $c_1=9.26 \times 10^{-4}$, $c_2=2250$  \cite{904}, ${\gamma}_0=52.5$ dB and $P_{\com}=5$ W.

Motivated by the above, energy-efficient UAV communications have been studied for different setups with a variety of practical constraints as given in Section~\ref{sec:ConstrUAVTraj}. In particular, \cite{904} firstly derived a rigorous mathematical model for the propulsion energy consumption  of fixed-wing UAVs in terms of the UAV velocity and acceleration, and based on the derived model, optimized the energy efficiency in bits/Joule for the point-to-point UAV-ground communication over a given finite time horizon. With time discretization approach presented in Section~\ref{sec:jointOptimization}, the SCA technique discussed above was extended to solve the non-convex energy efficiency maximization problem. By numerical simulations, it was revealed that the energy-efficient UAV trajectory has an interesting ``8'' shape around its communicating ground node. For fixed-wing UAV following a circular trajectory, both the spectrum efficiency and energy efficiency were derived  in \cite{961} by optimizing the circle radius and time allocation. However, the above results for fixed-wing UAVs cannot be directly applied for energy-efficient communication with  rotary-wing UAVs, due to their fundamentally different mechanical designs and hence drastically different energy consumption models, as discussed in Section~\ref{sec:energyModel}. This thus motivated the recent work \cite{980}, where the energy consumption model of rotary-wing UAVs was derived and used to design the UAV trajectory for minimizing its energy consumption, subject to the given communication rate requirements in a multiuser system. Apart from the more complicated energy consumption model as compared to \cite{904}, another major challenge  addressed in \cite{980} is to optimize  the mission completion time that is also a design variable. This thus renders the time discretization approach inapplicable. To address this issue, the path discretization approach has been proposed in \cite{980}, as discussed in Section~\ref{sec:jointOptimization}.

While the aforementioned works  focused on the energy consumption of either the ground nodes or the UAV,  an interesting tradeoff between them was revealed in \cite{960} for UAV-enabled data collection. Intuitively, the closer the UAV flies to each  GT, the less energy is needed for the GT to transmit its data with given package size. However, this usually comes at the cost of more UAV energy consumption. Such a tradeoff has been rigorously characterized in \cite{960} for fixed-wing UAV via jointly optimizing the transmit power of ground nodes, the mission completion time, and the UAV flying speed.  Note that while the energy consumptions for UAV and GTs are usually in different magnitude orders, the changes in terms of the percentage of energy consumption along the tradeoff curve are similar for them. This thus validates the practical value of such a tradeoff to save the energy of one while compromising that of the other, depending on their energy provisions and priorities in practical applications. Besides, the more general tradeoff between UAV energy consumption and other performance metrics, such as throughput and delay, has been studied in \cite{1074}.

\subsection{UAV-Assisted Communication via Intelligent Learning}\label{sec:learning}
The aforementioned works heavily rely on the assumed  channel models for UAV communications and/or the knowledge on CSI and locations of the GTs. In practice, the channel models discussed in Section~\ref{sec:channelModel} are mostly statistical, rendering them suitable only for average performance analysis and offline trajectory optimization  rather than providing guaranteed  performance in real time, which is affected by many practical factors such as mismatched model, imperfect knowledge, and realistic channel variations in space and time. For practical implementation of UAV-assisted communications, one promising approach to deal with the above issue is by letting the UAV learn the environment by intelligent sensing and data analytics and adapt its trajectory and communication resource allocation accordingly in real time. 

One useful information that could be learned for efficient UAV communication in urban environment is the {\it 3D city map}. In fact, once the accurate information of 3D  city map is available, for any pair of UAV-user locations, the LoS/NLoS condition can be inferred directly by e.g., ray tracing, instead of being modelled as a random event as in Section~\ref{sec:channelModel}. Exploiting 3D city map for UAV placement has been studied in \cite{1065,1064,1063,1066,1061}. For example, by using the 3D map of the environment together with the estimated channel parameters, an autonomous UAV placement algorithm was proposed and demonstrated experimentally in \cite{1061} for a flying UAV relay connecting an  LTE BS to a user terminal.

For scenarios where  3D city map is unavailable, the UAV can be deployed to learn the {\it radio map} by measuring the signal powers from  GTs at known locations \cite{1067,914,1062}. In \cite{1062}, the authors developed an approach to construct the radio map for UAV-enabled relaying based on the signal strength measurements from a limited number of locations. The main idea is to firstly partition the domain of all possible UAV-user position pairs into a finite number of disjoint segments, each of which  may have different propagation environment in terms of the channel modelling parameters such as path loss exponent, average channel power at the reference distance and shadowing variance. By using the set of measurement samples available, the corresponding parameters are then  estimated  based on the principle of maximum likelihood (ML). The radio map is then constructed by classifying each UAV position into one of the  segments, based on which the average channel strength can be obtained. The radio map thus offers useful information  for various  UAV placement or path planning designs. While the samples of power  measurement in \cite{1062} for radio map construction were assumed to be given, they actually depend on the selected UAV trajectory during the learning phase. Therefore, the authors in \cite{1063} extended the work \cite{1062} by  studying firstly the {\it learning} trajectory optimization problem to minimize the estimation error of channel model parameters, and then the {\it communication} trajectory design to maximize the communication throughput based on the learned channels.

While the main purpose of utilizing city map or radio map is to learn the channel indirectly or directly, another useful technique is to learn and adapt to the environment by directly interacting with it, for which {\it reinforcement learning} emerges as a powerful tool \cite{1084}. Reinforcement learning has been used in UAV networks for various purposes, e.g., navigation \cite{1058}, anti-jamming \cite{1059}, and communication rate maximization \cite{1057}. Specifically, the authors in \cite{1058} applied the deep reinforcement learning (DRL) technique for autonomous UAV navigation in complex environment to guide the UAV flying from a given initial location to the destination, using only sensory information, such as the UAV's orientation angle and the distances to obstacles and the destination. While the main objective of \cite{1058} was to find a feasible path without explicitly considering the communication performance, the authors in \cite{1057} studied the trajectory of a UAV BS serving multiple users to maximize the communication sum rate. By applying Q-learning, which is a model-free reinforcement learning method, the UAV acts as an autonomous agent to learn the trajectory to maximize the sum rate with multiple ground users, without assuming any explicit information about the environment (such as user locations and channels with them). By dividing the possible flying area into 15 by 15 grids, it was shown  that the UAV is able to interact with the environment to reach the location achieving  the maximum sum rate, and yet avoid flying through the shadowed area with obstacles and thus experiencing poor channel quality. However, as pointed out in \cite{1057}, one major limitation of the proposed Q-learning approach for trajectory optimization is the heavy learning time, which makes it infeasible even for moderate state spaces, e.g., 30 by 30 grids. Therefore, one promising future research direction is to reduce the complexity and learning time for machine learning based UAV trajectory and communication co-design. One possible approach is to combine the offline UAV trajectory designs as described in Section~\ref{sec:coDesign} for coarse initial trajectory planning and the online learning techniques to further refine the trajectory and optimize communication resource allocation in real time. Machine learning for UAV-assisted wireless communications is still in its infancy but anticipated to be a promising  avenue for future research and investigation.

%
%
%

%
%
%
%
%
%
%
%
%
%
%
%
%
%
%
%
%
%
%
%
%
%
%

\section{Cellular-Connected UAV}\label{sec:cellularConnected}
%


In this section, we focus on the other framework of cellular-connected UAV communications, where the UAVs are supported by cellular BSs as new aerial users. We first give a historical overview of the past efforts on  supporting aerial users in cellular networks, by highlighting the major field trials from 2G to 4G, including the latest standardization efforts by 3GPP. We then present some representative works on performance evaluation of  cellular-connected UAVs  by numerical simulations as well as  theoretical analysis. Last, we discuss some promising techniques to embrace the new aerial users in future cellular networks, for air-ground interference mitigation and QoS-aware UAV trajectory planning.

\subsection{Supporting Aerial Users: Field Trials From 2G to 4G}\label{sec:FieldTrials}
The attempt to support aerial users with cellular networks can be traced back to 2000's via 2G cellular networks, namely Global System for Mobile Communications (GSM) \cite{976,1044,1045}. A prototype system was developed in \cite{976} to test the remote UAV operation using General Packet Radio Service (GPRS), which is a transmission technology for GSM. Based on the flight test, it was concluded that GSM network infrastructures can provide a useful means as a complementary communication channel for UAV. In \cite{1044}, the aerial RSSI (Received Signal Strength Indicator) measurements were conducted over GSM networks to show the change of cellular coverage versus altitude. The results showed that RSSI increases with the altitude in urban environment, due to the reduced blockage, whereas it decreases with altitude in rural environment due to the increased link distance. The authors claimed that the experiment results provided the evidence of available RF coverage in altitude up to 500 m.

Later, UAV flight tests were  conducted over 3G UMTS (Universal Mobile Telecommunications System) network \cite{1042}. The measurement results showed good connections for the UAV altitude up to about 8000 feet (2438 m), beyond which the connection was lost. In addition, it was also shown that although the BS antenna orientations are optimized for ground users,  the average received power levels of the aerial users are 21\% stronger than those on the ground, with latency in the order of 500 ms. Based on such results, the authors concluded that the 3G UMTS network could provide a possible solution for non-safety-critical communications for aerial users with moderate speed and altitude (below  4000 feet or 1220 m).

While the research work on 2G/3G-supported UAVs was limited, the enthusiasm for supporting UAVs via the 4G LTE network has skyrocketed during the past few years, in both academia (see, e.g., \cite{981,949,950,952,945,978,984,1080}) and industry. This could be attributed to the significantly enhanced performance of LTE network over its predecessors, making it more promising to support aerial users, as well as the tremendous increase of UAV applications over the recent years.

In \cite{981}, flight tests with UAV altitude varying from 10 m to 100 m were conducted to compare the latency performance of cellular-supported UAVs with three different technologies: EDGE (Enhanced Data rates for GSM Evolution, regarded as pre-3G technology), HSPA+ (Evolved High Speed Packet Access), and 4G LTE. It was revealed that LTE achieved the best performance in terms of latency and jitter, with round-trip time (RTT) of 127 ms and standard deviation of 48 ms for the worst case scenario, and EDGE had the worst performance. Such results demonstrated the feasibility of (semi-)autonomous UAV operations over LTE network with low altitude  (say, up to 100 m). 

In \cite{987}, the possibility of using LTE for controlling multicopter was studied based on field measurement. The RSRP (Reference Signal Received Power) and RSRQ (Reference Signal Received Quality) were measured for an LTE-connected UAV moving vertically with a maximum altitude at 74 m, and with a building between the initial UAV location and the BS. It was shown that the RSRP firstly increases and then decreases with altitude, with the maximum value achieved at around 34 m. In contrast, the RSRQ has the trend of decreasing with the increase of altitude. This is due to that the increase of interference is more dominant than  the increase of RSRP. 

In \cite{949}, measurements were taken with the main goal to quantify the interference experienced by aerial users at different altitude. It was found that the number of detectable BSs increases as the UAV moves higher. However, the SINR of the best cell for  the aerial user at the measured altitude of 150 m or 300 m is much lower than that of ground user. This is due to the dramatic increase of downlink interference at higher altitude. Such  observations have been corroborated  by the extensive field trials for UAVs over commercial LTE networks by Qualcom, based on which a trial report on LTE UAS was released in May 2017 \cite{941}. It was found that although the BS antennas are downltilted towards the ground, satisfactory signal coverage can still be achieved for altitude up to 400 feet (122 m) in the studied test. In fact, the experiment showed that at 400 feet, the UAV is able to detect 18 BSs with the furthest one up to 11.5 miles (18.5 km) away. 
 Such observations have been corroborated by other field measurement campaigns in various setups \cite{978,984,990,1051,1054}.







%
%

\subsection{Recent Results by 3GPP Study}\label{sec:3GPP}
Realizing the great business opportunities for cellular operators with the fast growth of UAV industry, 3GPP approved the study item on enhanced LTE support for aerial vehicles in March 2017 \cite{1081}. The main objective of the study item is to investigate the feasibility and ability of serving aerial vehicles using LTE network with BS antennas downtilted mainly for terrestrial coverage. The study item was completed in December 2017 with the main results and findings reported in the Technical Report TR36.777 in Release 15 \cite{1012}. It was then followed by a new work item aiming to further improve the efficiency and robustness of terrestrial LTE network for serving UAVs.

In the Technical Report \cite{1012} resulted from the study item \cite{1081}, 3GPP has specified that the maximum height and the maximum horizontal speed for aerial vehicles are 300 m and 160 km/h, respectively. Among others, one of the main outputs from the study item is the comprehensive GBS-UAV channel model for three typical deployment scenarios, as presented  in Section~\ref{sec:channelModel}. The developed channel model extends the conventional terrestrial channel model for altitude up to 300 m, with detailed specifications on the path loss, LoS probability, shadowing, and small-scale fading. Such a channel model is very useful for detailed system level simulations for cellular networks with coexisting terrestrial and aerial suers. Furthermore, based on the extensive field measurements and system level simulations, 3GPP has identified some  main technical challenges in supporting aerial vehicles with cellular networks. While the detailed findings can be found in \cite{1012}, we provide a summary of them as follows to motivate future research.


{\bf Interference detection}: Detecting the interference levels to/from aerial UEs is necessary for identifying the strong interference scenarios and thereby implementing effective countermeasures for them, especially when the  UEs are potentially not certified for aerial usage. Interference detection can be achieved in practice via UE-based solutions and/or network-based solutions. For UE-based solutions, the interference can be detected based on measurement report by UE, on e.g., RSRP and RSRQ. Furthermore, other UE-side information such as mobility history report and speed estimation can be utilized to facilitate the interference detection. On the other hand, for network-based solutions, interference detection can be performed by exchanging information among BSs, such as their uplink scheduling information, and received measurement reports from UEs on their  RSRP, RSRQ, and CSI.

{\bf Uplink interference mitigation}: To mitigate the uplink interference caused by the transmission of aerial UEs to their non-associated BSs, 3GPP has suggested the following three techniques:

(i) Uplink power control:
 To deal with  the heterogeneous network with both terrestrial and aerial users, the existing uplink power control mechanism could be improved by e.g., introducing UE specific power control parameters. For example, in open loop power control for which the path loss of UEs is partially compensated, the UE's transmit power can be written as \cite{941}
 \begin{align}
 P_{\mathrm{tx}}=\min\{P_{\max}, 10\log_{10}(M_{\mathrm{RB}})+P_0+\alpha_{\mathrm{UE}}\cdot \mathrm{TPL}\},
 \end{align}
 where $P_{\max}$ is the maximum transmit power, $M_{\mathrm{RB}}$ is the number of RBs assigned, $P_0$ is a nominal value, $\alpha_{\mathrm{UE}}$ is the fractional path loss compensation factor, and $\mathrm{TPL}$ is the estimated total path loss. The simulation results in \cite{1012} showed that compared to the case where the same $\alpha_{\mathrm{UE}}$ is used for all UEs, significant performance gain can be attained by using height-dependent compensation factors, e.g., $\alpha_{\mathrm{UE}}=0.8$ for terrestrial UEs and aerial UEs below 100 m, and $\alpha_{\mathrm{UE}}=0.7$ for aerial UEs above 100 m.

(ii) FD-MIMO: With FD-MIMO (or 3D beamforming), BSs are equipped with full dimensional antenna arrays with active elements to achieve flexible beamforming in both azimuth and elevation dimensions. FD-MIMO has been  supported in LTE since Release 13, and is particularly promising to support aerial UEs for interference mitigation, as will be further elaborated  in Section~\ref{sec:performanceEvalu}.

(iii) Directional antenna at UE: Directional antennas can be used at aerial  UEs to focus the signal downward to their  associated cells while reducing the interference to other cells. Apparently, the performance of this technique critically depends on the ability to align the antenna main lobe with the direction of the serving BS. Depending on the directional antenna type as discussed in Section~\ref{sec:AntModel}, direction alignment can be achieved either mechanically  or electrically (via  phased array or digital beamforming).

{\bf Downlink interference mitigation}: For the mitigation of downlink interference from co-channel BSs to aerial UEs, the FD-MIMO and directional antenna at UE can be similarly applied. In addition, 3GPP has suggested three other  techniques: (i) intra-site joint transmission CoMP (JT CoMP), where multiple cells/sectors belonging to the same site jointly transmit to their served  UEs; (ii) coverage extension techniques to enhance synchronization and initial access for aerial UEs. This technique mainly aims to address the extremely severe interference scenario when even the minimum required SINR for the normal LTE control channels cannot be satisfied. The coverage extension introduced in Release 13 is achieved mainly via signal repetitions, which gives higher signal energy to mitigate interference through a processing gain \cite{1093};  and (iii) coordinated data and control transmission, where data and control signals are jointly transmitted to the UEs.

{\bf Mobility:} 3GPP has also briefly discussed the potential enhancement for mobility performance, by e.g., refining handover procedure and related parameters for aerial UEs based on their airborne status, location information, and flying path information, so as to avoid frequent handovers due to  high UAV mobility and BS antenna side-lobe gain variation.

As a summary, the extensive field measurement campaigns and 3GPP investigation   have provided strong evidence that the existing LTE networks should be able to support the initial UAV deployment with low density and low altitude, without the need of  major changes. On the other hand, they also revealed the more severe air-ground interference issue than that in the traditional terrestrial network. As the number of UAVs grow rapidly due to their more appealing applications, it is necessary to develop new techniques to enable cellular-connected UAVs for their larger-scale deployment, in terms of ubiquitous 3D aerial coverage, effective air-ground interference mitigation, as well as enhanced requirements for both CNPC and payload data communications in anticipation. In the following, we present some representative studies on the performance evaluation of cellular-connected UAVs to gain a deeper understanding of this new cellular system model, followed by  some promising and advanced techniques for performance enhancement.

\subsection{Performance Evaluation}\label{sec:performanceEvalu}
While field tests are very useful for feasibility studies, they are generally quite expensive and time-consuming to implement. Besides, the obtained results are typically dependent on the particular scenarios being tested. In parallel to the field tests discussed above, there have been research efforts on the performance evaluation of cellular-connected UAVs via numerical simulations  \cite{989,986,950,985,952} or theoretical analysis \cite{982,983,1046,1047}.

\begin{figure*}
\centering
\includegraphics[width=0.7\linewidth]{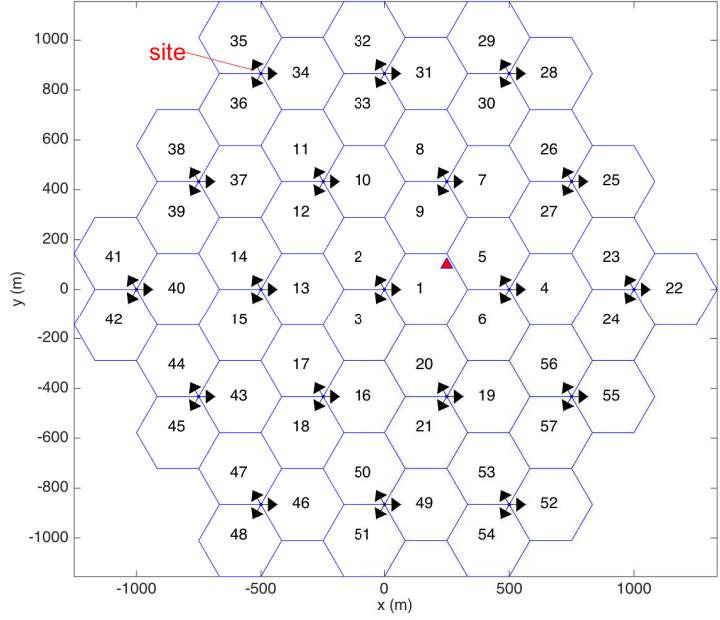}
\caption{Cell layout for numerical simulations of cellular-connected UAV. Arrows denote boresight of each cell \cite{952}.}\label{F:setup}
\end{figure*}

First, based on the simulation results reported in \cite{952}, we illustrate some new considerations that deserve particular attention in designing and implementing cellular-connected UAV communications. A simplified  cellular system with 19 sites is considered, each constituting 3 sectors/cells, with their cell IDs labelled in Fig.~\ref{F:setup}.  Two different BS array configurations  discussed in Section~\ref{sec:AntModel} are considered: {\it fixed pattern} versus {\it 3D beamforming}. For fixed pattern, a ULA of size  $(M_1,M_2)=(8,1)$ is employed at each sector, where $M_1$ and $M_2$ denote the number of antenna elements along the vertical and horizontal dimensions, respectively. For this configuration, the steering magnitude and phase of each antenna element are predetermined to achieve a $-10^\circ$ electrical downtilt. The synthesised array radiation pattern of this configuration is shown in Fig.~\ref{F:BSAntPattern}. On the other hand, with 3D beamforming, each sector is equipped with a UPA of size $(M_1,M_2)=(8,4)$, and the signal magnitude and phase by each antenna element can be flexibly designed to enable 3D beamforming.  

\begin{figure*}
\centering
\begin{subfigure}{0.48\textwidth}
\includegraphics[width=\linewidth]{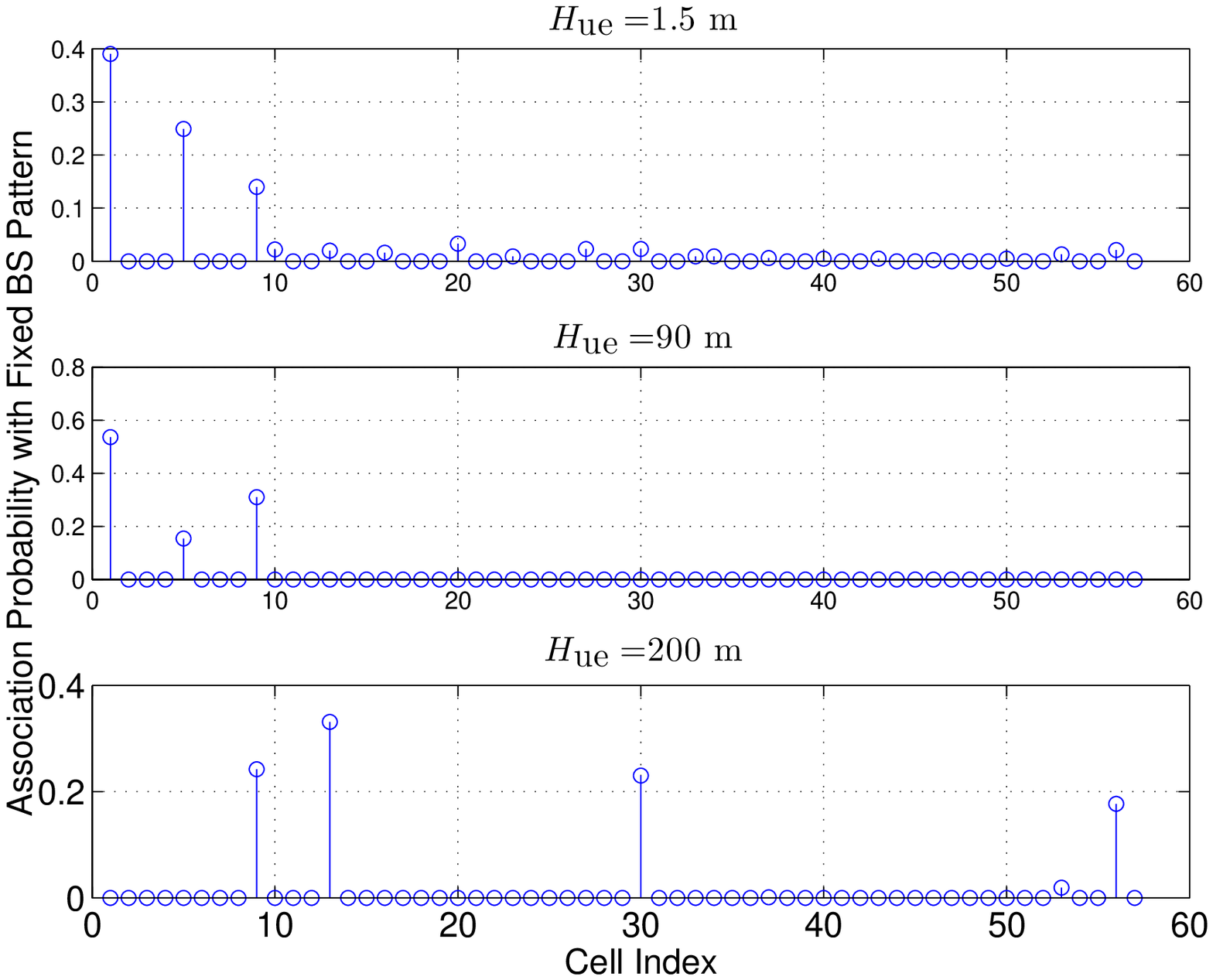}
\caption{Fixed BS pattern}
\end{subfigure}
\hspace{0.02\textwidth}
\begin{subfigure}{0.48\textwidth}
\centering
\includegraphics[width=\linewidth]{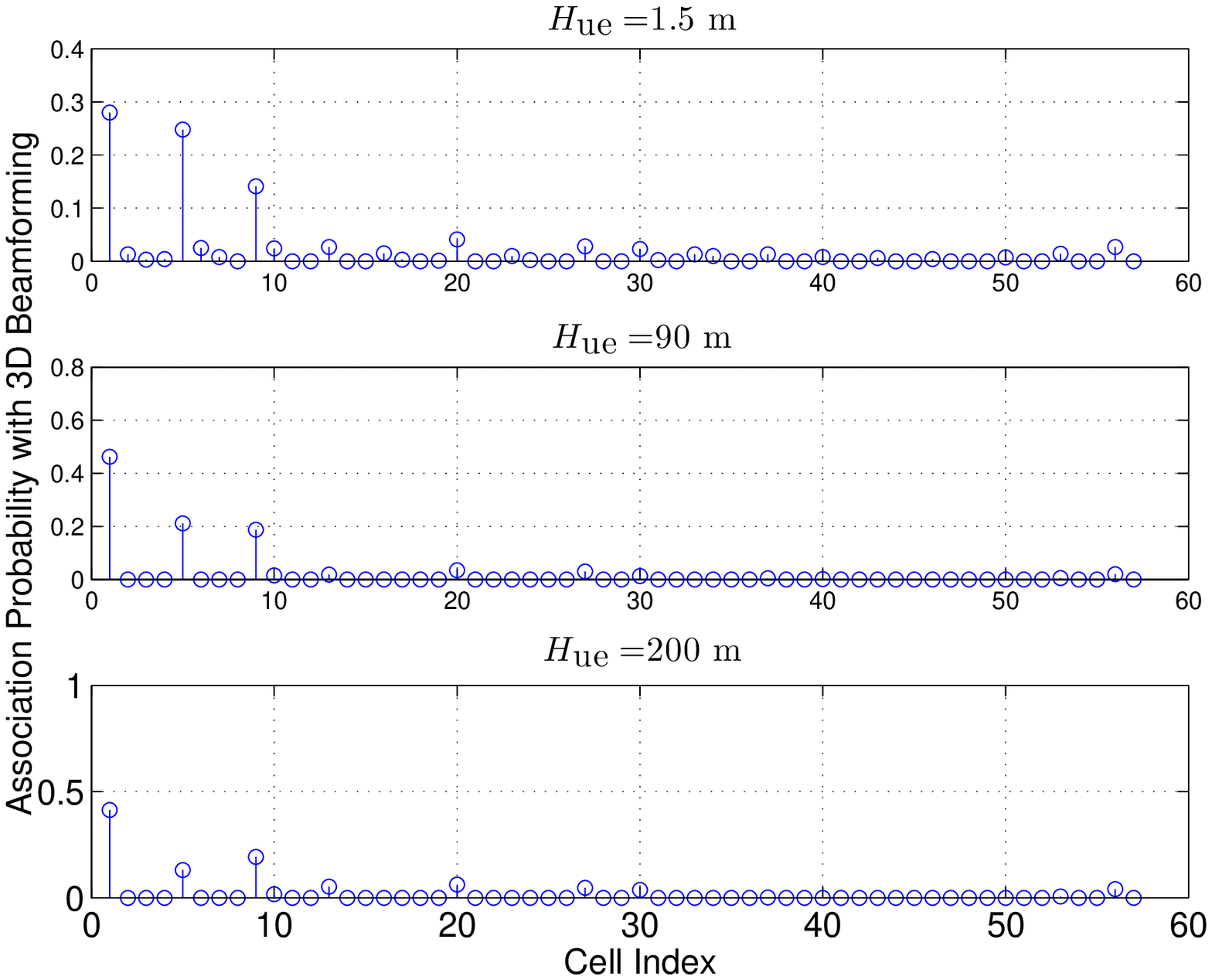}\\
\caption{3D Beamforming}
\end{subfigure}
\caption{Association probability at different UAV altitude \cite{952}.}\label{F:AssoProb}
\end{figure*}

Fig.~\ref{F:AssoProb} shows the empirical  cell association probability for a user with three different altitudes, while its horizontal location is fixed at (250 m, 100 m), as marked in red triangle in Fig.~\ref{F:setup}. The maximum RSRP-based association rule is used. It is observed from Fig.~\ref{F:AssoProb}(a) that with the fixed BS pattern, the UAV is most likely associated with the nearby cells when the altitude is low (e.g., cells 1, 5 and 9 for $H_{\mathrm{ue}}=$ 1.5 m and 90 m). However, as the altitude increases, it is more likely that the associated cell is far away from the UAV, e.g., cells 13, 30 and 56 for $H_{\mathrm{ue}}=$ 200 m. This is expected due to the downtilted antenna pattern as shown in Fig.~\ref{F:BSAntPattern}. Specifically, as the UAV moves higher, it is likely that it falls into the antenna nulls or weak side lobes of the nearby BSs. As a result, the UAV may need to  associate with more  distant cells via their stronger side lobes. In contrast, with 3D beamforming, Fig.~\ref{F:AssoProb}(b) shows that the UAV is almost surely associated with the nearby cells even for high altitude at $H_{\mathrm{ue}}=$ 200 m, thanks to the flexible beam adjustment to focus signals to the UAV with 3D beamforming.

\begin{figure*}
\centering
\includegraphics[width=0.48\linewidth]{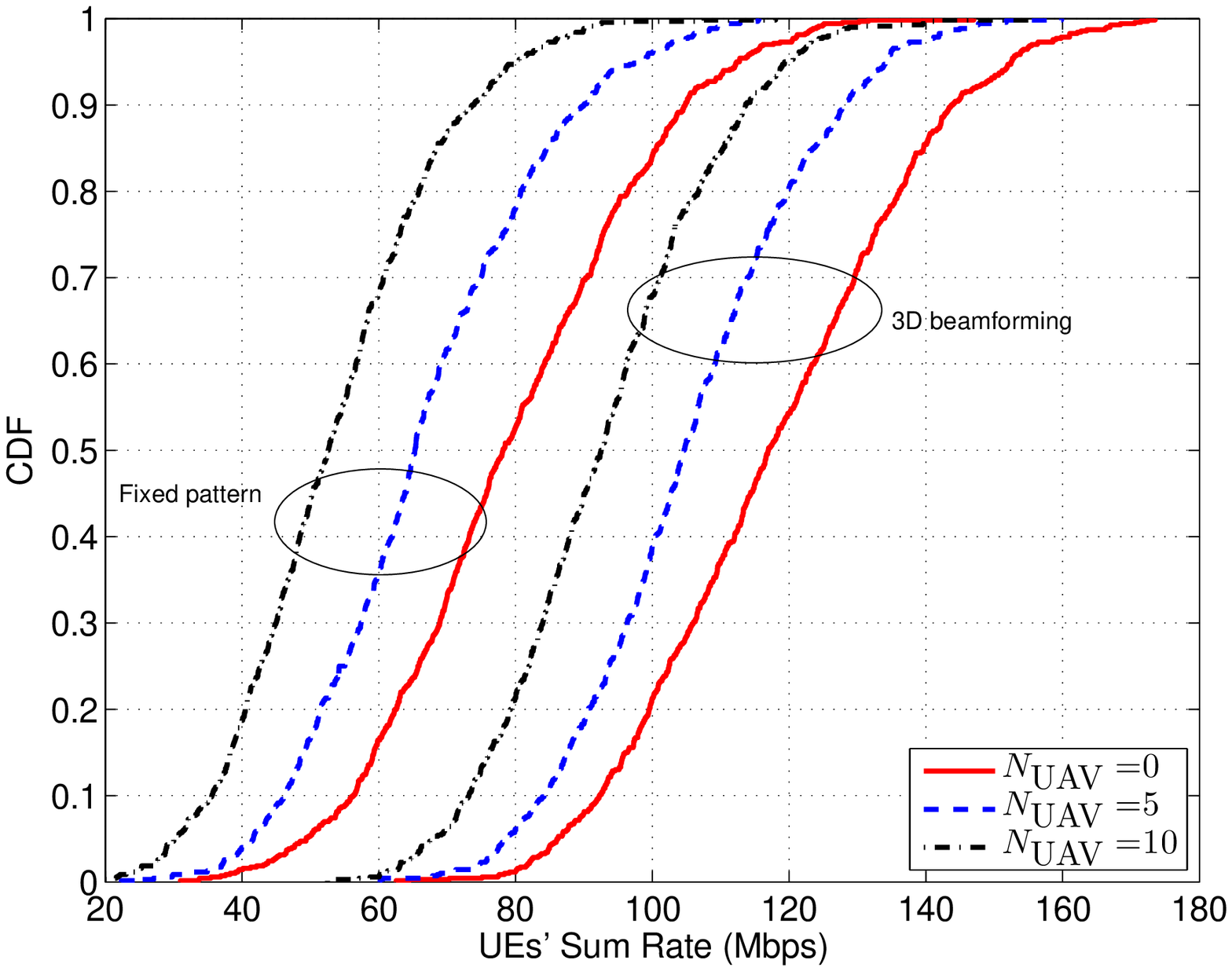}
\caption{Empirical CDF of UEs' achievable sum rate with different number of UAVs \cite{952}.}\label{F:SumRateCDF}
\end{figure*}

For a cellular network with a total of 15 aerial and ground users, Fig.~\ref{F:SumRateCDF} plots the empirical cumulative distribution function (CDF) of the users' achievable sum rate in the downlink as the number of UAVs changes. It is observed that for both array configurations, the overall system spectral efficiency degrades as the number of aerial users/UAVs increases. This is mainly due to  the stronger interference suffered by the aerial users as compared to ground users. On the other hand, Fig.~\ref{F:SumRateCDF} shows that by employing 3D beamforming, the system spectral efficiency can be significantly improved. This demonstrates the great potential of 3D beamforming for interference mitigation in cellular systems with coexisting aerial and ground users. Similar results and observations can be obtained for the uplink communication with the strong UAV interference to co-channel BSs.

Besides numerical simulations, there were also works  on theoretical performance analysis for cellular-connected UAVs. For example, in \cite{982}, based on  stochastic geometry with the ground BSs modelled by a HPPP,  the authors analyzed the downlink coverage probability for an aerial user coexisting with conventional ground users. For simplicity, the BS antenna was modelled as the two-lobe model given in \eqref{eq:twoLobe}, while the UAV was assumed to be associated with the nearest BS. Based on the derived coverage probability expression and numerical examples, it was concluded that lowering BS antenna height and increasing downtilt angle is beneficial. However, this result may not hold if the RSRP-based association is considered in practice as it has been shown in Fig.~\ref{F:AssoProb} that in general a UAV may not be associated with its nearest BS. Therefore, the analysis in \cite{982} was extended in \cite{983} by associating the UAV with the BS with the maximum power, instead of the nearest one. The authors further extended the analysis to the scenario that the UAV is also equipped with a directional antenna \cite{1046}, with the two-lobe antenna model shown in \eqref{eq:twoLobeUAV}. It was found that compared to the case of omnidirectional antennas at the UAV, the use of directional antennas with the optimum choice of antenna tilt can significantly improve the coverage probability and achievable throughput. The impact of using directional antenna at UAV for cellular UAV communications has also been studied in \cite{1047}, 
where the coverage performance was analyzed by assuming that the UAV can intelligently tilt its main lobe direction. A more comprehensive analysis for cellular-connected UAVs for both uplink and downlink communications with general directional BS and/or UAV antenna models has been given in \cite{1085}.

\subsection{Advanced Techniques for Air-Ground Interference Mitigation}\label{sec:InterfMitigation}
Existing studies based on field measurements, numerical simulations, and theoretical analysis all showed that cellular networks supporting aerial users will face the more severe interference issue. In the uplink transmission from UAV to BS, UAV could cause strong interference to a large number of co-channel BSs due to the high-probability LoS propagation at high altitude. On the other hand, in the downlink transmission, UAV is the victim that may suffer severe interference from many non-associated BSs. Thus, how to combat against the severe air-ground interference is of paramount importance for enhanced cellular support for aerial users. 
As summarized in Section~\ref{sec:3GPP}, 3GPP has suggested several practical interference mitigation techniques that are readily for use without radically changing the network infrastructure or specifications. In the following, we further elaborate several advanced interference mitigation techniques by highlighting their unique opportunities and challenges in cellular systems supporting both terrestrial and aerial users.

%
%
%
%
%
%
%
%
%
%
%
%

{\bf 3D beamforming:} Beamforming is an effective multi-antenna technique that dynamically adjusts the antenna radiation pattern based on user location or even instantaneous CSI. Furthermore, compared to conventional 2D beamforming, 3D beamforming (or FD-MIMO) offers the enhanced  capability of more refined angle resolutions in both azimuth and elevation dimensions. This thus significantly improves the interference mitigation capability by exploiting the elevation angle separation of users. 3D beamforming can only be achieved with full dimensional (FD) antenna array with active array elements, such as UPA/URA. Note that 3D beamforming is not new, which has received notable interest in conventional cellular networks \cite{946} and has been supported in LTE since Release 13. However, the integration of aerial users with dominant LoS BS-UAV channels offers a new {\it elevation angular diversity} that renders 3D beamforming particularly appealing in cellular-connected UAV systems. Specifically, compared to conventional cellular networks with terrestrial users only, it is more likely to find two users with sufficiently separated elevation angles in a hybrid  aerial-terrestrial cellular system, where 3D beamforming is more  effective. Note that similar angular diversity exists from the UAV perspective to sufficiently separate the ground BSs. Thus, 3D beamforming can also be quite effective at the UAV side. The preliminary studies in \cite{952} have demonstrated the promising gains of 3D beamforming over the conventional BS antenna configuration with fixed radiation pattern.

Of particular interest is the use of 3D beamforming under the massive MIMO paradigm \cite{373,374,497}, i.e., the number of BS antennas is much larger than the number of served users. There are some initial research efforts towards this direction for massive MIMO cellular UAV communications \cite{951,988,992}. For instance, in \cite{992}, via extensive numerical simulations based on the latest 3GPP channel models as discussed in Section~\ref{sec:channelModel} \cite{1012}, the authors provided a rather comprehensive and insightful performance comparison of the downlink UAV communication supported by the traditional cellular network versus the future massive MIMO network. The results showed that massive MIMO can dramatically enhance the reliability of the downlink UAV C\&C channel due to the better interference mitigation.

To practically enable 3D beamforming for cellular UAV communications, efficient channel/beam training and tracking techniques need to be developed to cope with the high UAV mobility, which may induce significant Doppler effect and channel phase variations. One possible approach is to exploit the LoS-dominant BS-UAV channel and the knowledge on the UAV trajectory/velocity, which can be acquired a priori or estimated in real time,  to reduce the pilot overhead. However, for 3D beamforming in cellular-connected UAV systems, both the azimuth and elevation beam directions need to be estimated and tracked, which thus calls for new and efficient designs.


{\bf Coordinated resource allocation:}
Coordinated resource allocation, or inter-cell interference coordination (ICIC) is the mechanism used in practice to mitigate inter-cell interference by jointly optimising the communication resources across different cells, which may include channel assignment, power allocation, beamforming, BS association, etc. To this end, the cooperating BSs usually need to exchange the CSI of their served users via cellular backhaul links. While ICIC has been extensively studied and standardized for LTE networks with terrestrial users, its performance  for the new  UAV users  deserves a further study. In particular, due to the LoS-dominant  propagation between UAV and BSs, the number of potential coordinating BSs is typically much larger than that for serving terrestrial users only. This brings new issues on the implementation complexity and latency.

 There has been some initial research effort on coordinated resource allocation for cellular-connected UAVs. In \cite{weidong2018GC} and \cite{1053}, the authors studied the ICIC designs for uplink UAV communications via jointly optimizing the UAV's uplink cell association and power allocation over multiple RBs.  To reduce the implementation complexity,  a decentralized ICIC scheme was proposed by dividing the cellular BSs into small-size clusters, where the information exchange is only needed between the UAV and cluster-head BSs by exploiting the LoS macro-diversity.

{\bf CoMP:}
Compared with coordinated resource allocation, one more effective technique for multi-cell cooperation is CoMP transmission/reception. In this case, the signals for each user are jointly transmitted/received by multiple cooperating BSs, which form a virtual distributed antenna array or network MIMO system \cite{211}. Different from coordinated resource allocation aiming to suppress the interfering links, CoMP essentially exploits the strong cross links for desired signal transmission by simultaneously associating each user with multiple BSs. This is especially appealing for cellular-connected UAVs, due to the larger macro-diversity gain available for aerial users as compared to terrestrial users. However, this also incurs more complexity and backhaul transmission delay as more cooperating BSs need to be involved. For low-complexity implementation, it is necessary to optimize the set of cooperating BSs so as to achieve a tradeoff between performance and complexity/delay, by taking into account the flying status such as UAV speed and altitude, as well as the BS-UAV channel models. For example, one possible approach is the {\it UAV-oriented} cell cooperation, where large-scale multi-cell cooperation is applied for those UAVs with low speed that induces slower channel variations, and/or at high altitude with potentially large macro-diversity gains. In addition, the impact of the additional delay due to CoMP on the performance of CNPC transmissions  needs to be critically evaluated.

There have been some recent research efforts on investigating low-complexity multi-cell cooperation for cellular-connected UAVs. For example, to reduce the backhaul delay of CoMP, the authors in \cite{1048} proposed a {\it cooperative interference cancellation} strategy for uplink cellular UAV MIMO communications. In this scheme, it is assumed that each UAV uses a RB that is occupied by ground users only at  some (not all) of the BSs, termed as {\it occupied BSs}, which is valid in practical cellular networks with fractional frequency reuse. Then those unoccupied BSs could be utilized to decode the UAV signals, and forward them to adjacent occupied BSs for interference cancellation. The proposed scheme achieves better performance than the conventional transmit beamforming without the cooperative interference cancellation, and on the other hand requires less complexity than CoMP since cooperation is limited only to adjacent BSs.

The above idea was further extended in \cite{1049}, leading to the novel scheme termed {\it cooperative NOMA (non-orthogonal multiple access)}. Different from  the cooperative interference cancellation between only non-occupied and occupied BSs as in \cite{1048}, the UAV signal might be also decoded at some of the occupied BSs, as long as their received UAV  signal  strengths  are sufficiently strong as compared to that of the  terrestrial users. Then the decoded UAV signal is forwarded to adjacent occupied  BSs for interference cancellation, even without using the non-occupied BSs. Compared to the conventional non-cooperative NOMA scheme with only local interference cancellation at occupied BSs, the proposed cooperative NOMA achieves significant performance gains. The extension of the above works for UAV downlink communication is more involved, which deserves further studies.



\subsection{QoS-Aware UAV Trajectory Optimization}
%
%
%
%

Different from the conventional terrestrial users that usually move sporadically and randomly, the mobility of UAV users is fully or at least partially controllable. This offers an additional  DoF for cellular-connected UAVs, via their communication QoS-aware trajectory design. For example, for areas where ubiquitous aerial coverage by cellular network has not been achieved yet, the UAV path can be deliberately planned to circumvent entering any coverage holes. However, it should be noted that the trajectory design for cellular-connected UAVs is different from that for UAV-assisted communications in the following aspects. Firstly, for cellular-connected UAVs, UAVs usually have their own missions such as inspection, delivery, photography, etc.,  which to a certain extent limit their flexibility in trajectory adaptation to enhance communication performance as compared to UAV-assisted communications, in which UAVs are dedicated BSs/relays/APs with fully controllably trajectories. Secondly, different from UAV-assisted communications where the trajectories of UAV BSs/relays/APs in general need to be designed to ensure the coverage of all their served users,  for cellular-connected UAVs, they are users and only need efficient  trajectories to fulfill their communication requirements with some BSs along the trajectories. As a result, the UAV trajectory designs for the above two cases are generally different.

\begin{figure*}
\centering
\includegraphics[scale=0.3]{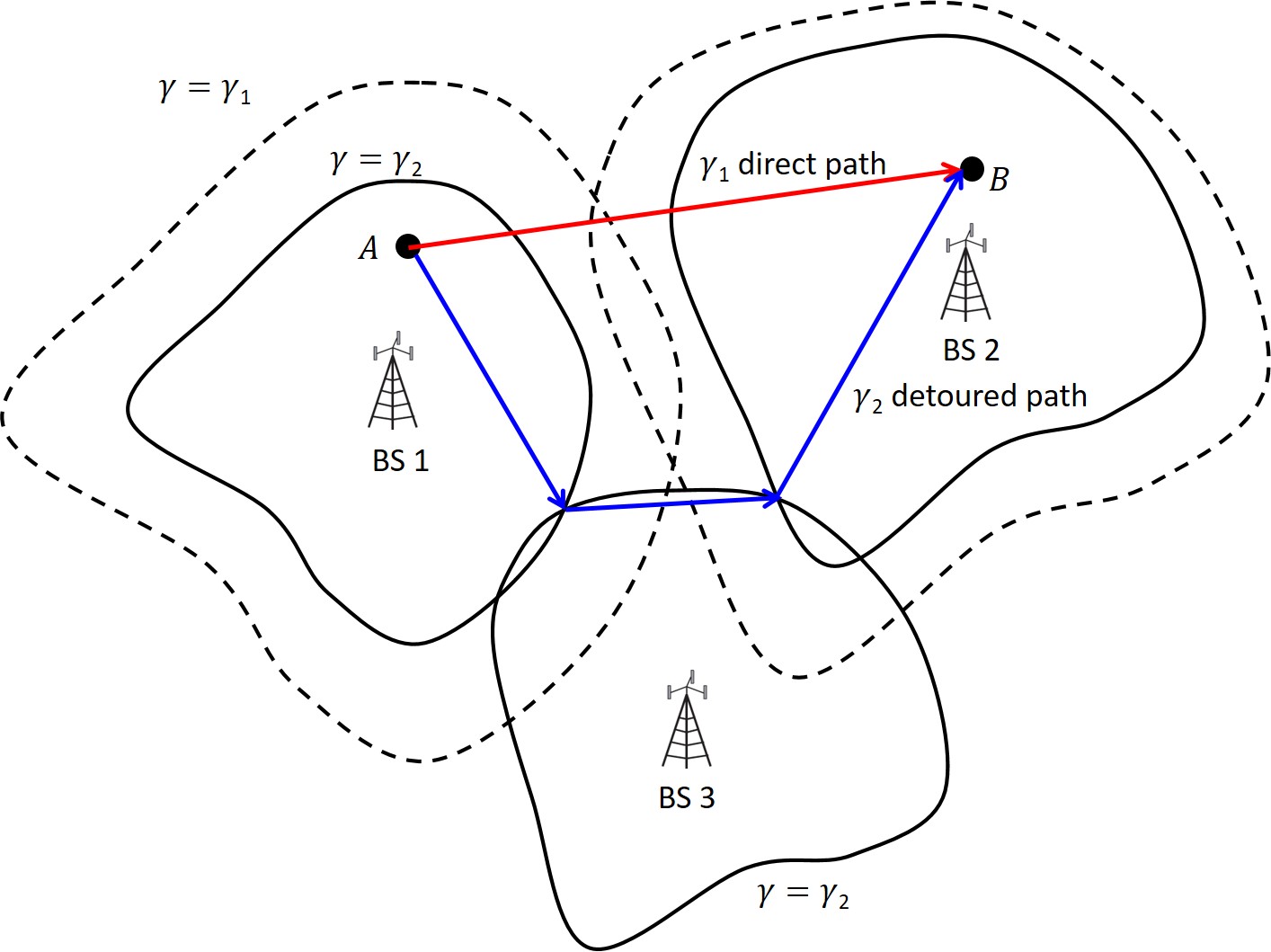}
\caption{An illustration of QoS-aware UAV path planning with target SNR $\gamma_1<\gamma_2$.}\label{F:QoSAwarePath}
\end{figure*}

 As an illustrating example, let's consider the scenario that a UAV aims  to deliver a package from an initial location $A$ to a destination $B$ with minimum time, while ensuring that it maintains good connection with at least one BS at any time along its trajectory. In practice, a good connection may be defined as follows: the outage probability that the SNR is below a target threshold $\gamma$ is less than some tolerable value $\epsilon$. Intuitively, for any given $\epsilon$,  the coverage region of each BS depends on $\gamma$, which in turn affects the UAV's optimal flying path. For simplicity, assuming that the UAV maintains a constant altitude $H_U$, the coverage areas of the BSs in the UAV's flying plane for two different $\gamma$ values are illustrated in Fig.~\ref{F:QoSAwarePath}. Note that the coverage area is in general of irregular shape that depends on the BS antenna radiation pattern and the random shadowing.  When $\gamma$  is small (i.e., $\gamma=\gamma_1$), it is possible that there exists a straight path from $A$ to $B$ satisfying the connectivity constraint with the flying distance minimized, as shown by the red path in Fig.~\ref{F:QoSAwarePath}. However, as  $\gamma$ increases (i.e., $\gamma=\gamma_2$), the UAV may have to detour its flying path to maintain the connection with BSs, as illustrated by the blue path in Fig.~\ref{F:QoSAwarePath}, and as a result, more traveling time is needed. 

There have been some recent works along the above  direction. For example, in \cite{1080}, the authors formulated the UAV trajectory optimization problem to minimize the UAV flying time, subject to the  stringent zero-outage constraint for the UAV at any time along its trajectory. By assuming the free-space LoS channel  model and isotropic antennas so that the coverage areas in Fig.~\ref{F:QoSAwarePath} reduce to circles, effective UAV trajectory solutions were obtained by utilizing the graph theory and convex optimization techniques. A similar problem was investigated in \cite{1008} and \cite{1091}, but with certain tolerance on the loss of cellular connection, provided that such disconnected duration  does not exceed a given threshold. Based on the field measurement of the uplink interference by UAVs, the authors in \cite{1051} suggested a possible solution to reduce uplink interference caused by aerial users by controlling their cruise height, though this usually compromises the UAV's link quality. A more general  trajectory optimization problem for UAV uplink communication subject to their interference power constraints at the terrestrial users has been studied in \cite{1087}.

Note that one practical challenge for optimal QoS-aware UAV path planning lies in how to obtain  the accurate 3D coverage maps of the BSs. Toward this end, the use of reinforcement learning for path planning of cellular-connected UAVs has been studied in \cite{1060}, where the UAV trajectory,  cell association, and power control were jointly optimized based on a noncooperative game formulation. Note that the research on machine learning empowered cellular-connected UAV communication and trajectory optimization  is still at an early stage. Similar to UAV-assisted communications, in order to reduce the learning time and complexity, trajectory planning for cellular-connected UAVs may also  require a combined offline and online design approach, which deserves further investigation. 

\section{Extensions}\label{sec:otherTopic}
Some other relevant topics to UAV communication that are worthy of further investigation  are discussed  as follows.
\subsection{Security}
Future wireless networks are expected to support massive user and device communications, which makes information security a more challenging task.  The network security issue can be tackled either at higher communication protocol  layers by using e.g. cryptographic methods or at the physical layer by exploiting the intrinsic characteristics of wireless channels. With the integration of UAVs into wireless networks, the LoS-dominant air-ground channel and high mobility of UAVs bring  new opportunities as well as  challenges for physical-layer  secure communications, depending on whether the UAVs are legitimate or malicious nodes  in the network \cite{wu2019safeguarding,1073,zhang2018securing,wang2017improving,zhang2017securing,cui2018robust,xiao2018secrecy,ye2018secure,cai2018dual,lee2018uav}. For example, thanks to  the high mobility,  legitimate UAV transmitters or receivers can move far away from ground eavesdroppers to reduce  information leakage to them  \cite{zhang2018securing,zhang2017securing,cui2018robust,xiao2018secrecy}. Besides, their high altitude also  helps detect the ground eavesdroppers' locations effectively  via UAV-mounted cameras/radars. More pro-actively, artificial noise can be sent  by dedicated UAV jammers deployed above ground eavesdroppers to interfere with them and thus prevent against their wiretapping   \cite{1073}.  In practice, using multiple cooperative UAVs with different roles can further improve the wireless communication security  \cite{cai2018dual,lee2018uav}.  On the other hand, if the UAVs are malicious nodes in the network, their aforementioned advantages turn out to be new threats to  the terrestrial secure communications as they can be  more easily eavesdropped and/or jammed by UAVs. Therefore,  effective  techniques to combat such  airborne eavesdropping and jamming are crucial \cite{wu2019safeguarding} and worth investigating in future work.

Besides information security, there are also other security issues for UAVs, such as how to detect and track malicious UAVs \cite{guvencc2017detection}, how to prevent the GPS spoofing attacks to the legitimate UAVs \cite{kerns2014unmanned}, etc., which are also crucial and deserve further investigation. For example, while active UAVs such as UAV jammers can be detected/localized by using conventional signal sensing and ranging techniques, passive UAVs such as UAV eavesdroppers generally require more sophisticated detection techniques  such as radar and/or  computer vision based methods.

\subsection{Caching}
Wireless caching is regarded as a promising solution to support the explosive growth of the mobile multimedia traffic arising from e.g., video streaming and mobile TV \cite{liu2016caching}. By leveraging the storage device  at BS/mobile terminal, the popular contents can be proactively cached during off-peak period so as to reduce the real-time  transmission delay and alleviate the network  backhaul burden. However, as each BS only has a finite storage space, only a certain amount of the contents can be cached at it. This makes it difficult to provide mobility support for users such as  vehicles in 5G applications that may move across different small cells rapidly. To resolve this issue, UAV-enabled caching is a potential solution thanks to the UAV's high mobility \cite{wang2018power,zhao2018caching}. Specifically,  UAVs can dynamically cache the popular contents and track the mobility pattern of the corresponding users so as to  effectively serve them. As compared to caching at fixed terrestrial  BSs, the UAV-enabled caching avoids the need of caching the same requested content at different BSs for serving a moving user and thus greatly saves  the storage resource. The results in \cite{chen2017caching} have shown that such a scheme achieves significant performance gains in terms of both the average transmit power and the percentage of the users with satisfied quality-of-experience (QoE) compared with the benchmark case without the use of UAV caching.

 However, the performance of the UAV-aided caching system is practically limited by the endurance of UAVs. To overcome this issue, \cite{xu2018overcoming} proposed a promising solution by jointly exploiting the D2D communications among the ground users  and their proactive caching.  Specifically, a UAV is dispatched to serve a group of ground users with random and asynchronous file requests and each service period is divided into two phases, i.e., the file caching phase and the file retrieval phase. In the first phase, the UAV proactively transmits each  file to a subset of selected users that cooperatively cache all the files of interest during that period, while in the second phase,  a requested file by a ground user can be retrieved either  from its own local cache directly or from its nearest neighbor via D2D communication. As such, the UAV is only needed in the first phase and the saved time  can be used for its battery charging or conducting other missions. 
\subsection{MmWave Communication}
 By exploiting the enormous chunks of new spectrum available at 30-300 GHz, mmWave communications are expected to  push  the  mobile  data  rates  to  tens  of  Gbps for  supporting  emerging  rate-demanding applications such as ultra-high definition video (UHDV) streaming  and  virtual/augmented reality (VR/AR)-based gaming. Although mmWave communication in general suffers high propagation  loss and is vulnerable to blockage, such issues are less severe when mmWave  is applied for UAV communications, thanks to the flexible UAV mobility and favorable  air-ground channel characteristics. For example, by exploiting the controllable UAV mobility, the communication distance can be significantly shortened, which not only reduces signal attenuation loss, but also enables high probability of LoS channels \cite{gapeyenko2018effects,yu2018capacity}. Furthermore, via smart positioning e.g., adjusting the  altitude, the UAV is able to bypass the obstacles such as high-rise buildings and trees that may induce blockage in the mmWave UAV communications.
Unfortunately, the high UAV mobility and the high operating carrier frequency make the Doppler frequency compensation a  critical issue for mmWave UAV communications.

Furthermore, although more antennas can be equipped at the UAV and/or ground node given the same size thanks to the smaller mmWave signal wavelength, the large-array beamforming gain is achievable only when  efficient channel estimation and tracking can be implemented.
The beam training with hierarchical beamforming codebooks has been shown to be an effective technique to achieve this  goal  \cite{948}, especially for LoS-dominant air-ground channels. However, the existing beam training algorithms are mostly designed for estimating the beam direction in azimuth domain only. Recently, a  channel tracking method  for the flight control system (FCS) was proposed in \cite{zhao2018channel} for UAV communications with mmWave MIMO. Specifically, a 3D geometry-based channel model was constructed by combining  the UAV movement state information and the channel gain information, where the former can be obtained by the sensor fusion of the FCS, while the latter can be estimated through the pilot signal. The proposed method has been  shown to have a much lower training overhead compared to the existing method without utilizing the UAV movement information.  Nevertheless, more research efforts are still needed to design the efficient channel/beam training and tracking  techniques catering  for 3D mmWave air-ground channels.
\subsection{Mobile  Edge Computing}
The concept  of mobile edge computing (MEC) was mainly motivated by the emerging new applications such as the VR/AR and autonomous driving, which usually demand ultra-low-latency communication, computation, and control among a large number of wireless devices. While the real-time computation tasks to be executed can be quite intensive, wireless devices are generally of small size and only have limited computation and data storage resources. As such, MEC has been considered as a key technology for enhancing  the computational capabilities of small devices by allowing them to offload the computation tasks to nearby MEC servers (e.g., APs and BSs). However, for users located at cell edge, such an offloading strategy may even cause more transmission energy and/or longer delay  than local computation due to the limited communication rate with the AP/BS.  To address this problem,  UAVs with highly controllable mobility can be used as the flying cloudlets to achieve more efficient computation offloading for the users by moving significantly closer to them \cite{jeong2016mobile,1052,zhou2018uav,hua2018energy,zhou2018computation,hu2018uav,zhang2018energy}.

On the other hand, in practice, small UAVs may also have the need to offload the computation tasks to ground BSs in cellular-connected UAVs. By exploiting its LoS dominant links with many ground BSs, a UAV user can  simultaneously connect with multiple GBSs to exploit their distributed computing resources to improve the computation offloading performance  \cite{cao2018mobile}.  In \cite{cao2018mobile}, it has been shown  that when the number of task-input bits is sufficiently large, the UAV should hover above its associated  GBSs  in order to achieve the most efficient computation offloading. However, if the UAV's propulsion  energy consumption is  taken into account, this result may not  hold, which thus requires further investigation.

\subsection{Wireless Power Transfer}
RF transmission enabled wireless power transfer (WPT) is envisioned as a promising solution to provide perpetual energy supplies for massive low-power devices in the forthcoming IoT networks \cite{clerckx2019fundamentals,888}. To compensate  the significant  signal attenuation over distance, a variety of techniques have been proposed to enhance the WPT efficiency,  including transmit beamforming/precoding, waveform optimization, energy scheduling, etc. However, the efficiency of WPT is still fundamentally limited by the distances between energy transmitters (ETs) and energy receivers (ERs) \cite{wu2016overview,qing15_wpcn_twc}.

To solve this problem, UAV-mounted ETs can be employed  to dramatically reduce the link distance by exploiting their highly controllable mobility in 3D space \cite{956,wu2018uav,yin2018uav,park2018minimum}. By moving close to the ERs with clear LoS links, the UAV-ET can significantly improve the efficiency of WPT to ERs, similarly like in wireless communication.  As the energy signals from the ET are broadcast to all ERs, the energy harvested at each ER critically depends on the UAV location/trajectory. In \cite{956}, it was shown  that to maximize the total harvested energy at all ERs, the UAV-ET with one single omnidirectional antenna should hover at one fixed location during the whole charging period. However, this may lead to unfair harvested energy among ERs due to their different distances from the UAV. To tackle this issue, the problem of maximizing the minimum energy harvested among all ERs was also considered in \cite{956}, where a successive hover-and-fly trajectory was shown to be optimal. However, how to extend the work \cite{956} to the more general setup with multiple and/or multi-antenna UAV-ETs is still not addressed yet.
To enable energy as well as information transfer,   single-antenna  UAV-enabled  wireless powered communication network (WPCN) and simultaneous wireless information and power transfer (SWIPT) system were studied in \cite{1072,yin2018uav} and \cite{park2018minimum}, respectively, all of which  have shown  that a joint design of the UAV trajectory and energy/communication scheduling can achieve significant performance gains as compared to the case with fixed UAV locations.
\section{Conclusions}\label{sec:conclusions}
In this paper, we provide a tutorial on UAV communication in  5G-and-beyond wireless systems, by addressing its main challenges due to the unique communication requirements and channel characteristics, as well as the new considerations such as UAV energy limitation, high altitude and high 3D mobility.  We first present the fundamental mathematical models useful for the performance analysis, evaluation and optimization  of UAV communication, including the channel and antenna models, UAV energy consumption models, as well as the mathematical optimization framework for UAV communication and trajectory co-design. The state-of-the-art results are then reviewed for the two main research and application paradigms of UAV communication, namely UAV-assisted terrestrial communications  and cellular-connected UAVs. We also highlight the promising directions in UAV communication and other related areas worthy of further investigation in future work. It is hoped that this paper will be a useful and inspiring resource for researchers working in this promising area to unlock the full potential of wireless communication meeting UAVs.

\bibliographystyle{IEEEtran}
\bibliography{IEEEabrv,IEEEfull,mybib}

\begin{IEEEbiography}[{\includegraphics[width=1in,height=1.25in,clip,keepaspectratio]{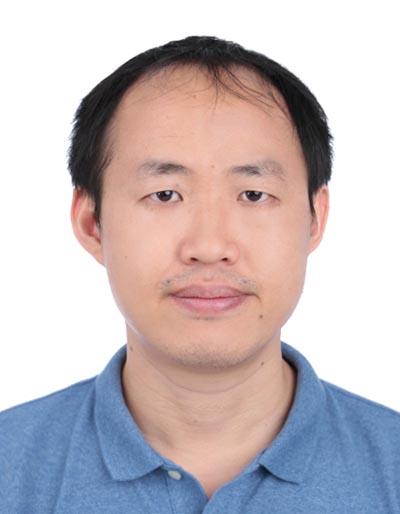}}]{Yong Zeng }
(S'12-M'14) is a Lecturer at the School of Electrical and Information Engineering, The University of Sydney, Australia. He received the Bachelor of Engineering (First-Class Honours) and Ph.D. degrees from the Nanyang Technological University, Singapore, in 2009 and 2014, respectively. From 2013 to 2018, he was a Research Fellow and Senior Research Fellow at the Department of Electrical and Computer Engineering, National University of Singapore. His research interests include UAV communications, wireless power transfer, massive MIMO and millimeter wave communications.

Dr. Zeng is the recipient of the Australia Research Council (ARC) Discovery Early Career Researcher Award (DECRA), the 2018 IEEE Communications Society Asia-Pacific Outstanding Young Researcher Award, 2017 IEEE Communications Society Heinrich Hertz Prize Paper Award, 2017 IEEE Transactions on Wireless Communications Best Reviewer, 2015 and 2017 IEEE Wireless Communications Letters Exemplary Reviewer, and the Best Paper Award for the 10th International Conference on Information, Communications and Signal Processing. He serves as an Associate Editor of IEEE Access, Leading Guest Editor of IEEE Wireless Communications on ``Integrating UAVs into 5G and Beyond'' and China Communications on ``Network-Connected UAV Communications''. He is the workshop co-chair for ICC 2018, ICC2019 workshop on UAV communications.

\end{IEEEbiography}

\begin{IEEEbiography}[{\includegraphics[width=1.8in,height=1.25in,clip,keepaspectratio]{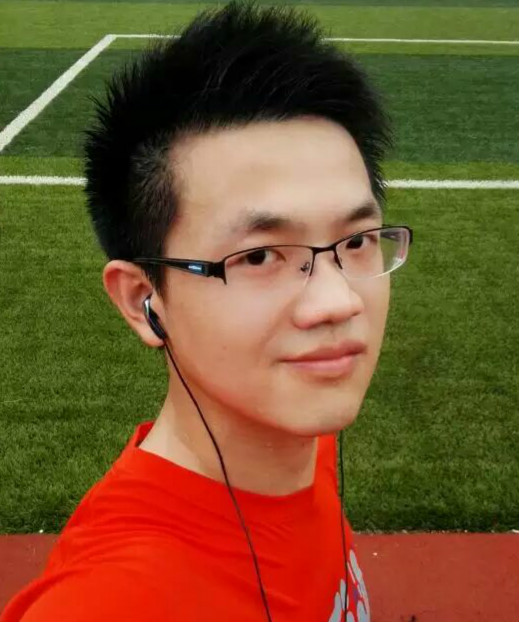}}]{Qingqing Wu}
(S'13-M'16) received the B.Eng. and the Ph.D. degrees in Electronic Engineering from South China University of Technology and Shanghai Jiao Tong University (SJTU) in 2012 and 2016, respectively.  He is currently a Research Fellow in the Department of Electrical and Computer Engineering at National University of Singapore. He was the recipient of the Outstanding Ph.D. Thesis Funding in SJTU in 2016 and the Outstanding Ph.D. Thesis Award of China Institute of Communications in 2017. He received the IEEE WCSP Best Paper Award in 2015, the Exemplary Reviewer of IEEE Wireless Communications Letters, IEEE Communications Letters, IEEE Transactions on Communications and IEEE Transactions on Wireless Communications.  His research interests include intelligent reflecting surface (IRS), unmanned aerial vehicle (UAV) communications,  energy-efficient wireless communications, and convex and nonconvex optimization.
\end{IEEEbiography}

\begin{IEEEbiography}[{\includegraphics[width=1in,height=1.25in,clip,keepaspectratio] {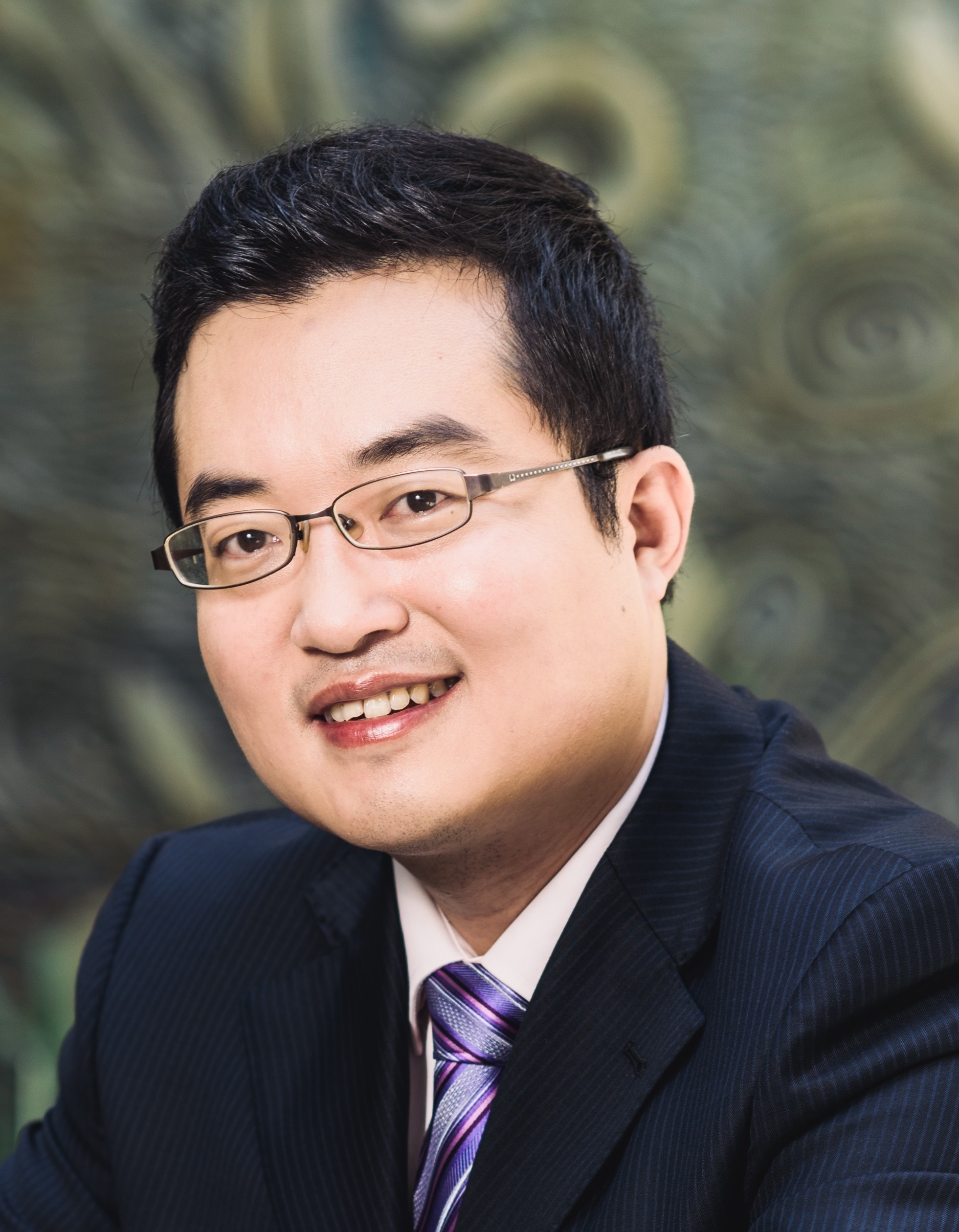}}]{Rui Zhang}
 (S'00-M'07-SM'15-F'17) received the B.Eng. (first-class Hons.) and M.Eng. degrees from the National University of Singapore, Singapore, and the Ph.D. degree from the Stanford University, Stanford, CA, USA, all in electrical engineering.

From 2007 to 2010, he worked as a Research Scientist with the Institute for Infocomm Research, ASTAR, Singapore. Since 2010, he has joined the Department of Electrical and Computer Engineering, National University of Singapore, where he is now a Dean's Chair Associate Professor in the Faculty of Engineering. He has authored over 300 papers. He has been listed as a Highly Cited Researcher (also known as the World's Most Influential Scientific Minds), by Thomson Reuters since 2015. His research interests include wireless information and power transfer, drone communication, wireless eavesdropping and spoofing, energy-efficient and energy-harvesting-enabled wireless communication, multiuser MIMO, cognitive radio, and optimization methods.

He was the recipient of the 6th IEEE Communications Society Asia-Pacific Region Best Young Researcher Award in 2011, and the Young Researcher Award of National University of Singapore in 2015. He was the co-recipient of the IEEE Marconi Prize Paper Award in Wireless Communications in 2015, the IEEE Communications Society Asia-Pacific Region Best Paper Award in 2016, the IEEE Signal Processing Society Best Paper Award in 2016, the IEEE Communications Society Heinrich Hertz Prize Paper Award in 2017, the IEEE Signal Processing Society Donald G. Fink Overview Paper Award in 2017, and the IEEE Technical Committee on Green Communications \& Computing (TCGCC) Best Journal Paper Award in 2017. His coauthored paper received the IEEE Signal Processing Society Young Author Best Paper Award in 2017. He served for over 30 international conferences as TPC Co-Chair or Organizing Committee Member, and as the guest editor for 3 special issues in IEEE Journal of Selected Topics in Signal Processing and IEEE Journal on Selected Areas in Communications. He was an elected member of the IEEE Signal Processing Society SPCOM (2012-2017) and SAM (2013-2015) Technical Committees, and served as the Vice Chair of the IEEE Communications Society Asia-Pacific Board Technical Affairs Committee (2014-2015). He served as an Editor for the IEEE TRANSACTIONS ON WIRELESS COMMUNICATIONS (2012-2016), the IEEE JOURNAL ON SELECTED AREAS IN COMMUNICATIONS: Green Communications and Networking Series (2015-2016), and the IEEE TRANSACTIONS ON SIGNAL PROCESSING (2013-2017). He is now an Editor for the IEEE TRANSACTIONS ON COMMUNICATIONS and the IEEE TRANSACTIONS ON GREEN COMMUNICATIONS AND NETWORKING. He serves as a member of the Steering Committee of the IEEE Wireless Communications Letters. \end{IEEEbiography}

\end{document}